\documentclass[%
 reprint,
 superscriptaddress,
 amsmath,amssymb,
 aps,longbibliography
]{revtex4-1}

\usepackage{graphicx}
\usepackage{amsmath}
\usepackage{xcolor}

\newcommand{\ket}[1]{\mathinner{|#1\rangle}}

\begin{document}

\title{Universal non-adiabatic control of small-gap superconducting qubits}

\author{Daniel L. Campbell}
\email{Daniel.Campbell.22@us.af.mil}
\altaffiliation{Present address: Air Force Research Laboratory, Information Directorate, Rome, NY, 13441, USA}
\affiliation{Research Laboratory of Electronics, Massachusetts Institute of Technology, Cambridge, MA 02139, USA}
\author{Yun-Pil Shim}
\altaffiliation{Present address: Department of Physics, University of Texas at El Paso, El Paso, Texas 79968, USA}
\affiliation{Laboratory for Physical Sciences, College Park, MD 20740, USA}
\affiliation{Department of Physics, University of Maryland, College Park, MD 20740, USA}
\author{Bharath Kannan}
\affiliation{Research Laboratory of Electronics, Massachusetts Institute of Technology, Cambridge, MA 02139, USA}
\affiliation{Department of Electrical Engineering and Computer Science, Massachusetts Institute of Technology, Cambridge, MA 02139, USA}
\author{Roni Winik}
\affiliation{Research Laboratory of Electronics, Massachusetts Institute of Technology, Cambridge, MA 02139, USA}
\author{David K. Kim}
\affiliation{MIT Lincoln Laboratory, 244 Wood Street, Lexington, MA 02420, USA}
\author{Alexander Melville}
\affiliation{MIT Lincoln Laboratory, 244 Wood Street, Lexington, MA 02420, USA}
\author{Bethany M. Niedzielski}
\affiliation{MIT Lincoln Laboratory, 244 Wood Street, Lexington, MA 02420, USA}
\author{Jonilyn L. Yoder}
\affiliation{MIT Lincoln Laboratory, 244 Wood Street, Lexington, MA 02420, USA}
\author{Charles Tahan}
\affiliation{Laboratory for Physical Sciences, College Park, MD 20740, USA}
\author{Simon Gustavsson}
\affiliation{Research Laboratory of Electronics, Massachusetts Institute of Technology, Cambridge, MA 02139, USA}
\author{William D. Oliver}
\email{william.oliver@mit.edu}
\affiliation{Research Laboratory of Electronics, Massachusetts Institute of Technology, Cambridge, MA 02139, USA}
\affiliation{Department of Electrical Engineering and Computer Science, Massachusetts Institute of Technology, Cambridge, MA 02139, USA}
\affiliation{MIT Lincoln Laboratory, 244 Wood Street, Lexington, MA 02420, USA}
\affiliation{Department of Physics, Massachusetts Institute of Technology, Cambridge, MA 02139, USA}

\date{\today}

\begin{abstract}

Resonant transverse driving of a two-level system as viewed in the rotating frame couples two degenerate states at the Rabi frequency, an equivalence that emerges in quantum mechanics.
While successful at controlling natural and artificial quantum systems, certain limitations may arise (e.g., the achievable gate speed) due to non-idealities like the counter-rotating term.
We introduce a superconducting composite qubit (CQB), formed from two capacitively coupled transmon qubits, which features a small avoided crossing -- smaller than the environmental temperature -- between two energy levels.
We control this low-frequency CQB using solely baseband pulses, non-adiabatic transitions, and coherent Landau-Zener interference to achieve fast, high-fidelity, single-qubit operations with Clifford fidelities exceeding 99.7\%.
We also perform coupled qubit operations between two low-frequency CQBs.
This work demonstrates that universal non-adiabatic control of low-frequency qubits is feasible using solely baseband pulses.

\end{abstract}
\maketitle

Variations on the transmon qubit~\cite{Koch2007} and the capacitively shunted flux qubit~\cite{Yan2016} have come to form the foundation for contemporary superconducting quantum computing~\cite{Devoret_review2013,Oliver_review2013,Gambetta_review2017,Wendin_review2017,Kjaergaard_review2019} and explorations of quantum mechanics in solid-state systems.
In the context of quantum control, we generally view superconducting qubits as ``artificial atoms'': electrical circuits that exhibit quantum states and energy levels similar in many respects to those present in natural atoms.
It is then a straightforward extension to use a resonant, transverse field -- typically at microwave frequencies -- to drive transitions between states and thereby perform qubit operations.
For superconducting qubits~\cite{Krantz_apr2019}, with their large electric or magnetic dipole moments, this approach has worked remarkably well, enabling single-qubit gate fidelities that exceed 99.9\% and two-qubit fidelities that are not far behind~\cite{Kjaergaard_review2019,Arute2019,Kjaergaard2020}.
However, as architectures scale and qubit numbers increase, it becomes increasingly challenging to route microwave control signals in higher-density circuits while avoiding unwanted crosstalk.
Reducing the qubit frequency -- and thereby the resonant drive frequency -- helps mitigate capacitive crosstalk, but at the expense of the achievable Rabi frequency (gate speed) before non-idealities like the counter-rotating term come into play.
Furthermore, for qubit frequencies below the environmental temperature, one may question whether such operation is even practically feasible due to the excess excited-state population in equilibrium and the resulting need for fast gates to polarize (initialize) the qubit in its ground state.

Spin-based qubits in semiconductors offer an alternative path forward.
Resonant single-qubit operations based on magnetic-field driving of spin qubits are relatively slow, necessitating large driving amplitudes~\cite{Koppens2006,Veldhorst2014} and, in conjunction with the relatively small qubit size (high qubit density), result in excessive microwave crosstalk.
Consequently, from the very beginning, the spin-qubit community has instead largely relied on effective, encoded qubits comprising two or more individual spins and their exchange interactions~\cite{divincenzo_bacon_nature2000}.
The exchange interaction enables fast encoded qubit gates controlled solely using baseband pulses, alleviating the need for pulsed-microwave control signals, whose shortcomings include expense, speed limitations associated with qubit anharmonicity, and frequency-dependent cross-talk compensation.
Additionally, the qubit encoding features a degree of immunity to global field fluctuations~\cite{Bacon2000,Kempe2001}.
This physics, which occurs naturally for spin systems\textcolor{black}{, and was in fact used in several early demonstrations with superconducting charge qubits~\cite{Nakamura1999,Pashkin2003,Yamamoto2003},} motivates us to explore analogous forms of encoding and quantum control for small-gap superconducting systems~\cite{Shim2016}.

\begin{figure*}
\includegraphics[width=7in]{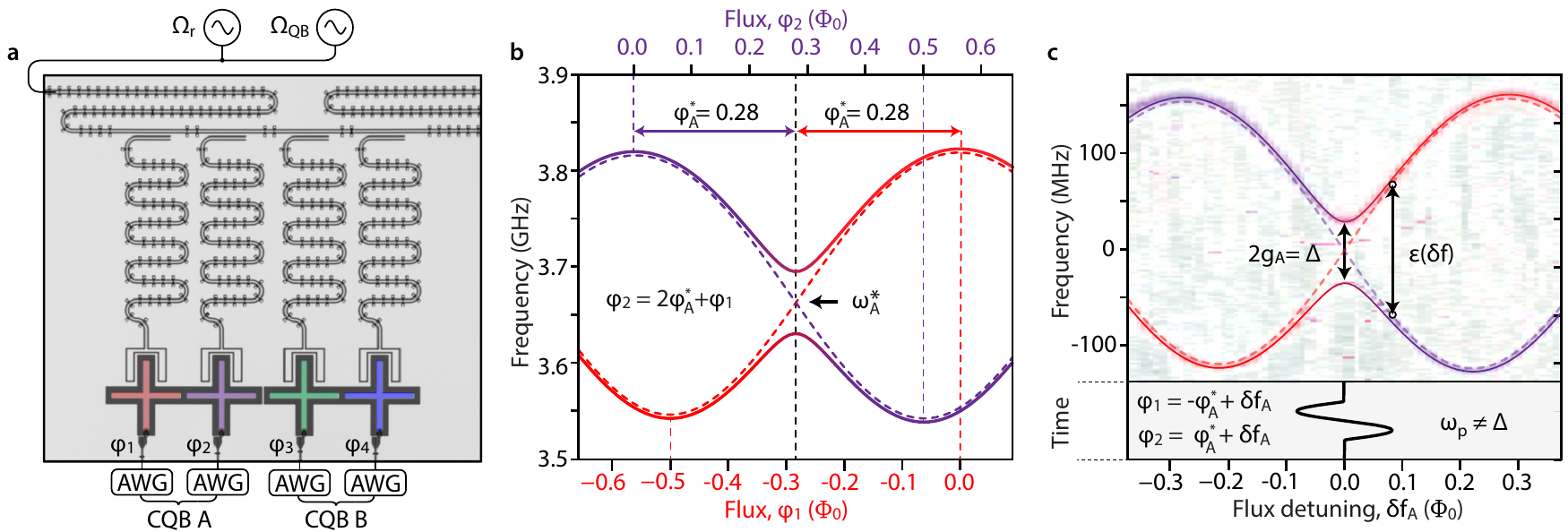}
    \caption{\label{fig:fig1}
    \textbf{Device and control. a)} Optical micrograph of two composite qubits (CQB-A and CQB-B) comprising four transmon qubits (1, 2, 3, and 4) with nearest neighbor capacitive coupling. A microwave feedline allows frequency multiplexed readout and state preparation via the microwave driving fields $\Omega_{\mathrm{r}}$, and $\Omega_{\mathrm{QB}}$, respectively.
    \textbf{b)} Eigenenergies of \textcolor{black}{individual (dashed lines) and coupled (solid lines) asymmetric-junction transmons} 1 and 2 (CQB-A). An avoided level crossing occurs when the coupled transmons are resonantly biased at $\varphi_{1,2} = \pm \varphi^*_{\mathrm{A}}$.
    \textbf{c)} (Upper panel) Corresponding measured excited-state spectroscopy centered at zero frequency (offset in frequency) to form a two-level system model with parameters $\Delta$ and $\varepsilon(\delta f)$. Near the avoided crossing region, $\varepsilon$ is proportional to the flux detuning $\delta f$, realized by simultaneously biasing $\varphi_1$ and $\varphi_2$. (Lower panel) Non-adiabatic control is implemented by applying a single period of a non-resonant ($\omega_p \neq \Delta$) sinusoidal excursion about the avoided crossing.
    }
\end{figure*}

In this work, we demonstrate universal control of a superconducting composite qubit (CQB) using solely baseband pulses reliant on non-adiabatic, Landau-Zener transitions and quantum interference~\cite{Oliver_Science2005,Sillanpaa2006,LZ_PhysRep2010}.
The CQB comprises two coupled transmon qubits and features a gap ($\Delta / 2 \pi \approx 65$ MHz) that is appropriately sized for such baseband control.
The small gap reduces the relaxation rate in the computational basis~\cite{Oliver_Science2005,Yan2016}, and the composite nature of the CQB design features resilience to both environmental flux noise~\cite{Vion2002,Yoshihara2006,Bylander2011} and photon shot noise from the readout resonator~\cite{Schuster2005,Sears2012,Yan2018}.
We present a tune-up protocol for single-CQB and two-CQB gates and benchmark their performance, achieving 99.7\% single-qubit average Clifford fidelity.
Although demonstrated with CQBs, the non-adiabatic control protocols
demonstrated here are generally applicable to
quantum systems featuring small gaps.

Our test device comprises four asymmetric superconducting transmon qubits~\cite{Hutchings2017} of the ``xmon'' geometry~\cite{xmon} with fixed, nearest-neighbor capacitive coupling (Fig.~\ref{fig:fig1}a).
Pairs of transmons are grouped to form the composite qubits used here, denoted ``CQB-A'' and ``CQB-B''.
Qubit spectroscopy of CQB-A (Fig.~\ref{fig:fig1}b) shows the constituent transmon spectra of the ground-state $\ket{g_i}$ to excited-state $\ket{e_i}$ transitions for $i=1,2$ as a function of the reduced flux biases $\varphi_{i}\equiv \Phi_i/\Phi_0$, where $\Phi_i$ is the magnetic flux and $\Phi_0$ is the superconducting flux quantum.
Similar spectra are observed for CQB-B and transmons $i=3,4$~\cite{supp-mat}.

When the transmons are biased at the same frequency, $\omega_1 = \omega_2 \equiv \omega_{\textrm{A}}^*$, an avoided crossing $\Delta = 2g_{\mathrm{A}}$ opens due to the fixed capacitive coupling within CQB-A of strength $g_{\mathrm{A}}$.
The size of the avoided crossing, $\Delta/2 \pi \approx 65$ MHz, is determined predominantly by the value of the coupling capacitance, but its location -- centered at frequency $\omega_{\textrm{A}}^*(\varphi^*_{\mathrm{A}})$ -- can be generally chosen along the transmon spectra at flux biases $\varphi_{1,2} = \mp \varphi^*_{\mathrm{A}}$.
In Fig.~\ref{fig:fig1}b, we have chosen $\varphi^*_{\mathrm{A}}=0.28$.
More generally, CQB-A can be flux-biased over its entire frequency range using the individual transmon biases $\varphi_1$ and $\varphi_2 = 2\varphi^*_{\mathrm{A}} + \varphi_1$, and similarly for CQB-B and its transmons.

\begin{figure*}
\includegraphics[width=7in]{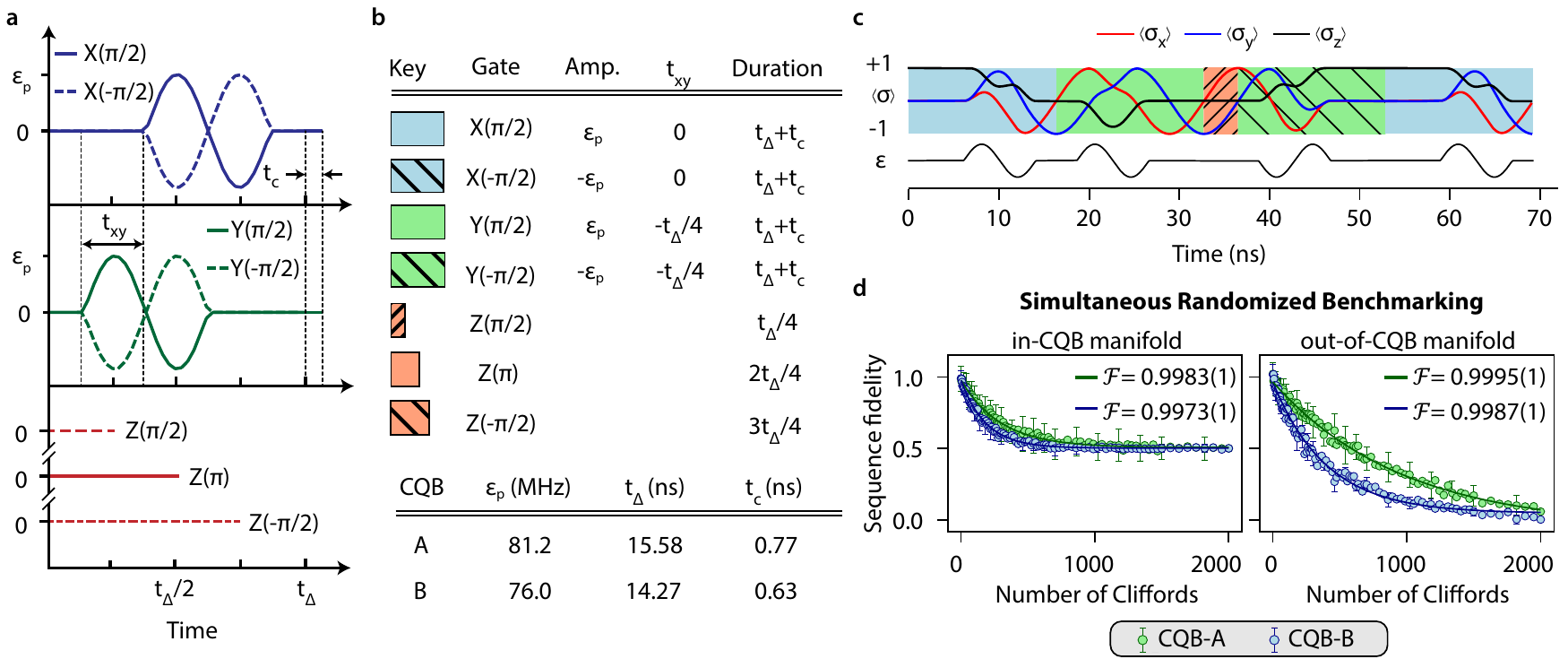}
    \caption{\label{fig:fig2}
    \textbf{Single-CQB gates. a)} Pulses for X, Y, and Z gates. For $X(\pm \pi/2)$ and $Y(\pm \pi/2)$ gates, the sinusoidal pulse is applied within a time window with a relative shift $t_{\mathrm{xy}} = t_{\Delta}/4 = 2\pi / 4\Delta$, which establishes the $x$ and $y$ axes. The $X(\pm \pi/2)$ and $Y(\pm \pi/2)$ gates require an additional $Z$ rotation of duration $t_{\mathrm{c}}$ that corrects for small parasitic $Z$ evolution during the gate. $Z$ gates are realized by idling at the avoided crossing for the appropriate fraction of the precession period $1/t_{\Delta}$.
    \textbf{b)} Table of parameters $\varepsilon_{\mathrm{p}}$, $t_{\mathrm{xy}}$, $t_{\Delta}$, and $t_{\mathrm{c}}$ for the various $X$, $Y$, and $Z$ gates, and the calibrated values for CQB-A and CQB-B.
\textbf{c)}
     Examples of concatenated gates and simulations of the resulting Bloch vector projections on $\langle\sigma_x\rangle$, $\langle\sigma_y\rangle$, and $\langle\sigma_z\rangle$.
    \textbf{d)} Simultaneous randomized benchmarking traces corresponding to Clifford fidelity (errors within the computational subspace) and leakage Clifford fidelity (errors that leave the computational subspace).}
\end{figure*}
The CQB subspace, given by the avoided crossing in Fig. 1c, is described by the standard two-level-system Hamiltonian
\begin{equation}
    \hat{H}/ \hbar = -\frac{1}{2} \left[
        \Delta \hat{\sigma}_z + \varepsilon \hat{\sigma}_x \right],
    \label{singleEQ}
\end{equation}
where $\hbar=h/2\pi$ with $h$ being Planck's constant, and $\hat{\sigma}_x$ and $\hat{\sigma}_z$ are Pauli operators.
For highly asymmetric transmons, the parameter $\varepsilon = 2 \delta \omega \sin (2 \pi \varphi^*_{\mathrm{A}}) \sin (2 \pi \delta f_{\mathrm{A}})$ is the difference between the bare transmon frequencies referenced with respect to the avoided crossing through the flux detuning $\delta f_{\mathrm{A}} \equiv \varphi_{1,2} \mp \varphi^*_{\mathrm{A}}$, with $\delta \omega$ defined as the difference between maximum and minimum transmon frequencies~\cite{Gramajo2019}.
Near the avoided crossing, $\varepsilon \approx 4 \pi \delta\omega \sin (\varphi^*_{\mathrm{A}}) \delta f_{\mathrm{A}}$ is approximately a linear function of $\delta f_{\mathrm{A}}$, reminiscent of the persistent current flux qubit~\cite{Mooij1999,Orlando1999} (see also supplementary material~\cite{supp-mat}).
Although $\Delta$ transversally couples the bare (diabatic) transmon states, we have elected to associate $\Delta$ with $\hat{\sigma}_z$, as the computational basis $\{\ket{0},\ket{1}\}$ is defined at the avoided crossing.
At this bias point, the coupling hybridizes the bare transmon states to form the CQB computational states
$\ket{0} , \ket{1}= \ket{g_1,e_2} \pm \ket{e_1,g_2}$.

Initializing the CQB in state $\ket{0}$ does not require a precise knowledge of $\omega^*_{A,B}$ or microwave mixer calibration. Beginning with both transmons in their ground states $\ket{g_1,g_2}$, the CQB is biased far from $\varepsilon=0$ (transmon frequency degeneracy). Then, in the presence of a continuous-wave microwave drive, the system is further detuned such that one of the transmons (e.g., transmon 1) adiabatically passes through resonance with the drive, which excites the CQB to diabatic state $\ket{e_1,g_2}$. The field is then turned off, and the CQB is adiabatically ramped back to the degeneracy point, initializing the qubit in state $\ket{0}$~\cite{supp-mat}.

\begin{figure}
\includegraphics[width=3.3in]{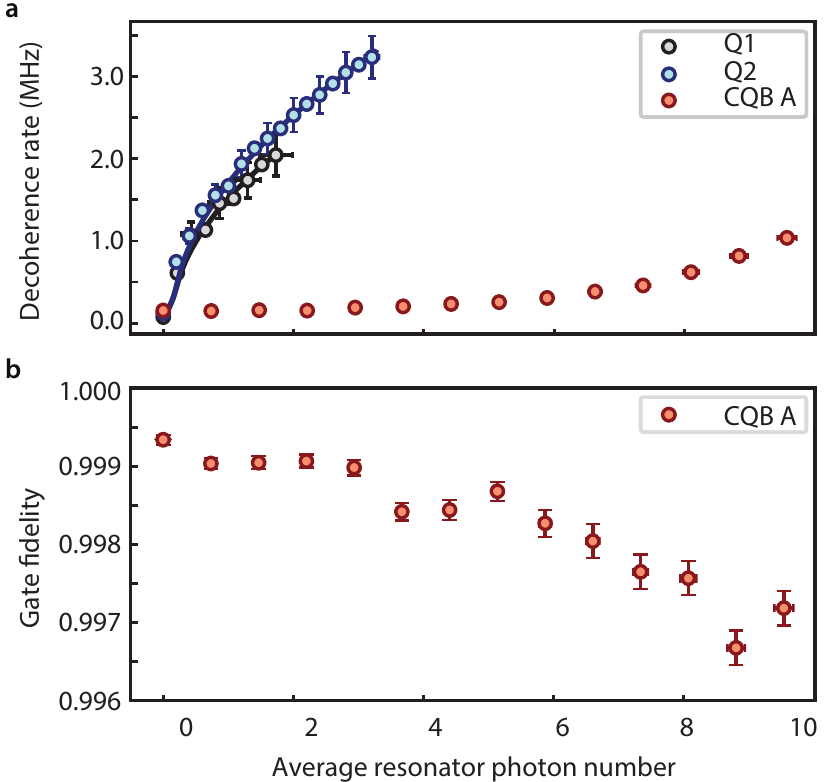}
    \caption{\label{fig:fig4}
    \textbf{Photon shot noise characterization.
    a)} Ramsey decoherence rate. The CQB is protected by design against photon shot noise, exhibiting a much lower decay rate (see text).
    \textbf{b)} This protection extends to operations applied to the encoded qubit, as indicated by the slow falloff of the randomized benchmarking Clifford fidelity with the number of photons in the resonator.}
\end{figure}

CQB readout is performed by adiabatically detuning the CQB away from the avoided crossing, such that the hybridized computational states $\ket{0}$ and $\ket{1}$ are uniquely mapped onto the bare (diabatic) transmon states $\ket{e_1,g_2}$ and $\ket{g_1,e_2}$. Doing so enables CQB readout using standard dispersive readout on the underlying transmons~\cite{supp-mat}. As we describe in the discussion surrounding Fig.~\ref{fig:fig4}, although dispersive transmon measurement performed at degeneracy cannot distinguish CQB states $\ket{0}$ and $\ket{1}$, it has the useful property that it can be used to detect leakage out of the CQB subspace without destroying the CQB quantum information.

When applied to small-gap qubits, resonant excitation in the perturbative Rabi-driving regime $\varepsilon(\delta f) \ll \Delta$ results in nutation periods $\tau \gg 1/\Delta \approx 15 \text{ ns}$ and leads to prohibitively slow qubit gates.
A better approach is to use a \textrm{non-resonant} baseband pulse that sweeps the parameter $\varepsilon$ of a qubit around and through a transverse avoided crossing of size $\Delta$.
For sufficiently large driving amplitudes $\varepsilon(\delta f_j) > \Delta$, these excursions cause coherent, non-adiabatic transitions, which in conjunction with quantum interference, lead to controllable state transitions on a time-scale that can approach the speed limit for the system, $\tau \sim 1/\Delta$.
This effect, known as Landau-Zener-Stueckelberg interference, has been demonstrated in both natural and artificial atomic systems~\cite{LZ_PhysRep2010}, including demonstrations of St\"{u}ckelberg interferometry~\cite{Oliver_Science2005,Sillanpaa2006,Berns2006}, qubit cooling~\cite{Valenzuela2006}, amplitude spectroscopy~\cite{Berns2008}, temporal oscillations~\cite{Berns2008,Bylander2009}, and its use in the quantum simulation of universal conductance fluctuations~\cite{Gustavsson2013} and weak localization~\cite{Gramajo2019}.
In this ``strong-driving'' regime, the trajectory of the Bloch vector is no longer a simple function of the amplitude, frequency, or phase of a sinusoidal drive (as it is in the Rabi-driving case), necessitating an alternative gate-calibration protocol. Our approach begins with the CQB prepared in state $\ket{0}$ at the avoided crossing, where it is first-order protected from flux noise~\cite{Vion2002}, and we use a single-period sinusoidal pulse to implement quantum control (see Fig~\ref{fig:fig1}c).

To gain intuition, we first note that a \textcolor{black}{large}-amplitude, solely-diabatic excursion away from the avoided crossing effectively performs a 50:50 beamsplitting operation, projecting state $\ket{0}$ on to an equal superposition of the diabatic states $\ket{g_1,e_2}$ and $\ket{e_1,g_2}$ (dashed lines).
Away from the avoided crossing, the higher-energy diabatic state accrues a relative azimuthal phase at a rate proportional to the energy separation $\varepsilon(\delta f)$.
Rapidly returning to the avoided crossing region performs a second ``beamsplitter''-type operation which again mixes the states, resulting in a general superposition state $\alpha \ket{0} + \beta \ket{1}$ depending on the accrued phase and quantum interference.
\textcolor{black}{This is conceptually similar to the Larmor control of early charge qubits~\cite{Nakamura1999,Pashkin2003,Yamamoto2003}. In those experiments, a qubit starting in a diabatic state far from its avoided crossing was rapidly pulsed to the avoided crossing region, where it underwent Larmor precession, and was then rapidly returned to its starting point.} 

In practice, we use one period of a finite amplitude sinusoid (Fig.~\ref{fig:fig1}c), $\delta f = A_{\mathrm{p}} \sin (\omega_{\mathrm{p}} t)$, that features partially diabatic excursions and incorporates the mixing and quantum interference associated with leaving, traversing, and returning to the avoided crossing region.
Due to the proximity to the avoided crossing, $\varepsilon$ is proportional to $\delta f$, and we can similarly parameterize $\varepsilon = \varepsilon_{\mathrm{p}} \sin (\omega_{\mathrm{p}} t)$ without loss of generality.
The symmetric driving protocol has the added benefit of canceling dc components associated with pulse transients, creating a ``dynamic sweet spot''.
Although this driving protocol does not likewise protect a CQB from the direct flux control crosstalk of another CQB, the fact that the pulses are essentially in the quasi-static limit results in a straightforward and essentially frequency-independent calibration matrix, highlighting a further advantage to eliminating microwave control.
The calibration protocol is then to scan the pulse amplitude and frequency to realize high-fidelity single-qubit gates (see supplementary materials for details of the procedure~\cite{supp-mat}).

$Z$-gates are realized as idling operations:
$Z(\phi(t_{\mathrm{d}}))=\exp{(-i \hat{\sigma}_z \phi(t_{\mathrm{d}})/2 )}$  where $\phi(t_{\mathrm{d}}) = \Delta t_{\mathrm{d}}$ for a gate of duration $t_{\mathrm{d}}$, as shown in Fig.~\ref{fig:fig2}a .
The gate duration $t_{\mathrm{d}}$ determines the type of $Z$ gate along a continua: increments of quarter periods in the precession period $t_{\mathrm{\Delta}} \equiv 2 \pi / \Delta$ at the avoided crossing yield the familiar gates $I$, $Z(\pm \pi/2)$ and $Z(\pi)$ (see table in Fig.~\ref{fig:fig2}b).
The timing jitter associated the baseband pulse generator is less than 2 ps, compared with the precession period $2 \pi/ \Delta \approx 15$ ns, corresponding to an error rate less than 0.02\%. This may be compared with the baseband envelope of a microwave pulse for microwave gates of similar duration.

We use $t_{\mathrm{d}}$ as the basic clocking unit for $X(\pm \pi/2)$ and $Y(\pm \pi/2)$ gates, compatible with our selected pulse frequency $\omega_{\mathrm{p}}/2\pi = 125\text{ MHz}$, such that the gates can be completed within the time window and are sufficiently nonadiabatic (see Fig.~\ref{fig:fig2}a).
The start of the $X(\pm \pi/2)$ pulse within the window is in principle arbitrary, but once chosen, it establishes the $x$ axis for the Bloch sphere.
The $y$ axis then corresponds to a $\pi/2$ phase shift, implemented by advancing the onset of the $Y(\pm \pi/2)$ gate by an amount $t_{\mathrm{xy}} = t_{\Delta}/4$, a quarter of the precession period at the avoided crossing.
We elect to start the $X(\pm \pi/2)$ and $Y(\pm \pi/2)$ pulses symmetrically about the mid-point of the pulse window $t_{\mathrm{d}}$, as shown in Fig.~\ref{fig:fig2}a.
During the operations, the $X(\pm \pi/2)$ and $Y(\pm \pi/2)$ gates may accumulate a small parasitic $Z$-component, which we can correct by padding the gate with corrective $Z$-rotations of duration $t_{\mathrm{c}}$, such that the total duration becomes $t_{\mathrm{d}} = t_{\Delta} + t_{\mathrm{c}}$.
The calibration parameters for both CQB-A and CQB-B are shown in Fig.~\ref{fig:fig2}b.

We apply these gates to benchmark the coherence properties of the CQB.
Within the CQB subspace, the standard coherence metrics are a relaxation time $T_{\textrm{1}}>2$ms, Ramsey time $T_{\textrm{2R}} \approx 8 \; \mu$s, and Hahn echo time $T_{\textrm{2E}} \approx 25 \; \mu$s.
Monte Carlo simulations of the CQB system are consistent with these times using a noise amplitude of approximately $5\mu\Phi_0/\sqrt{\text{Hz}}$ for each transmon.
The long $T_1$ time is a general feature of all small-gap qubits~\cite{Oliver_Science2005,Kerman2010,Manucharyan2009,Pop2014,Nguyen2019,Gyenis2019}, and it can be understood in the context of Fermi's Golden Rule, where the smaller gap (matrix element that couples the qubit states) translates to a reduced decay rate.
In the specific case of excitation or relaxation within the computational subspace of a CQB, a correlated two-photon interaction with the environment is needed, resulting in a relatively low decay rate.
Thus, fast, non-adiabatic control is consistent with robust qubit state initialization and operation, despite the presence of relatively hot environmental bath.
Because we can independently read out each individual transmon, we can also extract metrics accounting for leakage to states outside the CQB computational subspace (predominantly the ground state $\ket{g_1, g_2}$).
Leakage occurs on a time scale $T_{\textrm{1,leakage}}\approx 30 \; \mu$s and is comparable with the bare-transmon $T_1$.
While this is certainly an area for improvement, error correction protocols exist to address leakage errors (in any system), and as we describe below, the CQB readout affords an efficient means to detect leakage while protecting the CQB quantum information.
The coherence properties of the CQBs and their constituent transmons are tabulated in the supplementary material~\cite{supp-mat}.

The gates shown in Fig.~\ref{fig:fig2}a are concatenated sequentially in a ``back-to-back'' or ``bonded'' manner to implement multi-pulse, non-adiabatic control, realizing encoded operations on the CQB.
An example of a sequence of gates is shown in Fig.~\ref{fig:fig2}c, along with numerical simulations of the CQB Bloch vector, to illustrate the operability of this approach.
The simulations indicate that high-fidelity universal control is achievable on times scales approaching the inverse coupling strength $1/\Delta$,
which is similar in duration to state-of-the-art single-qubit microwave gates in this test sample, and much faster than could be achieved by resonant Rabi driving. The CQB single-qubit gate duration is not limited by the transmon anharmonicity and may therefore be further reduced (baseband pulse generator bandwidth permitting) by increasing $\Delta$.
We then obtain the average Clifford fidelity of these non-adiabatic gates using simultaneous randomized benchmarking (RB) on CQB-A and CQB-B, shown in Fig.~\ref{fig:fig2}d.
Both Clifford fidelities exceed 99.7\%, near state-of-the-art for conventional single-qubit microwave gates~\cite{Barends_Martinis_nature2014,Kjaergaard_review2019}, and they are approximately coherence limited.

Next, we investigate the CQBs susceptibility to various forms of noise via Eq.~\ref{singleEQ}.
The CQB is in principle linearly sensitive to fluctuations in $\Delta$, but since this frequency is generated predominantly by a lithographically defined
fixed capacitive coupling between transmons, its noise contribution is small.
Due to the avoided crossing, the CQB also exhibits the familiar first-order insensitivity to low-frequency magnetic flux noise, which enters via the transverse frequency $\epsilon$.
As a result, the CQB flux insensitivity is substantially stronger than that of the individual transmons biased at the corresponding point, $\varphi_{1,2} = \mp \phi*$. The CQB exhibits Hahn echo times exceeding 23 $\mu\text{s}$, compared to around $3 \mu\text{s}$ for the individual transmons. Furthermore, the CQB second-order sensitivity to flux noise is inversely proportional to $\Delta$ (the transmon-transmon coupling $g$) \cite{supp-mat}. This implies that, in addition to enabling faster gates, increasing $\Delta$ will also improve CQB coherence. For the circuit under consideration, the $\Delta/2\pi$ that yields an optimal balance between $T_1$ and coherence is likely in the low hundreds of MHz.

More substantially, the CQB is first-order insensitive to \textit{any} such fluctuations in the bare transmon frequencies (\textit{i.e.,} fluctuations in $\epsilon$), such as those that arise from photon number fluctuations in the readout resonator.
In the dispersive regime, resonator photon fluctuations dephase transmons through an AC Stark shift, which leads to a photon-number-dependent frequency shift $\chi$ of the qubit.
The spectrum and amplitude of such photon noise that arises from coherent driving of a resonator is well understood~\cite{Gambetta2006}.
In Fig.~\ref{fig:fig4}a, the Ramsey decoherence rate as a function of the average number of coherent photons in the resonator of CQB-A is compared with that of its bare transmons.
The CQB is substantially less sensitive to these photon number fluctuations compared with the bare transmons, and so its coherence is largely preserved.

The CQB insensitivity to photon noise in the resonator implies that the resonator cannot be used for its readout when biased at the avoided level crossing.
This is reminiscent of SQUID-based measurements of persistent current flux qubits biased at degeneracy: hybridization of the clockwise and counter-clockwise circulating currents from strong tunnel coupling prevents a relatively slow readout SQUID magnetometer from being able to distinguish between the diabatic circulating current states~\cite{Orlando1999,Chiorescu2003}.
Rather, the SQUID is sensitive to the average circulating current in the energy eigenbasis.
Similarly, when the tunnel coupling $\Delta$ between the transmons is much stronger than the resonator readout speeds, the resonators are unable to distinguish the diabatic states that hybridize into the CQB subspace.
However, importantly, the resonators \textit{are} capable of discriminating between states that are within and outside the CQB subspace.
As a result, the resonators can be continuously monitored to detect leakage \textit{without} reducing the CQB gate fidelity.
We demonstrate this resilience to a continuous readout tone in Fig.~\ref{fig:fig4}b, where the gate fidelities $\mathcal{F}$ remain nearly constant for up to 3 photons in the resonator.
Leakage detection raises the possibility for postselection or error correction  on  the  leakage  channel, allowing the $T_1$ and $T_2$ within the CQB subspace to dictate the operational fidelity (which could in turn also be further error corrected in the conventional manner).

\begin{figure}
\includegraphics[width=3.3in]{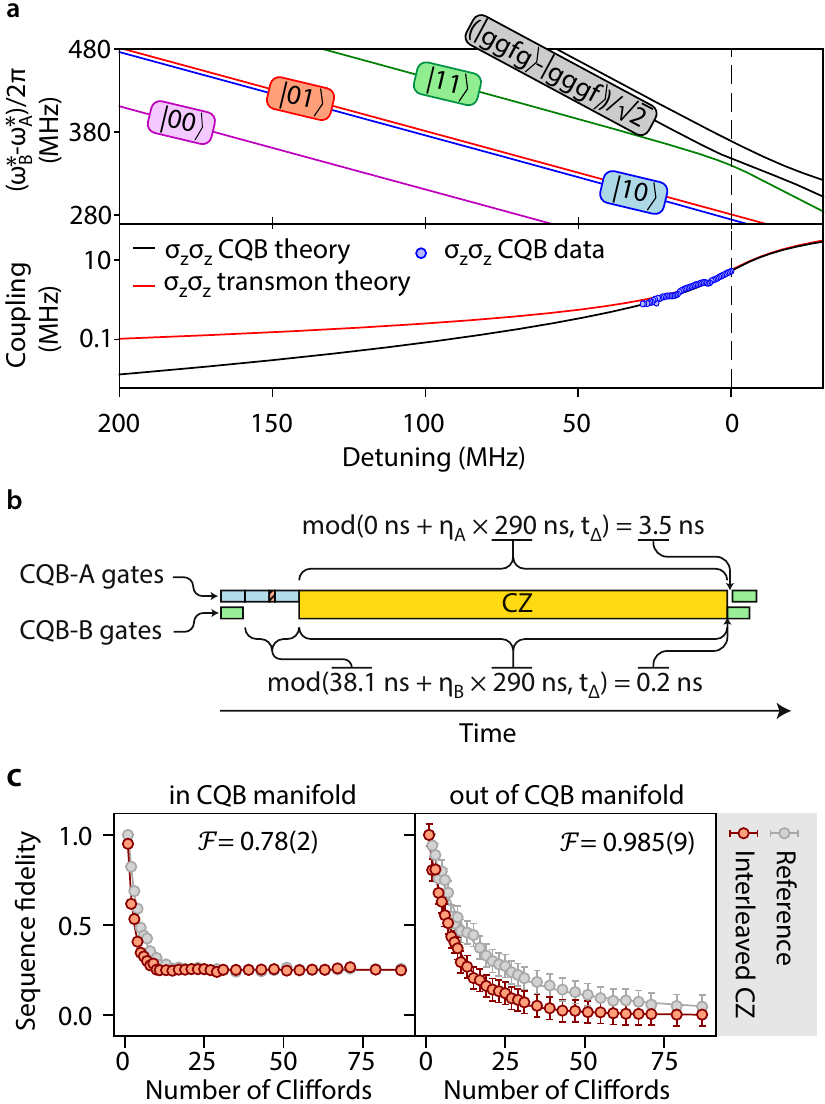}
\caption{\label{fig:fig5} \textbf{Two-CQB gates.
    a)} Controlled-Z (CZ) gate between CQB-A and CQB-B (upper panel) is performed by adiabatically ramping the CQB diabatic frequencies (at the transmon energy degeneracies)  $\omega^*_{\textrm{A}}$ and $\omega^*_{\textrm{B}}$~\cite{supp-mat}, to the avoided level crossing between $|11\rangle$ and the hybridized second excited states $(|g_1,g_2,f_3,g_4\rangle-|g_1,g_2,g_3,f_4\rangle)/\sqrt{2}$ of CQB-B's bare transmons. The CQB CZ avoided level crossing occurs when $\omega^*_{\textrm{B}} \approx \omega^*_{\textrm{A}} + (\Delta_A+\Delta_B)/2 + E_C/h$.
    (lower panel) Effective $\sigma_z\sigma_z$ coupling as a function of detuning from the avoided level crossing.
    \textbf{b)} Syncing single-CQB gates after a CZ gate is achieved by applying a compensatory Z gate, which takes into account the time of the last single-CQB gate, and phase evolution rate $\eta_{A,B}$ during the CZ.
    \textbf{c)} Two-CQB randomized benchmarking within and out of the CQB manifold.}
\end{figure}

We complete the universal gate-set for quantum computation with a CQB architecture by demonstrating a two-CQB gate.
Conventionally, a controlled-Z (CZ) gate between two transmons is realized by adiabatically tuning one of their frequencies such that its second excited state $\ket{g_1,f_2}$ hybridizes with $\ket{e_1,e_2}$, inducing a joint $ZZ$ operation~\cite{DiCarlo_nature2009}.
This operation is similarly implemented in our transmon-based CQB architecture by dynamically adjusting $\varphi^*_{A,B}$ to hybridize $\ket{11}$ with a non-computational state, as shown in Fig.~\ref{fig:fig5}a.
An important distinction, however, is that the phase between two CQBs can desynchronize from the always-on Z rotations when idling at the avoided crossing.
To keep the them synchronized, we apply corrective $Z$ operations after each CZ gate.
These corrections are easily computed given the pulse sequence, shown in Fig.~\ref{fig:fig5}b.

During the CZ gate, the CQBs are kept fully hybridized ($\delta f_{\textrm{A,B}} = 0$) such that they remain insensitive to the frequency fluctuations of their constituent transmons.
However, the system becomes sensitive to noise at the relative detuning between the two CQBs during a CZ gate.
Our CZ time was 290 ns (not including single-qubit gates), corresponding to an optimal interaction time $2\pi \times 4/g_{23} = 250\text{ ns}$ with additional 20 ns Gaussian ramps to and from the $ZZ$ operation point (Fig.~\ref{fig:fig5}\textbf{b}).
In conjunction with single-qubit gates, the measured CZ-gate fidelity was $\mathcal{F} = 0.77$, obtained by interleaved randomized benchmarking (Fig.~\ref{fig:fig5}\textbf{c}).
This admittedly moderate fidelity was due in small part to coherence limitations~\cite{supp-mat}, but was primarily related to an insufficient CZ-gate calibration. In either case, it is not due to a fundamental limitation.
The CZ gate time between two CQBs is increased by a factor of 4 relative to that between two bare transmons with the same coupling strengths.
%
Increasing the coupling between transmons 2 and 3 in Fig.~\ref{fig:fig1}a will reduce this gate time and thereby 
improve the error rate~\cite{Martinis2014}.
More importantly, since performing this work, we have focused on developing and automating the calibration methods needed to implement high-fidelity CZ gates with two transmon qubits, having recently achieved two-qubit fidelities of $99.7\%$~\cite{Kjaergaard_review2019,Kjaergaard_DME2020}.
With these calibration techniques in place and using optimized device parameters, we expect two-CQB CZ gates will achieve state of the art fidelities, as its underlying mechanics are nominally identical to a CZ gate between two bare transmons.
In fact, because CQBs are kept at a noise insensitive point throughout the CZ operation (unlike CZ gates with two bare transmons), the CQBs will experience lower dephasing rates during the CZ gate as compared to performing the same operation in conventional transmon architectures.
In conjunction with the higher coherence times within the CQB subspace, this holds the promise for even higher gate fidelities for CQBs.

Our results demonstrate that the CQB and other small-gapped qubits can serve as a building blocks for quantum computing architectures.
Using CQBs can reduce sensitivity to many common forms of noise in transmons including always-on $\sigma_z\sigma_z$ crosstalk with other qubits (see Fig.~\ref{fig:fig5}a lower panel), allowing for stronger CQB-CQB coupling and thereby enabling faster gates in future designs.
Furthermore, while transmons are susceptible to TLSs appearing near their first-order insensitive point, and cannot be detuned without admitting additional flux noise,
the transmon frequencies at which the CQB avoided crossing occurs is broadly tunable.
While this CQB architecture introduces a source of incoherent leakage out of the computational basis (via relaxation to $\ket{g_1,g_2}$), we have demonstrated that such leakage can be detected in real-time without sacrificing gate fidelity, unlike bare-transmon architectures.

More generally, qubits with small gaps -- including composite qubits -- need not compromise between control speed and protection from decoherence.
The non-adiabatic procedures demonstrated here enables the universal control of small-gap systems where conventional Rabi driving is impractical or even infeasible.
For example, our demonstration complements a recent parallel work~\cite{Zhang2020} with a small-gap fluxonium qubit~\cite{Manucharyan2009,Pop2014,Nguyen2019}, and it may be useful for other small-gap superconducting qubits such as the metastable flux qubit~\cite{Kerman2010} and the $0-\pi$ qubit~\cite{Brooks2013,Gyenis2019}.
Similarly, other systems with small or stable gaps, such as semiconductor-based spin qubits~\cite{Petta2010,Shim2016,Andrews2019}, neutral atomic systems~\cite{Bernien2017}, polar molecules~\cite{DeMille2009, Yu2019}, and laser-dressed NV-centers~\cite{Avinadav2014}, may also be controllable using these strong driving techniques.

\begin{acknowledgments}
This research was funded in part by the U.S. Army Research Office Grant No. W911NF-18-1-0116 and by the Assistant Secretary of Defense for Research \& Engineering via MIT Lincoln Laboratory under Air Force Contract No. FA8721-05-C-0002. D.L.C. gratefully acknowledges support by an appointment to the Intelligence Community Postdoctoral Research Fellowship Program at the Massachusetts Institute of Technology, administered by Oak Ridge Institute for Science and Education through an interagency agreement between the U.S. Department of Energy and the Office of the Director of National Intelligence. B.K. gratefully acknowledges support from the National Defense Science and Engineering Graduate Fellowship program.
\end{acknowledgments}

\bibliography{references}

\begin{thebibliography}{62}%
\makeatletter
\providecommand \@ifxundefined [1]{%
 \@ifx{#1\undefined}
}%
\providecommand \@ifnum [1]{%
 \ifnum #1\expandafter \@firstoftwo
 \else \expandafter \@secondoftwo
 \fi
}%
\providecommand \@ifx [1]{%
 \ifx #1\expandafter \@firstoftwo
 \else \expandafter \@secondoftwo
 \fi
}%
\providecommand \natexlab [1]{#1}%
\providecommand \enquote  [1]{``#1''}%
\providecommand \bibnamefont  [1]{#1}%
\providecommand \bibfnamefont [1]{#1}%
\providecommand \citenamefont [1]{#1}%
\providecommand \href@noop [0]{\@secondoftwo}%
\providecommand \href [0]{\begingroup \@sanitize@url \@href}%
\providecommand \@href[1]{\@@startlink{#1}\@@href}%
\providecommand \@@href[1]{\endgroup#1\@@endlink}%
\providecommand \@sanitize@url [0]{\catcode `\\12\catcode `\$12\catcode
  `\&12\catcode `\#12\catcode `\^12\catcode `\_12\catcode `\%12\relax}%
\providecommand \@@startlink[1]{}%
\providecommand \@@endlink[0]{}%
\providecommand \url  [0]{\begingroup\@sanitize@url \@url }%
\providecommand \@url [1]{\endgroup\@href {#1}{\urlprefix }}%
\providecommand \urlprefix  [0]{URL }%
\providecommand \Eprint [0]{\href }%
\providecommand \doibase [0]{http://dx.doi.org/}%
\providecommand \selectlanguage [0]{\@gobble}%
\providecommand \bibinfo  [0]{\@secondoftwo}%
\providecommand \bibfield  [0]{\@secondoftwo}%
\providecommand \translation [1]{[#1]}%
\providecommand \BibitemOpen [0]{}%
\providecommand \bibitemStop [0]{}%
\providecommand \bibitemNoStop [0]{.\EOS\space}%
\providecommand \EOS [0]{\spacefactor3000\relax}%
\providecommand \BibitemShut  [1]{\csname bibitem#1\endcsname}%
\let\auto@bib@innerbib\@empty
\bibitem [{\citenamefont {Koch}\ \emph {et~al.}(2007)\citenamefont {Koch},
  \citenamefont {Yu}, \citenamefont {Gambetta}, \citenamefont {Houck},
  \citenamefont {Schuster}, \citenamefont {Majer}, \citenamefont {Blais},
  \citenamefont {Devoret}, \citenamefont {Girvin},\ and\ \citenamefont
  {Schoelkopf}}]{Koch2007}%
  \BibitemOpen
  \bibfield  {author} {\bibinfo {author} {\bibfnamefont {Jens}\ \bibnamefont
  {Koch}}, \bibinfo {author} {\bibfnamefont {Terri~M.}\ \bibnamefont {Yu}},
  \bibinfo {author} {\bibfnamefont {Jay}\ \bibnamefont {Gambetta}}, \bibinfo
  {author} {\bibfnamefont {A.~A.}\ \bibnamefont {Houck}}, \bibinfo {author}
  {\bibfnamefont {D.~I.}\ \bibnamefont {Schuster}}, \bibinfo {author}
  {\bibfnamefont {J.}~\bibnamefont {Majer}}, \bibinfo {author} {\bibfnamefont
  {Alexandre}\ \bibnamefont {Blais}}, \bibinfo {author} {\bibfnamefont {M.~H.}\
  \bibnamefont {Devoret}}, \bibinfo {author} {\bibfnamefont {S.~M.}\
  \bibnamefont {Girvin}}, \ and\ \bibinfo {author} {\bibfnamefont {R.~J.}\
  \bibnamefont {Schoelkopf}},\ }\bibfield  {title} {\enquote {\bibinfo {title}
  {Charge-insensitive qubit design derived from the cooper pair box},}\
  }\href@noop {} {\bibfield  {journal} {\bibinfo  {journal} {Phys. Rev. A}\
  }\textbf {\bibinfo {volume} {76}},\ \bibinfo {pages} {042319} (\bibinfo
  {year} {2007})}\BibitemShut {NoStop}%
\bibitem [{\citenamefont {Yan}\ \emph {et~al.}(2016)\citenamefont {Yan},
  \citenamefont {Gustavsson}, \citenamefont {Kamal}, \citenamefont {Birenbaum},
  \citenamefont {Sears}, \citenamefont {Hover}, \citenamefont {Gudmundsen},
  \citenamefont {Rosenberg}, \citenamefont {Samach}, \citenamefont {Weber},
  \citenamefont {Yoder}, \citenamefont {Orlando}, \citenamefont {Clarke},
  \citenamefont {Kerman},\ and\ \citenamefont {Oliver}}]{Yan2016}%
  \BibitemOpen
  \bibfield  {author} {\bibinfo {author} {\bibfnamefont {Fei}\ \bibnamefont
  {Yan}}, \bibinfo {author} {\bibfnamefont {Simon}\ \bibnamefont {Gustavsson}},
  \bibinfo {author} {\bibfnamefont {Archana}\ \bibnamefont {Kamal}}, \bibinfo
  {author} {\bibfnamefont {Jeffery}\ \bibnamefont {Birenbaum}}, \bibinfo
  {author} {\bibfnamefont {Adam~P.}\ \bibnamefont {Sears}}, \bibinfo {author}
  {\bibfnamefont {David}\ \bibnamefont {Hover}}, \bibinfo {author}
  {\bibfnamefont {Ted~J.}\ \bibnamefont {Gudmundsen}}, \bibinfo {author}
  {\bibfnamefont {Danna}\ \bibnamefont {Rosenberg}}, \bibinfo {author}
  {\bibfnamefont {Gabriel}\ \bibnamefont {Samach}}, \bibinfo {author}
  {\bibfnamefont {S}~\bibnamefont {Weber}}, \bibinfo {author} {\bibfnamefont
  {Jonilyn~L.}\ \bibnamefont {Yoder}}, \bibinfo {author} {\bibfnamefont
  {Terry~P.}\ \bibnamefont {Orlando}}, \bibinfo {author} {\bibfnamefont {John}\
  \bibnamefont {Clarke}}, \bibinfo {author} {\bibfnamefont {Andrew~J.}\
  \bibnamefont {Kerman}}, \ and\ \bibinfo {author} {\bibfnamefont {William~D.}\
  \bibnamefont {Oliver}},\ }\bibfield  {title} {\enquote {\bibinfo {title} {The
  flux qubit revisited to enhance coherence and reproducibility},}\ }\href@noop
  {} {\bibfield  {journal} {\bibinfo  {journal} {Nat. Commun.}\ }\textbf
  {\bibinfo {volume} {7}} (\bibinfo {year} {2016})}\BibitemShut {NoStop}%
\bibitem [{\citenamefont {Devoret}\ and\ \citenamefont
  {Schoelkopf}(2013)}]{Devoret_review2013}%
  \BibitemOpen
  \bibfield  {author} {\bibinfo {author} {\bibfnamefont {M.~H.}\ \bibnamefont
  {Devoret}}\ and\ \bibinfo {author} {\bibfnamefont {R.~J.}\ \bibnamefont
  {Schoelkopf}},\ }\bibfield  {title} {\enquote {\bibinfo {title}
  {Superconducting circuits for quantum information: An outlook},}\ }\href@noop
  {} {\bibfield  {journal} {\bibinfo  {journal} {Science}\ }\textbf {\bibinfo
  {volume} {339}},\ \bibinfo {pages} {1169} (\bibinfo {year}
  {2013})}\BibitemShut {NoStop}%
\bibitem [{\citenamefont {Oliver}\ and\ \citenamefont
  {Welander}(2013)}]{Oliver_review2013}%
  \BibitemOpen
  \bibfield  {author} {\bibinfo {author} {\bibfnamefont {William~D}\
  \bibnamefont {Oliver}}\ and\ \bibinfo {author} {\bibfnamefont {Paul~B}\
  \bibnamefont {Welander}},\ }\bibfield  {title} {\enquote {\bibinfo {title}
  {{Materials in superconducting quantum bits}},}\ }\href {\doibase DOI:
  10.1557/mrs.2013.229} {\bibfield  {journal} {\bibinfo  {journal} {MRS
  Bulletin}\ }\textbf {\bibinfo {volume} {38}},\ \bibinfo {pages} {816--825}
  (\bibinfo {year} {2013})}\BibitemShut {NoStop}%
\bibitem [{\citenamefont {Gambetta}\ \emph {et~al.}(2017)\citenamefont
  {Gambetta}, \citenamefont {Chow},\ and\ \citenamefont
  {Steffen}}]{Gambetta_review2017}%
  \BibitemOpen
  \bibfield  {author} {\bibinfo {author} {\bibfnamefont {Jay~M.}\ \bibnamefont
  {Gambetta}}, \bibinfo {author} {\bibfnamefont {Jerry~M.}\ \bibnamefont
  {Chow}}, \ and\ \bibinfo {author} {\bibfnamefont {Matthias}\ \bibnamefont
  {Steffen}},\ }\bibfield  {title} {\enquote {\bibinfo {title} {Building
  logical qubits in a superconducting quantum computing system},}\ }\href@noop
  {} {\bibfield  {journal} {\bibinfo  {journal} {npj Quant. Inf.}\ }\textbf
  {\bibinfo {volume} {3}},\ \bibinfo {pages} {2} (\bibinfo {year}
  {2017})}\BibitemShut {NoStop}%
\bibitem [{\citenamefont {Wendin}(2017)}]{Wendin_review2017}%
  \BibitemOpen
  \bibfield  {author} {\bibinfo {author} {\bibfnamefont {G.}~\bibnamefont
  {Wendin}},\ }\bibfield  {title} {\enquote {\bibinfo {title} {Quantum
  information processing with superconducting circuits: a review},}\
  }\href@noop {} {\bibfield  {journal} {\bibinfo  {journal} {Rep. Prog. Phys.}\
  }\textbf {\bibinfo {volume} {80}},\ \bibinfo {pages} {106001} (\bibinfo
  {year} {2017})}\BibitemShut {NoStop}%
\bibitem [{\citenamefont {Kjaergaard}\ \emph
  {et~al.}(2020{\natexlab{a}})\citenamefont {Kjaergaard}, \citenamefont
  {Schwartz}, \citenamefont {Braumüller}, \citenamefont {Krantz},
  \citenamefont {Wang}, \citenamefont {Gustavsson},\ and\ \citenamefont
  {Oliver}}]{Kjaergaard_review2019}%
  \BibitemOpen
  \bibfield  {author} {\bibinfo {author} {\bibfnamefont {Morten}\ \bibnamefont
  {Kjaergaard}}, \bibinfo {author} {\bibfnamefont {Mollie~E.}\ \bibnamefont
  {Schwartz}}, \bibinfo {author} {\bibfnamefont {Jochen}\ \bibnamefont
  {Braumüller}}, \bibinfo {author} {\bibfnamefont {Philip}\ \bibnamefont
  {Krantz}}, \bibinfo {author} {\bibfnamefont {Joel I-Jan}\ \bibnamefont
  {Wang}}, \bibinfo {author} {\bibfnamefont {Simon}\ \bibnamefont
  {Gustavsson}}, \ and\ \bibinfo {author} {\bibfnamefont {William~D.}\
  \bibnamefont {Oliver}},\ }\bibfield  {title} {\enquote {\bibinfo {title}
  {Superconducting qubits: Current state of play},}\ }\href@noop {} {\bibfield
  {journal} {\bibinfo  {journal} {Annual Review of Condensed Matter Physics}\
  }\textbf {\bibinfo {volume} {11}},\ \bibinfo {pages} {369--395} (\bibinfo
  {year} {2020}{\natexlab{a}})}\BibitemShut {NoStop}%
\bibitem [{\citenamefont {Krantz}\ \emph {et~al.}(2019)\citenamefont {Krantz},
  \citenamefont {Kjaergaard}, \citenamefont {Yan}, \citenamefont {Orlando},
  \citenamefont {Gustavsson},\ and\ \citenamefont {Oliver}}]{Krantz_apr2019}%
  \BibitemOpen
  \bibfield  {author} {\bibinfo {author} {\bibfnamefont {P.}~\bibnamefont
  {Krantz}}, \bibinfo {author} {\bibfnamefont {M.}~\bibnamefont {Kjaergaard}},
  \bibinfo {author} {\bibfnamefont {F.}~\bibnamefont {Yan}}, \bibinfo {author}
  {\bibfnamefont {T.~P.}\ \bibnamefont {Orlando}}, \bibinfo {author}
  {\bibfnamefont {S.}~\bibnamefont {Gustavsson}}, \ and\ \bibinfo {author}
  {\bibfnamefont {W.~D.}\ \bibnamefont {Oliver}},\ }\bibfield  {title}
  {\enquote {\bibinfo {title} {A quantum engineer's guide to superconducting
  qubits},}\ }\href@noop {} {\bibfield  {journal} {\bibinfo  {journal} {Applied
  Physics Reviews}\ }\textbf {\bibinfo {volume} {6}},\ \bibinfo {pages}
  {021318} (\bibinfo {year} {2019})}\BibitemShut {NoStop}%
\bibitem [{\citenamefont {Arute}\ \emph {et~al.}(2019)\citenamefont {Arute},
  \citenamefont {Arya}, \citenamefont {Babbush}, \citenamefont {Bacon},
  \citenamefont {Bardin}, \citenamefont {Barends}, \citenamefont {Biswas},
  \citenamefont {Boixo}, \citenamefont {Brandao}, \citenamefont {Buell},
  \citenamefont {Burkett}, \citenamefont {Chen}, \citenamefont {Chen},
  \citenamefont {Chiaro}, \citenamefont {Collins}, \citenamefont {Courtney},
  \citenamefont {Dunsworth}, \citenamefont {Farhi}, \citenamefont {Foxen},
  \citenamefont {Fowler}, \citenamefont {Gidney}, \citenamefont {Giustina},
  \citenamefont {Graff}, \citenamefont {Guerin}, \citenamefont {Habegger},
  \citenamefont {Harrigan}, \citenamefont {Hartmann}, \citenamefont {Ho},
  \citenamefont {Hoffmann}, \citenamefont {Huang}, \citenamefont {Humble},
  \citenamefont {Isakov}, \citenamefont {Jeffrey}, \citenamefont {Jiang},
  \citenamefont {Kafri}, \citenamefont {Kechedzhi}, \citenamefont {Kelly},
  \citenamefont {Klimov}, \citenamefont {Knysh}, \citenamefont {Korotkov},
  \citenamefont {Kostritsa}, \citenamefont {Landhuis}, \citenamefont
  {Lindmark}, \citenamefont {Lucero}, \citenamefont {Lyakh}, \citenamefont
  {Mandr{\`a}}, \citenamefont {McClean}, \citenamefont {McEwen}, \citenamefont
  {Megrant}, \citenamefont {Mi}, \citenamefont {Michielsen}, \citenamefont
  {Mohseni}, \citenamefont {Mutus}, \citenamefont {Naaman}, \citenamefont
  {Neeley}, \citenamefont {Neill}, \citenamefont {Niu}, \citenamefont {Ostby},
  \citenamefont {Petukhov}, \citenamefont {Platt}, \citenamefont {Quintana},
  \citenamefont {Rieffel}, \citenamefont {Roushan}, \citenamefont {Rubin},
  \citenamefont {Sank}, \citenamefont {Satzinger}, \citenamefont {Smelyanskiy},
  \citenamefont {Sung}, \citenamefont {Trevithick}, \citenamefont
  {Vainsencher}, \citenamefont {Villalonga}, \citenamefont {White},
  \citenamefont {Yao}, \citenamefont {Yeh}, \citenamefont {Zalcman},
  \citenamefont {Neven},\ and\ \citenamefont {Martinis}}]{Arute2019}%
  \BibitemOpen
  \bibfield  {author} {\bibinfo {author} {\bibfnamefont {Frank}\ \bibnamefont
  {Arute}}, \bibinfo {author} {\bibfnamefont {Kunal}\ \bibnamefont {Arya}},
  \bibinfo {author} {\bibfnamefont {Ryan}\ \bibnamefont {Babbush}}, \bibinfo
  {author} {\bibfnamefont {Dave}\ \bibnamefont {Bacon}}, \bibinfo {author}
  {\bibfnamefont {Joseph~C.}\ \bibnamefont {Bardin}}, \bibinfo {author}
  {\bibfnamefont {Rami}\ \bibnamefont {Barends}}, \bibinfo {author}
  {\bibfnamefont {Rupak}\ \bibnamefont {Biswas}}, \bibinfo {author}
  {\bibfnamefont {Sergio}\ \bibnamefont {Boixo}}, \bibinfo {author}
  {\bibfnamefont {Fernando G. S.~L.}\ \bibnamefont {Brandao}}, \bibinfo
  {author} {\bibfnamefont {David~A.}\ \bibnamefont {Buell}}, \bibinfo {author}
  {\bibfnamefont {Brian}\ \bibnamefont {Burkett}}, \bibinfo {author}
  {\bibfnamefont {Yu}~\bibnamefont {Chen}}, \bibinfo {author} {\bibfnamefont
  {Zijun}\ \bibnamefont {Chen}}, \bibinfo {author} {\bibfnamefont {Ben}\
  \bibnamefont {Chiaro}}, \bibinfo {author} {\bibfnamefont {Roberto}\
  \bibnamefont {Collins}}, \bibinfo {author} {\bibfnamefont {William}\
  \bibnamefont {Courtney}}, \bibinfo {author} {\bibfnamefont {Andrew}\
  \bibnamefont {Dunsworth}}, \bibinfo {author} {\bibfnamefont {Edward}\
  \bibnamefont {Farhi}}, \bibinfo {author} {\bibfnamefont {Brooks}\
  \bibnamefont {Foxen}}, \bibinfo {author} {\bibfnamefont {Austin}\
  \bibnamefont {Fowler}}, \bibinfo {author} {\bibfnamefont {Craig}\
  \bibnamefont {Gidney}}, \bibinfo {author} {\bibfnamefont {Marissa}\
  \bibnamefont {Giustina}}, \bibinfo {author} {\bibfnamefont {Rob}\
  \bibnamefont {Graff}}, \bibinfo {author} {\bibfnamefont {Keith}\ \bibnamefont
  {Guerin}}, \bibinfo {author} {\bibfnamefont {Steve}\ \bibnamefont
  {Habegger}}, \bibinfo {author} {\bibfnamefont {Matthew~P.}\ \bibnamefont
  {Harrigan}}, \bibinfo {author} {\bibfnamefont {Michael~J.}\ \bibnamefont
  {Hartmann}}, \bibinfo {author} {\bibfnamefont {Alan}\ \bibnamefont {Ho}},
  \bibinfo {author} {\bibfnamefont {Markus}\ \bibnamefont {Hoffmann}}, \bibinfo
  {author} {\bibfnamefont {Trent}\ \bibnamefont {Huang}}, \bibinfo {author}
  {\bibfnamefont {Travis~S.}\ \bibnamefont {Humble}}, \bibinfo {author}
  {\bibfnamefont {Sergei~V.}\ \bibnamefont {Isakov}}, \bibinfo {author}
  {\bibfnamefont {Evan}\ \bibnamefont {Jeffrey}}, \bibinfo {author}
  {\bibfnamefont {Zhang}\ \bibnamefont {Jiang}}, \bibinfo {author}
  {\bibfnamefont {Dvir}\ \bibnamefont {Kafri}}, \bibinfo {author}
  {\bibfnamefont {Kostyantyn}\ \bibnamefont {Kechedzhi}}, \bibinfo {author}
  {\bibfnamefont {Julian}\ \bibnamefont {Kelly}}, \bibinfo {author}
  {\bibfnamefont {Paul~V.}\ \bibnamefont {Klimov}}, \bibinfo {author}
  {\bibfnamefont {Sergey}\ \bibnamefont {Knysh}}, \bibinfo {author}
  {\bibfnamefont {Alexander}\ \bibnamefont {Korotkov}}, \bibinfo {author}
  {\bibfnamefont {Fedor}\ \bibnamefont {Kostritsa}}, \bibinfo {author}
  {\bibfnamefont {David}\ \bibnamefont {Landhuis}}, \bibinfo {author}
  {\bibfnamefont {Mike}\ \bibnamefont {Lindmark}}, \bibinfo {author}
  {\bibfnamefont {Erik}\ \bibnamefont {Lucero}}, \bibinfo {author}
  {\bibfnamefont {Dmitry}\ \bibnamefont {Lyakh}}, \bibinfo {author}
  {\bibfnamefont {Salvatore}\ \bibnamefont {Mandr{\`a}}}, \bibinfo {author}
  {\bibfnamefont {Jarrod~R.}\ \bibnamefont {McClean}}, \bibinfo {author}
  {\bibfnamefont {Matthew}\ \bibnamefont {McEwen}}, \bibinfo {author}
  {\bibfnamefont {Anthony}\ \bibnamefont {Megrant}}, \bibinfo {author}
  {\bibfnamefont {Xiao}\ \bibnamefont {Mi}}, \bibinfo {author} {\bibfnamefont
  {Kristel}\ \bibnamefont {Michielsen}}, \bibinfo {author} {\bibfnamefont
  {Masoud}\ \bibnamefont {Mohseni}}, \bibinfo {author} {\bibfnamefont {Josh}\
  \bibnamefont {Mutus}}, \bibinfo {author} {\bibfnamefont {Ofer}\ \bibnamefont
  {Naaman}}, \bibinfo {author} {\bibfnamefont {Matthew}\ \bibnamefont
  {Neeley}}, \bibinfo {author} {\bibfnamefont {Charles}\ \bibnamefont {Neill}},
  \bibinfo {author} {\bibfnamefont {Murphy~Yuezhen}\ \bibnamefont {Niu}},
  \bibinfo {author} {\bibfnamefont {Eric}\ \bibnamefont {Ostby}}, \bibinfo
  {author} {\bibfnamefont {Andre}\ \bibnamefont {Petukhov}}, \bibinfo {author}
  {\bibfnamefont {John~C.}\ \bibnamefont {Platt}}, \bibinfo {author}
  {\bibfnamefont {Chris}\ \bibnamefont {Quintana}}, \bibinfo {author}
  {\bibfnamefont {Eleanor~G.}\ \bibnamefont {Rieffel}}, \bibinfo {author}
  {\bibfnamefont {Pedram}\ \bibnamefont {Roushan}}, \bibinfo {author}
  {\bibfnamefont {Nicholas~C.}\ \bibnamefont {Rubin}}, \bibinfo {author}
  {\bibfnamefont {Daniel}\ \bibnamefont {Sank}}, \bibinfo {author}
  {\bibfnamefont {Kevin~J.}\ \bibnamefont {Satzinger}}, \bibinfo {author}
  {\bibfnamefont {Vadim}\ \bibnamefont {Smelyanskiy}}, \bibinfo {author}
  {\bibfnamefont {Kevin~J.}\ \bibnamefont {Sung}}, \bibinfo {author}
  {\bibfnamefont {Matthew~D.}\ \bibnamefont {Trevithick}}, \bibinfo {author}
  {\bibfnamefont {Amit}\ \bibnamefont {Vainsencher}}, \bibinfo {author}
  {\bibfnamefont {Benjamin}\ \bibnamefont {Villalonga}}, \bibinfo {author}
  {\bibfnamefont {Theodore}\ \bibnamefont {White}}, \bibinfo {author}
  {\bibfnamefont {Z.~Jamie}\ \bibnamefont {Yao}}, \bibinfo {author}
  {\bibfnamefont {Ping}\ \bibnamefont {Yeh}}, \bibinfo {author} {\bibfnamefont
  {Adam}\ \bibnamefont {Zalcman}}, \bibinfo {author} {\bibfnamefont {Hartmut}\
  \bibnamefont {Neven}}, \ and\ \bibinfo {author} {\bibfnamefont {John~M.}\
  \bibnamefont {Martinis}},\ }\bibfield  {title} {\enquote {\bibinfo {title}
  {Quantum supremacy using a programmable superconducting processor},}\
  }\href@noop {} {\bibfield  {journal} {\bibinfo  {journal} {Nature}\ }\textbf
  {\bibinfo {volume} {574}},\ \bibinfo {pages} {505--510} (\bibinfo {year}
  {2019})}\BibitemShut {NoStop}%
\bibitem [{\citenamefont {Kjaergaard}\ \emph {et~al.}()\citenamefont
  {Kjaergaard}, \citenamefont {Schwartz}, \citenamefont {Greene}, \citenamefont
  {Samach}, \citenamefont {Bengtsson}, \citenamefont {O'Keeffe}, \citenamefont
  {McNally}, \citenamefont {Braumüller}, \citenamefont {Kim}, \citenamefont
  {Krantz}, \citenamefont {Marvian}, \citenamefont {Melville}, \citenamefont
  {Niedzielski}, \citenamefont {Sung}, \citenamefont {Winik}, \citenamefont
  {Yoder}, \citenamefont {Rosenberg}, \citenamefont {Obenland}, \citenamefont
  {Lloyd}, \citenamefont {Orlando}, \citenamefont {Marvian}, \citenamefont
  {Gustavsson},\ and\ \citenamefont {Oliver}}]{Kjaergaard2020}%
  \BibitemOpen
  \bibfield  {author} {\bibinfo {author} {\bibfnamefont {M.}~\bibnamefont
  {Kjaergaard}}, \bibinfo {author} {\bibfnamefont {M.E.}\ \bibnamefont
  {Schwartz}}, \bibinfo {author} {\bibfnamefont {A.}~\bibnamefont {Greene}},
  \bibinfo {author} {\bibfnamefont {G.O.}\ \bibnamefont {Samach}}, \bibinfo
  {author} {\bibfnamefont {A.}~\bibnamefont {Bengtsson}}, \bibinfo {author}
  {\bibfnamefont {M.}~\bibnamefont {O'Keeffe}}, \bibinfo {author}
  {\bibfnamefont {C.M.}\ \bibnamefont {McNally}}, \bibinfo {author}
  {\bibfnamefont {J.}~\bibnamefont {Braumüller}}, \bibinfo {author}
  {\bibfnamefont {D.K.}\ \bibnamefont {Kim}}, \bibinfo {author} {\bibfnamefont
  {P.}~\bibnamefont {Krantz}}, \bibinfo {author} {\bibfnamefont
  {M.}~\bibnamefont {Marvian}}, \bibinfo {author} {\bibfnamefont
  {A.}~\bibnamefont {Melville}}, \bibinfo {author} {\bibfnamefont {B.M.}\
  \bibnamefont {Niedzielski}}, \bibinfo {author} {\bibfnamefont
  {Y.}~\bibnamefont {Sung}}, \bibinfo {author} {\bibfnamefont {R.}~\bibnamefont
  {Winik}}, \bibinfo {author} {\bibfnamefont {J.L.}\ \bibnamefont {Yoder}},
  \bibinfo {author} {\bibfnamefont {D.}~\bibnamefont {Rosenberg}}, \bibinfo
  {author} {\bibfnamefont {K.}~\bibnamefont {Obenland}}, \bibinfo {author}
  {\bibfnamefont {S.}~\bibnamefont {Lloyd}}, \bibinfo {author} {\bibfnamefont
  {T.P.}\ \bibnamefont {Orlando}}, \bibinfo {author} {\bibfnamefont
  {I.}~\bibnamefont {Marvian}}, \bibinfo {author} {\bibfnamefont
  {S.}~\bibnamefont {Gustavsson}}, \ and\ \bibinfo {author} {\bibfnamefont
  {W.D.}\ \bibnamefont {Oliver}},\ }\href@noop {} {\enquote {\bibinfo {title}
  {A quantum instruction set implemented on a superconducting quantum
  processor},}\ }\bibinfo {note} {ArXiv:2001.08838}\BibitemShut {NoStop}%
\bibitem [{\citenamefont {Koppens}\ \emph {et~al.}(2006)\citenamefont
  {Koppens}, \citenamefont {Buizert}, \citenamefont {Tielrooij}, \citenamefont
  {Vink}, \citenamefont {Nowack}, \citenamefont {Meunier}, \citenamefont
  {Kouwenhoven},\ and\ \citenamefont {Vandersypen}}]{Koppens2006}%
  \BibitemOpen
  \bibfield  {author} {\bibinfo {author} {\bibfnamefont {F.~H.~L.}\
  \bibnamefont {Koppens}}, \bibinfo {author} {\bibfnamefont {C.}~\bibnamefont
  {Buizert}}, \bibinfo {author} {\bibfnamefont {K.~J.}\ \bibnamefont
  {Tielrooij}}, \bibinfo {author} {\bibfnamefont {I.~T.}\ \bibnamefont {Vink}},
  \bibinfo {author} {\bibfnamefont {K.~C.}\ \bibnamefont {Nowack}}, \bibinfo
  {author} {\bibfnamefont {T.}~\bibnamefont {Meunier}}, \bibinfo {author}
  {\bibfnamefont {L.~P.}\ \bibnamefont {Kouwenhoven}}, \ and\ \bibinfo {author}
  {\bibfnamefont {L.~M.~K.}\ \bibnamefont {Vandersypen}},\ }\bibfield  {title}
  {\enquote {\bibinfo {title} {Driven coherent oscillations of a single
  electron spin in a quantum dot},}\ }\href@noop {} {\bibfield  {journal}
  {\bibinfo  {journal} {Nature}\ }\textbf {\bibinfo {volume} {442}},\ \bibinfo
  {pages} {766--771} (\bibinfo {year} {2006})}\BibitemShut {NoStop}%
\bibitem [{\citenamefont {Veldhorst}\ \emph {et~al.}(2014)\citenamefont
  {Veldhorst}, \citenamefont {Hwang}, \citenamefont {Yang}, \citenamefont
  {Leenstra}, \citenamefont {de~Ronde}, \citenamefont {Dehollain},
  \citenamefont {Muhonen}, \citenamefont {Hudson}, \citenamefont {Itoh},
  \citenamefont {Morello},\ and\ \citenamefont {Dzurak}}]{Veldhorst2014}%
  \BibitemOpen
  \bibfield  {author} {\bibinfo {author} {\bibfnamefont {M.}~\bibnamefont
  {Veldhorst}}, \bibinfo {author} {\bibfnamefont {J.~C.~C.}\ \bibnamefont
  {Hwang}}, \bibinfo {author} {\bibfnamefont {C.~H.}\ \bibnamefont {Yang}},
  \bibinfo {author} {\bibfnamefont {A.~W.}\ \bibnamefont {Leenstra}}, \bibinfo
  {author} {\bibfnamefont {B.}~\bibnamefont {de~Ronde}}, \bibinfo {author}
  {\bibfnamefont {J.~P.}\ \bibnamefont {Dehollain}}, \bibinfo {author}
  {\bibfnamefont {J.~T.}\ \bibnamefont {Muhonen}}, \bibinfo {author}
  {\bibfnamefont {F.~E.}\ \bibnamefont {Hudson}}, \bibinfo {author}
  {\bibfnamefont {K.~M.}\ \bibnamefont {Itoh}}, \bibinfo {author}
  {\bibfnamefont {A.}~\bibnamefont {Morello}}, \ and\ \bibinfo {author}
  {\bibfnamefont {A.~S.}\ \bibnamefont {Dzurak}},\ }\bibfield  {title}
  {\enquote {\bibinfo {title} {An addressable quantum dot qubit with
  fault-tolerant control-fidelity},}\ }\href@noop {} {\bibfield  {journal}
  {\bibinfo  {journal} {Nat. Nanotechnol.}\ }\textbf {\bibinfo {volume} {9}},\
  \bibinfo {pages} {981–985} (\bibinfo {year} {2014})}\BibitemShut {NoStop}%
\bibitem [{\citenamefont {DiVincenzo}\ \emph {et~al.}(2000)\citenamefont
  {DiVincenzo}, \citenamefont {Bacon}, \citenamefont {Kempe}, \citenamefont
  {Burkard},\ and\ \citenamefont {Whaley}}]{divincenzo_bacon_nature2000}%
  \BibitemOpen
  \bibfield  {author} {\bibinfo {author} {\bibfnamefont {D.~P.}\ \bibnamefont
  {DiVincenzo}}, \bibinfo {author} {\bibfnamefont {D.}~\bibnamefont {Bacon}},
  \bibinfo {author} {\bibfnamefont {J.}~\bibnamefont {Kempe}}, \bibinfo
  {author} {\bibfnamefont {G.}~\bibnamefont {Burkard}}, \ and\ \bibinfo
  {author} {\bibfnamefont {K.~B.}\ \bibnamefont {Whaley}},\ }\bibfield  {title}
  {\enquote {\bibinfo {title} {Universal quantum computation with the exchange
  interaction},}\ }\href@noop {} {\bibfield  {journal} {\bibinfo  {journal}
  {Nature (London)}\ }\textbf {\bibinfo {volume} {408}},\ \bibinfo {pages}
  {339--342} (\bibinfo {year} {2000})}\BibitemShut {NoStop}%
\bibitem [{\citenamefont {Bacon}\ \emph {et~al.}(2000)\citenamefont {Bacon},
  \citenamefont {Kempe}, \citenamefont {Lidar},\ and\ \citenamefont
  {Whaley}}]{Bacon2000}%
  \BibitemOpen
  \bibfield  {author} {\bibinfo {author} {\bibfnamefont {D.}~\bibnamefont
  {Bacon}}, \bibinfo {author} {\bibfnamefont {J.}~\bibnamefont {Kempe}},
  \bibinfo {author} {\bibfnamefont {D.~A.}\ \bibnamefont {Lidar}}, \ and\
  \bibinfo {author} {\bibfnamefont {K.~B.}\ \bibnamefont {Whaley}},\ }\bibfield
   {title} {\enquote {\bibinfo {title} {Universal fault-tolerant quantum
  computation on decoherence-free subspaces},}\ }\href@noop {} {\bibfield
  {journal} {\bibinfo  {journal} {Phys. Rev. Lett.}\ }\textbf {\bibinfo
  {volume} {85}},\ \bibinfo {pages} {1758--1761} (\bibinfo {year}
  {2000})}\BibitemShut {NoStop}%
\bibitem [{\citenamefont {Kempe}\ \emph {et~al.}(2001)\citenamefont {Kempe},
  \citenamefont {Bacon}, \citenamefont {Lidar}, ,\ and\ \citenamefont
  {Whaley}}]{Kempe2001}%
  \BibitemOpen
  \bibfield  {author} {\bibinfo {author} {\bibfnamefont {J.}~\bibnamefont
  {Kempe}}, \bibinfo {author} {\bibfnamefont {D.}~\bibnamefont {Bacon}},
  \bibinfo {author} {\bibfnamefont {D.~A.}\ \bibnamefont {Lidar}}, , \ and\
  \bibinfo {author} {\bibfnamefont {K.~B.}\ \bibnamefont {Whaley}},\ }\bibfield
   {title} {\enquote {\bibinfo {title} {Theory of decoherence-free
  fault-tolerant universal quantum computation},}\ }\href@noop {} {\bibfield
  {journal} {\bibinfo  {journal} {Phys. Rev. A}\ }\textbf {\bibinfo {volume}
  {63}},\ \bibinfo {pages} {042307} (\bibinfo {year} {2001})}\BibitemShut
  {NoStop}%
\bibitem [{\citenamefont {Nakamura}\ \emph {et~al.}(1999)\citenamefont
  {Nakamura}, \citenamefont {Pashkin},\ and\ \citenamefont
  {Tsai}}]{Nakamura1999}%
  \BibitemOpen
  \bibfield  {author} {\bibinfo {author} {\bibfnamefont {Y}~\bibnamefont
  {Nakamura}}, \bibinfo {author} {\bibfnamefont {Yu.~A.}\ \bibnamefont
  {Pashkin}}, \ and\ \bibinfo {author} {\bibfnamefont {J.~S.}\ \bibnamefont
  {Tsai}},\ }\bibfield  {title} {\enquote {\bibinfo {title} {Coherent control
  of macroscopic quantum states in a single-cooper-pair box},}\ }\href@noop {}
  {\bibfield  {journal} {\bibinfo  {journal} {Nature}\ }\textbf {\bibinfo
  {volume} {398}},\ \bibinfo {pages} {786--788} (\bibinfo {year}
  {1999})}\BibitemShut {NoStop}%
\bibitem [{\citenamefont {Pashkin}\ \emph {et~al.}(2003)\citenamefont
  {Pashkin}, \citenamefont {Yamamoto}, \citenamefont {Astafiev}, \citenamefont
  {Nakamura}, \citenamefont {Averin},\ and\ \citenamefont
  {Tsai}}]{Pashkin2003}%
  \BibitemOpen
  \bibfield  {author} {\bibinfo {author} {\bibfnamefont {Yu.~A.}\ \bibnamefont
  {Pashkin}}, \bibinfo {author} {\bibfnamefont {T.}~\bibnamefont {Yamamoto}},
  \bibinfo {author} {\bibfnamefont {O.}~\bibnamefont {Astafiev}}, \bibinfo
  {author} {\bibfnamefont {Y}~\bibnamefont {Nakamura}}, \bibinfo {author}
  {\bibfnamefont {D.~V.}\ \bibnamefont {Averin}}, \ and\ \bibinfo {author}
  {\bibfnamefont {J.~S.}\ \bibnamefont {Tsai}},\ }\bibfield  {title} {\enquote
  {\bibinfo {title} {Quantum oscillations in two coupled charge qubits},}\
  }\href@noop {} {\bibfield  {journal} {\bibinfo  {journal} {Nature}\ }\textbf
  {\bibinfo {volume} {421}},\ \bibinfo {pages} {823--826} (\bibinfo {year}
  {2003})}\BibitemShut {NoStop}%
\bibitem [{\citenamefont {Yamamoto}\ \emph {et~al.}(2003)\citenamefont
  {Yamamoto}, \citenamefont {Pashkin}, \citenamefont {Astafiev}, \citenamefont
  {Nakamura},\ and\ \citenamefont {Tsai}}]{Yamamoto2003}%
  \BibitemOpen
  \bibfield  {author} {\bibinfo {author} {\bibfnamefont {T.}~\bibnamefont
  {Yamamoto}}, \bibinfo {author} {\bibfnamefont {Yu.~A.}\ \bibnamefont
  {Pashkin}}, \bibinfo {author} {\bibfnamefont {O.}~\bibnamefont {Astafiev}},
  \bibinfo {author} {\bibfnamefont {Y}~\bibnamefont {Nakamura}}, \ and\
  \bibinfo {author} {\bibfnamefont {J.~S.}\ \bibnamefont {Tsai}},\ }\bibfield
  {title} {\enquote {\bibinfo {title} {Demonstration of conditional gate
  operation using superconducting charge qubits},}\ }\href@noop {} {\bibfield
  {journal} {\bibinfo  {journal} {Nature}\ }\textbf {\bibinfo {volume} {425}},\
  \bibinfo {pages} {941--944} (\bibinfo {year} {2003})}\BibitemShut {NoStop}%
\bibitem [{\citenamefont {Shim}\ and\ \citenamefont {Tahan}(2016)}]{Shim2016}%
  \BibitemOpen
  \bibfield  {author} {\bibinfo {author} {\bibfnamefont {Yun-Pil}\ \bibnamefont
  {Shim}}\ and\ \bibinfo {author} {\bibfnamefont {Charles}\ \bibnamefont
  {Tahan}},\ }\bibfield  {title} {\enquote {\bibinfo {title}
  {{Semiconductor-inspired design principles for superconducting quantum
  computing}},}\ }\href {\doibase 10.1038/ncomms11059} {\bibfield  {journal}
  {\bibinfo  {journal} {Nat. Commun.}\ }\textbf {\bibinfo {volume} {7}},\
  \bibinfo {pages} {11059} (\bibinfo {year} {2016})}\BibitemShut {NoStop}%
\bibitem [{\citenamefont {Oliver}\ \emph {et~al.}(2005)\citenamefont {Oliver},
  \citenamefont {Yu}, \citenamefont {Lee}, \citenamefont {Berggren},
  \citenamefont {Levitov},\ and\ \citenamefont {Orlando}}]{Oliver_Science2005}%
  \BibitemOpen
  \bibfield  {author} {\bibinfo {author} {\bibfnamefont {William~D.}\
  \bibnamefont {Oliver}}, \bibinfo {author} {\bibfnamefont {Yang}\ \bibnamefont
  {Yu}}, \bibinfo {author} {\bibfnamefont {Janice~C.}\ \bibnamefont {Lee}},
  \bibinfo {author} {\bibfnamefont {Karl~K.}\ \bibnamefont {Berggren}},
  \bibinfo {author} {\bibfnamefont {Leonid~S.}\ \bibnamefont {Levitov}}, \ and\
  \bibinfo {author} {\bibfnamefont {Terry~P.}\ \bibnamefont {Orlando}},\
  }\bibfield  {title} {\enquote {\bibinfo {title} {Mach-zehnder interferometry
  in a strongly driven superconducting qubit},}\ }\href@noop {} {\bibfield
  {journal} {\bibinfo  {journal} {Science}\ }\textbf {\bibinfo {volume}
  {310}},\ \bibinfo {pages} {1653--1657} (\bibinfo {year} {2005})}\BibitemShut
  {NoStop}%
\bibitem [{\citenamefont {Sillanp\"a\"a}\ \emph {et~al.}(2006)\citenamefont
  {Sillanp\"a\"a}, \citenamefont {Lehtinen}, \citenamefont {Paila},
  \citenamefont {Makhlin},\ and\ \citenamefont {Hakonen}}]{Sillanpaa2006}%
  \BibitemOpen
  \bibfield  {author} {\bibinfo {author} {\bibfnamefont {Mika}\ \bibnamefont
  {Sillanp\"a\"a}}, \bibinfo {author} {\bibfnamefont {Teijo}\ \bibnamefont
  {Lehtinen}}, \bibinfo {author} {\bibfnamefont {Antti}\ \bibnamefont {Paila}},
  \bibinfo {author} {\bibfnamefont {Yuriy}\ \bibnamefont {Makhlin}}, \ and\
  \bibinfo {author} {\bibfnamefont {Pertti}\ \bibnamefont {Hakonen}},\
  }\bibfield  {title} {\enquote {\bibinfo {title} {Continuous-time monitoring
  of landau-zener interference in a cooper-pair box},}\ }\href {\doibase
  10.1103/PhysRevLett.96.187002} {\bibfield  {journal} {\bibinfo  {journal}
  {Phys. Rev. Lett.}\ }\textbf {\bibinfo {volume} {96}},\ \bibinfo {pages}
  {187002} (\bibinfo {year} {2006})}\BibitemShut {NoStop}%
\bibitem [{\citenamefont {Shevchenko}\ \emph {et~al.}(2010)\citenamefont
  {Shevchenko}, \citenamefont {Ashhab},\ and\ \citenamefont
  {Nori}}]{LZ_PhysRep2010}%
  \BibitemOpen
  \bibfield  {author} {\bibinfo {author} {\bibfnamefont {S.N.}\ \bibnamefont
  {Shevchenko}}, \bibinfo {author} {\bibfnamefont {S.}~\bibnamefont {Ashhab}},
  \ and\ \bibinfo {author} {\bibfnamefont {Franco}\ \bibnamefont {Nori}},\
  }\bibfield  {title} {\enquote {\bibinfo {title}
  {Landau–zener–st{\"u}ckelberg interferometry},}\ }\href@noop {}
  {\bibfield  {journal} {\bibinfo  {journal} {Phys. Rep.}\ }\textbf {\bibinfo
  {volume} {492}},\ \bibinfo {pages} {1--30} (\bibinfo {year}
  {2010})}\BibitemShut {NoStop}%
\bibitem [{\citenamefont {Vion}\ \emph {et~al.}(2002)\citenamefont {Vion},
  \citenamefont {Aassime}, \citenamefont {Cottet}, \citenamefont {Joyez},
  \citenamefont {Pothier}, \citenamefont {Urbina}, \citenamefont {Esteve},\
  and\ \citenamefont {Devoret}}]{Vion2002}%
  \BibitemOpen
  \bibfield  {author} {\bibinfo {author} {\bibfnamefont {D.}~\bibnamefont
  {Vion}}, \bibinfo {author} {\bibfnamefont {A.}~\bibnamefont {Aassime}},
  \bibinfo {author} {\bibfnamefont {A.}~\bibnamefont {Cottet}}, \bibinfo
  {author} {\bibfnamefont {P.}~\bibnamefont {Joyez}}, \bibinfo {author}
  {\bibfnamefont {H.}~\bibnamefont {Pothier}}, \bibinfo {author} {\bibfnamefont
  {C.}~\bibnamefont {Urbina}}, \bibinfo {author} {\bibfnamefont
  {D.}~\bibnamefont {Esteve}}, \ and\ \bibinfo {author} {\bibfnamefont {M.H.}\
  \bibnamefont {Devoret}},\ }\bibfield  {title} {\enquote {\bibinfo {title}
  {Manipulating the quantum state of an electrical circuit},}\ }\href@noop {}
  {\bibfield  {journal} {\bibinfo  {journal} {Science}\ }\textbf {\bibinfo
  {volume} {296}},\ \bibinfo {pages} {886} (\bibinfo {year}
  {2002})}\BibitemShut {NoStop}%
\bibitem [{\citenamefont {Yoshihara}\ \emph {et~al.}(2006)\citenamefont
  {Yoshihara}, \citenamefont {Harrabi}, \citenamefont {Niskanen}, \citenamefont
  {Nakamura},\ and\ \citenamefont {Tsai}}]{Yoshihara2006}%
  \BibitemOpen
  \bibfield  {author} {\bibinfo {author} {\bibfnamefont {Y.}~\bibnamefont
  {Yoshihara}}, \bibinfo {author} {\bibfnamefont {K.}~\bibnamefont {Harrabi}},
  \bibinfo {author} {\bibfnamefont {A.O.}\ \bibnamefont {Niskanen}}, \bibinfo
  {author} {\bibfnamefont {Y.}~\bibnamefont {Nakamura}}, \ and\ \bibinfo
  {author} {\bibfnamefont {J.S.}\ \bibnamefont {Tsai}},\ }\bibfield  {title}
  {\enquote {\bibinfo {title} {Manipulating the quantum state of an electrical
  circuit},}\ }\href@noop {} {\bibfield  {journal} {\bibinfo  {journal} {Phys.
  Rev. Lett.}\ }\textbf {\bibinfo {volume} {97}},\ \bibinfo {pages} {167001}
  (\bibinfo {year} {2006})}\BibitemShut {NoStop}%
\bibitem [{\citenamefont {Bylander}\ \emph {et~al.}(2011)\citenamefont
  {Bylander}, \citenamefont {Gustavsson}, \citenamefont {Yan}, \citenamefont
  {Yoshihara}, \citenamefont {Harrabi}, \citenamefont {Fitch}, \citenamefont
  {Cory}, \citenamefont {Nakamura}, \citenamefont {Tsai},\ and\ \citenamefont
  {Oliver}}]{Bylander2011}%
  \BibitemOpen
  \bibfield  {author} {\bibinfo {author} {\bibfnamefont {Jonas}\ \bibnamefont
  {Bylander}}, \bibinfo {author} {\bibfnamefont {Simon}\ \bibnamefont
  {Gustavsson}}, \bibinfo {author} {\bibfnamefont {Fei}\ \bibnamefont {Yan}},
  \bibinfo {author} {\bibfnamefont {Fumiki}\ \bibnamefont {Yoshihara}},
  \bibinfo {author} {\bibfnamefont {Khalil}\ \bibnamefont {Harrabi}}, \bibinfo
  {author} {\bibfnamefont {George}\ \bibnamefont {Fitch}}, \bibinfo {author}
  {\bibfnamefont {David~G.}\ \bibnamefont {Cory}}, \bibinfo {author}
  {\bibfnamefont {Yasunobu}\ \bibnamefont {Nakamura}}, \bibinfo {author}
  {\bibfnamefont {Jaw-Shen}\ \bibnamefont {Tsai}}, \ and\ \bibinfo {author}
  {\bibfnamefont {William~D.}\ \bibnamefont {Oliver}},\ }\bibfield  {title}
  {\enquote {\bibinfo {title} {{Noise spectroscopy through dynamical decoupling
  with a superconducting flux qubit}},}\ }\href {\doibase 10.1038/nphys1994}
  {\bibfield  {journal} {\bibinfo  {journal} {Nature Physics}\ }\textbf
  {\bibinfo {volume} {7}},\ \bibinfo {pages} {565--570} (\bibinfo {year}
  {2011})}\BibitemShut {NoStop}%
\bibitem [{\citenamefont {Schuster}\ \emph {et~al.}(2005)\citenamefont
  {Schuster}, \citenamefont {Wallraff}, \citenamefont {Blais}, \citenamefont
  {Frunzio}, \citenamefont {Huang}, \citenamefont {Majer}, \citenamefont
  {Girvin},\ and\ \citenamefont {Schoelkopf}}]{Schuster2005}%
  \BibitemOpen
  \bibfield  {author} {\bibinfo {author} {\bibfnamefont {D.~I.}\ \bibnamefont
  {Schuster}}, \bibinfo {author} {\bibfnamefont {A.}~\bibnamefont {Wallraff}},
  \bibinfo {author} {\bibfnamefont {A.}~\bibnamefont {Blais}}, \bibinfo
  {author} {\bibfnamefont {L.}~\bibnamefont {Frunzio}}, \bibinfo {author}
  {\bibfnamefont {R.-S.}\ \bibnamefont {Huang}}, \bibinfo {author}
  {\bibfnamefont {J.}~\bibnamefont {Majer}}, \bibinfo {author} {\bibfnamefont
  {S.~M.}\ \bibnamefont {Girvin}}, \ and\ \bibinfo {author} {\bibfnamefont
  {R.~J.}\ \bibnamefont {Schoelkopf}},\ }\bibfield  {title} {\enquote {\bibinfo
  {title} {ac stark shift and dephasing of a superconducting qubit strongly
  coupled to a cavity field},}\ }\href {\doibase 10.1103/PhysRevLett.94.123602}
  {\bibfield  {journal} {\bibinfo  {journal} {Phys. Rev. Lett.}\ }\textbf
  {\bibinfo {volume} {94}},\ \bibinfo {pages} {123602} (\bibinfo {year}
  {2005})}\BibitemShut {NoStop}%
\bibitem [{\citenamefont {Sears}\ \emph {et~al.}(2012)\citenamefont {Sears},
  \citenamefont {Petrenko}, \citenamefont {Catelani}, \citenamefont {Sun},
  \citenamefont {Paik}, \citenamefont {Kirchmair}, \citenamefont {Frunzio},
  \citenamefont {Glazman}, \citenamefont {Girvin},\ and\ \citenamefont
  {Schoelkopf}}]{Sears2012}%
  \BibitemOpen
  \bibfield  {author} {\bibinfo {author} {\bibfnamefont {A.~P.}\ \bibnamefont
  {Sears}}, \bibinfo {author} {\bibfnamefont {A.}~\bibnamefont {Petrenko}},
  \bibinfo {author} {\bibfnamefont {G.}~\bibnamefont {Catelani}}, \bibinfo
  {author} {\bibfnamefont {L.}~\bibnamefont {Sun}}, \bibinfo {author}
  {\bibfnamefont {Hanhee}\ \bibnamefont {Paik}}, \bibinfo {author}
  {\bibfnamefont {G.}~\bibnamefont {Kirchmair}}, \bibinfo {author}
  {\bibfnamefont {L.}~\bibnamefont {Frunzio}}, \bibinfo {author} {\bibfnamefont
  {L.~I.}\ \bibnamefont {Glazman}}, \bibinfo {author} {\bibfnamefont {S.~M.}\
  \bibnamefont {Girvin}}, \ and\ \bibinfo {author} {\bibfnamefont {R.~J.}\
  \bibnamefont {Schoelkopf}},\ }\bibfield  {title} {\enquote {\bibinfo {title}
  {Photon shot noise dephasing in the strong-dispersive limit of circuit
  qed},}\ }\href {\doibase 10.1103/PhysRevB.86.180504} {\bibfield  {journal}
  {\bibinfo  {journal} {Phys. Rev. B}\ }\textbf {\bibinfo {volume} {86}},\
  \bibinfo {pages} {180504} (\bibinfo {year} {2012})}\BibitemShut {NoStop}%
\bibitem [{\citenamefont {Yan}\ \emph {et~al.}(2018)\citenamefont {Yan},
  \citenamefont {Campbell}, \citenamefont {Krantz}, \citenamefont {Kjaergaard},
  \citenamefont {Kim}, \citenamefont {Yoder}, \citenamefont {Hover},
  \citenamefont {Sears}, \citenamefont {Kerman}, \citenamefont {Orlando},
  \citenamefont {Gustavsson},\ and\ \citenamefont {Oliver}}]{Yan2018}%
  \BibitemOpen
  \bibfield  {author} {\bibinfo {author} {\bibfnamefont {Fei}\ \bibnamefont
  {Yan}}, \bibinfo {author} {\bibfnamefont {Dan}\ \bibnamefont {Campbell}},
  \bibinfo {author} {\bibfnamefont {Philip}\ \bibnamefont {Krantz}}, \bibinfo
  {author} {\bibfnamefont {Morten}\ \bibnamefont {Kjaergaard}}, \bibinfo
  {author} {\bibfnamefont {David}\ \bibnamefont {Kim}}, \bibinfo {author}
  {\bibfnamefont {Jonilyn~L.}\ \bibnamefont {Yoder}}, \bibinfo {author}
  {\bibfnamefont {David}\ \bibnamefont {Hover}}, \bibinfo {author}
  {\bibfnamefont {Adam}\ \bibnamefont {Sears}}, \bibinfo {author}
  {\bibfnamefont {Andrew~J.}\ \bibnamefont {Kerman}}, \bibinfo {author}
  {\bibfnamefont {Terry~P.}\ \bibnamefont {Orlando}}, \bibinfo {author}
  {\bibfnamefont {Simon}\ \bibnamefont {Gustavsson}}, \ and\ \bibinfo {author}
  {\bibfnamefont {William~D.}\ \bibnamefont {Oliver}},\ }\bibfield  {title}
  {\enquote {\bibinfo {title} {Distinguishing coherent and thermal photon noise
  in a circuit quantum electrodynamical system},}\ }\href {\doibase
  10.1103/PhysRevLett.120.260504} {\bibfield  {journal} {\bibinfo  {journal}
  {Phys. Rev. Lett.}\ }\textbf {\bibinfo {volume} {120}},\ \bibinfo {pages}
  {260504} (\bibinfo {year} {2018})}\BibitemShut {NoStop}%
\bibitem [{\citenamefont {Hutchings}\ \emph {et~al.}(2017)\citenamefont
  {Hutchings}, \citenamefont {Hertzberg}, \citenamefont {Liu}, \citenamefont
  {Bronn}, \citenamefont {Keefe}, \citenamefont {Brink}, \citenamefont {Chow},\
  and\ \citenamefont {Plourde}}]{Hutchings2017}%
  \BibitemOpen
  \bibfield  {author} {\bibinfo {author} {\bibfnamefont {M.~D.}\ \bibnamefont
  {Hutchings}}, \bibinfo {author} {\bibfnamefont {J.~B.}\ \bibnamefont
  {Hertzberg}}, \bibinfo {author} {\bibfnamefont {Y.}~\bibnamefont {Liu}},
  \bibinfo {author} {\bibfnamefont {N.~T.}\ \bibnamefont {Bronn}}, \bibinfo
  {author} {\bibfnamefont {G.~A.}\ \bibnamefont {Keefe}}, \bibinfo {author}
  {\bibfnamefont {Markus}\ \bibnamefont {Brink}}, \bibinfo {author}
  {\bibfnamefont {Jerry~M.}\ \bibnamefont {Chow}}, \ and\ \bibinfo {author}
  {\bibfnamefont {B.~L.~T.}\ \bibnamefont {Plourde}},\ }\bibfield  {title}
  {\enquote {\bibinfo {title} {Tunable superconducting qubits with
  flux-independent coherence},}\ }\href {\doibase
  10.1103/PhysRevApplied.8.044003} {\bibfield  {journal} {\bibinfo  {journal}
  {Phys. Rev. Applied}\ }\textbf {\bibinfo {volume} {8}},\ \bibinfo {pages}
  {044003} (\bibinfo {year} {2017})}\BibitemShut {NoStop}%
\bibitem [{\citenamefont {Barends}\ \emph {et~al.}(2013)\citenamefont
  {Barends}, \citenamefont {Kelly}, \citenamefont {Megrant}, \citenamefont
  {Sank}, \citenamefont {Jeffrey}, \citenamefont {Chen}, \citenamefont {Yin},
  \citenamefont {Chiaro}, \citenamefont {Mutus}, \citenamefont {Neill},
  \citenamefont {O’Malley}, \citenamefont {Roushan}, \citenamefont {Wenner},
  \citenamefont {White}, \citenamefont {Cleland},\ and\ \citenamefont
  {Martinis}}]{xmon}%
  \BibitemOpen
  \bibfield  {author} {\bibinfo {author} {\bibfnamefont {R.}~\bibnamefont
  {Barends}}, \bibinfo {author} {\bibfnamefont {J.}~\bibnamefont {Kelly}},
  \bibinfo {author} {\bibfnamefont {A.}~\bibnamefont {Megrant}}, \bibinfo
  {author} {\bibfnamefont {D.}~\bibnamefont {Sank}}, \bibinfo {author}
  {\bibfnamefont {E.}~\bibnamefont {Jeffrey}}, \bibinfo {author} {\bibfnamefont
  {Y.}~\bibnamefont {Chen}}, \bibinfo {author} {\bibfnamefont {Y.}~\bibnamefont
  {Yin}}, \bibinfo {author} {\bibfnamefont {B.}~\bibnamefont {Chiaro}},
  \bibinfo {author} {\bibfnamefont {J.}~\bibnamefont {Mutus}}, \bibinfo
  {author} {\bibfnamefont {C.}~\bibnamefont {Neill}}, \bibinfo {author}
  {\bibfnamefont {P.}~\bibnamefont {O’Malley}}, \bibinfo {author}
  {\bibfnamefont {P.}~\bibnamefont {Roushan}}, \bibinfo {author} {\bibfnamefont
  {J.}~\bibnamefont {Wenner}}, \bibinfo {author} {\bibfnamefont {T.~C.}\
  \bibnamefont {White}}, \bibinfo {author} {\bibfnamefont {A.~N.}\ \bibnamefont
  {Cleland}}, \ and\ \bibinfo {author} {\bibfnamefont {John~M.}\ \bibnamefont
  {Martinis}},\ }\bibfield  {title} {\enquote {\bibinfo {title} {Coherent
  josephson qubit suitable for scalable quantum integrated circuits},}\
  }\href@noop {} {\bibfield  {journal} {\bibinfo  {journal} {Phys. Rev. Lett.}\
  }\textbf {\bibinfo {volume} {111}},\ \bibinfo {pages} {080502} (\bibinfo
  {year} {2013})}\BibitemShut {NoStop}%
\bibitem [{sup()}]{supp-mat}%
  \BibitemOpen
  \href@noop {} {\bibinfo  {journal} {See supplementary material}\
  }\BibitemShut {NoStop}%
\bibitem [{\citenamefont {Gramajo}\ \emph {et~al.}()\citenamefont {Gramajo},
  \citenamefont {Campbell}, \citenamefont {Kannan}, \citenamefont {Kim},
  \citenamefont {Melville}, \citenamefont {Niedzielski}, \citenamefont {Yoder},
  \citenamefont {S\'anchez}, \citenamefont {Dom\'inguez}, \citenamefont
  {Gustavsson},\ and\ \citenamefont {Oliver}}]{Gramajo2019}%
  \BibitemOpen
\bibfield  {journal} {  }\bibfield  {author} {\bibinfo {author} {\bibfnamefont
  {Ana~Laura}\ \bibnamefont {Gramajo}}, \bibinfo {author} {\bibfnamefont {Dan}\
  \bibnamefont {Campbell}}, \bibinfo {author} {\bibfnamefont {Bharath}\
  \bibnamefont {Kannan}}, \bibinfo {author} {\bibfnamefont {David~K.}\
  \bibnamefont {Kim}}, \bibinfo {author} {\bibfnamefont {Alexander}\
  \bibnamefont {Melville}}, \bibinfo {author} {\bibfnamefont {Bethany~M.}\
  \bibnamefont {Niedzielski}}, \bibinfo {author} {\bibfnamefont {Jonilyn~L.}\
  \bibnamefont {Yoder}}, \bibinfo {author} {\bibfnamefont {Mar\'ia~Jos\'e}\
  \bibnamefont {S\'anchez}}, \bibinfo {author} {\bibfnamefont {Daniel}\
  \bibnamefont {Dom\'inguez}}, \bibinfo {author} {\bibfnamefont {Simon}\
  \bibnamefont {Gustavsson}}, \ and\ \bibinfo {author} {\bibfnamefont
  {William~D.}\ \bibnamefont {Oliver}},\ }\href@noop {} {\enquote {\bibinfo
  {title} {Quantum simulation of coherent backscattering in a system of
  superconducting qubits},}\ }\bibinfo {note} {ArXiv:1912.12488}\BibitemShut
  {NoStop}%
\bibitem [{\citenamefont {Mooij}\ \emph {et~al.}(1999)\citenamefont {Mooij},
  \citenamefont {Orlando}, \citenamefont {Levitov}, \citenamefont {Tian},
  \citenamefont {van~der Wal},\ and\ \citenamefont {Lloyd}}]{Mooij1999}%
  \BibitemOpen
  \bibfield  {author} {\bibinfo {author} {\bibfnamefont {J.~E.}\ \bibnamefont
  {Mooij}}, \bibinfo {author} {\bibfnamefont {T.~P.}\ \bibnamefont {Orlando}},
  \bibinfo {author} {\bibfnamefont {L.}~\bibnamefont {Levitov}}, \bibinfo
  {author} {\bibfnamefont {Lin}\ \bibnamefont {Tian}}, \bibinfo {author}
  {\bibfnamefont {Caspar~H.}\ \bibnamefont {van~der Wal}}, \ and\ \bibinfo
  {author} {\bibfnamefont {Seth}\ \bibnamefont {Lloyd}},\ }\bibfield  {title}
  {\enquote {\bibinfo {title} {Josephson persistent-current qubit},}\ }\href
  {\doibase 10.1126/science.285.5430.1036} {\bibfield  {journal} {\bibinfo
  {journal} {Science}\ }\textbf {\bibinfo {volume} {285}},\ \bibinfo {pages}
  {1036--1039} (\bibinfo {year} {1999})}\BibitemShut {NoStop}%
\bibitem [{\citenamefont {Orlando}\ \emph {et~al.}(1999)\citenamefont
  {Orlando}, \citenamefont {Mooij}, \citenamefont {Tian}, \citenamefont
  {van~der Wal}, \citenamefont {Levitov}, \citenamefont {Lloyd},\ and\
  \citenamefont {Mazo}}]{Orlando1999}%
  \BibitemOpen
  \bibfield  {author} {\bibinfo {author} {\bibfnamefont {T.~P.}\ \bibnamefont
  {Orlando}}, \bibinfo {author} {\bibfnamefont {J.~E.}\ \bibnamefont {Mooij}},
  \bibinfo {author} {\bibfnamefont {Lin}\ \bibnamefont {Tian}}, \bibinfo
  {author} {\bibfnamefont {Caspar~H.}\ \bibnamefont {van~der Wal}}, \bibinfo
  {author} {\bibfnamefont {L.~S.}\ \bibnamefont {Levitov}}, \bibinfo {author}
  {\bibfnamefont {Seth}\ \bibnamefont {Lloyd}}, \ and\ \bibinfo {author}
  {\bibfnamefont {J.~J.}\ \bibnamefont {Mazo}},\ }\bibfield  {title} {\enquote
  {\bibinfo {title} {Superconducting persistent-current qubit},}\ }\href
  {\doibase 10.1103/PhysRevB.60.15398} {\bibfield  {journal} {\bibinfo
  {journal} {Phys. Rev. B}\ }\textbf {\bibinfo {volume} {60}},\ \bibinfo
  {pages} {15398--15413} (\bibinfo {year} {1999})}\BibitemShut {NoStop}%
\bibitem [{\citenamefont {Berns}\ \emph {et~al.}(2006)\citenamefont {Berns},
  \citenamefont {Oliver}, \citenamefont {Valenzuela}, \citenamefont {Shytov},
  \citenamefont {Berggren}, \citenamefont {Levitov},\ and\ \citenamefont
  {Orlando}}]{Berns2006}%
  \BibitemOpen
  \bibfield  {author} {\bibinfo {author} {\bibfnamefont {D.~M.}\ \bibnamefont
  {Berns}}, \bibinfo {author} {\bibfnamefont {W.~D.}\ \bibnamefont {Oliver}},
  \bibinfo {author} {\bibfnamefont {S.~O.}\ \bibnamefont {Valenzuela}},
  \bibinfo {author} {\bibfnamefont {A.~V.}\ \bibnamefont {Shytov}}, \bibinfo
  {author} {\bibfnamefont {K.~K.}\ \bibnamefont {Berggren}}, \bibinfo {author}
  {\bibfnamefont {L.~S.}\ \bibnamefont {Levitov}}, \ and\ \bibinfo {author}
  {\bibfnamefont {T.~P.}\ \bibnamefont {Orlando}},\ }\bibfield  {title}
  {\enquote {\bibinfo {title} {Coherent quasiclassical dynamics of a persistent
  current qubit},}\ }\href {\doibase 10.1103/PhysRevLett.97.150502} {\bibfield
  {journal} {\bibinfo  {journal} {Phys. Rev. Lett.}\ }\textbf {\bibinfo
  {volume} {97}},\ \bibinfo {pages} {150502} (\bibinfo {year}
  {2006})}\BibitemShut {NoStop}%
\bibitem [{\citenamefont {Valenzuela}\ \emph {et~al.}(2006)\citenamefont
  {Valenzuela}, \citenamefont {Oliver}, \citenamefont {Berns}, \citenamefont
  {Berggren}, \citenamefont {Levitov},\ and\ \citenamefont
  {Orlando}}]{Valenzuela2006}%
  \BibitemOpen
  \bibfield  {author} {\bibinfo {author} {\bibfnamefont {Sergio~O.}\
  \bibnamefont {Valenzuela}}, \bibinfo {author} {\bibfnamefont {William~D.}\
  \bibnamefont {Oliver}}, \bibinfo {author} {\bibfnamefont {David~M.}\
  \bibnamefont {Berns}}, \bibinfo {author} {\bibfnamefont {Karl~K.}\
  \bibnamefont {Berggren}}, \bibinfo {author} {\bibfnamefont {Leonid~S.}\
  \bibnamefont {Levitov}}, \ and\ \bibinfo {author} {\bibfnamefont {Terry~P.}\
  \bibnamefont {Orlando}},\ }\bibfield  {title} {\enquote {\bibinfo {title}
  {Microwave-induced cooling of a superconducting qubit},}\ }\href@noop {}
  {\bibfield  {journal} {\bibinfo  {journal} {Science}\ }\textbf {\bibinfo
  {volume} {314}},\ \bibinfo {pages} {1589--1592} (\bibinfo {year}
  {2006})}\BibitemShut {NoStop}%
\bibitem [{\citenamefont {Berns}\ \emph {et~al.}(2008)\citenamefont {Berns},
  \citenamefont {Rudner}, \citenamefont {Valenzuela}, \citenamefont {Berggren},
  \citenamefont {Oliver}, \citenamefont {Levitov},\ and\ \citenamefont
  {Orlando}}]{Berns2008}%
  \BibitemOpen
  \bibfield  {author} {\bibinfo {author} {\bibfnamefont {David~M}\ \bibnamefont
  {Berns}}, \bibinfo {author} {\bibfnamefont {Mark~S}\ \bibnamefont {Rudner}},
  \bibinfo {author} {\bibfnamefont {Sergio~O}\ \bibnamefont {Valenzuela}},
  \bibinfo {author} {\bibfnamefont {Karl~K}\ \bibnamefont {Berggren}}, \bibinfo
  {author} {\bibfnamefont {William~D}\ \bibnamefont {Oliver}}, \bibinfo
  {author} {\bibfnamefont {Leonid~S}\ \bibnamefont {Levitov}}, \ and\ \bibinfo
  {author} {\bibfnamefont {Terry~P}\ \bibnamefont {Orlando}},\ }\bibfield
  {title} {\enquote {\bibinfo {title} {{Amplitude spectroscopy of a solid-state
  artificial atom}},}\ }\href {https://doi.org/10.1038/nature07262
  http://10.0.4.14/nature07262
  https://www.nature.com/articles/nature07262{\#}supplementary-information}
  {\bibfield  {journal} {\bibinfo  {journal} {Nature}\ }\textbf {\bibinfo
  {volume} {455}},\ \bibinfo {pages} {51} (\bibinfo {year} {2008})}\BibitemShut
  {NoStop}%
\bibitem [{\citenamefont {Bylander}\ \emph {et~al.}(2009)\citenamefont
  {Bylander}, \citenamefont {Rudner}, \citenamefont {Shytov}, \citenamefont
  {Valenzuela}, \citenamefont {Berns}, \citenamefont {Berggren}, \citenamefont
  {Levitov}, ,\ and\ \citenamefont {Oliver}}]{Bylander2009}%
  \BibitemOpen
  \bibfield  {author} {\bibinfo {author} {\bibfnamefont {J}~\bibnamefont
  {Bylander}}, \bibinfo {author} {\bibfnamefont {Mark~S}\ \bibnamefont
  {Rudner}}, \bibinfo {author} {\bibfnamefont {Andrey~V.}\ \bibnamefont
  {Shytov}}, \bibinfo {author} {\bibfnamefont {Sergio~O}\ \bibnamefont
  {Valenzuela}}, \bibinfo {author} {\bibfnamefont {David~M}\ \bibnamefont
  {Berns}}, \bibinfo {author} {\bibfnamefont {Karl~K}\ \bibnamefont
  {Berggren}}, \bibinfo {author} {\bibfnamefont {Leonid~S}\ \bibnamefont
  {Levitov}}, , \ and\ \bibinfo {author} {\bibfnamefont {William~D}\
  \bibnamefont {Oliver}},\ }\bibfield  {title} {\enquote {\bibinfo {title}
  {{Pulse imaging and nonadiabatic control of solid-state artificial atoms}},}\
  }\href@noop {} {\bibfield  {journal} {\bibinfo  {journal} {Phys. Rev. B}\
  }\textbf {\bibinfo {volume} {80}},\ \bibinfo {pages} {220506(R)} (\bibinfo
  {year} {2009})}\BibitemShut {NoStop}%
\bibitem [{\citenamefont {Gustavsson}\ \emph {et~al.}(2013)\citenamefont
  {Gustavsson}, \citenamefont {Bylander},\ and\ \citenamefont
  {Oliver}}]{Gustavsson2013}%
  \BibitemOpen
  \bibfield  {author} {\bibinfo {author} {\bibfnamefont {Simon}\ \bibnamefont
  {Gustavsson}}, \bibinfo {author} {\bibfnamefont {Jonas}\ \bibnamefont
  {Bylander}}, \ and\ \bibinfo {author} {\bibfnamefont {William~D.}\
  \bibnamefont {Oliver}},\ }\bibfield  {title} {\enquote {\bibinfo {title}
  {{Time-reversal symmetry and universal conductance fluctuations in a driven
  two-level system}},}\ }\href {\doibase 10.1103/PhysRevLett.110.016603}
  {\bibfield  {journal} {\bibinfo  {journal} {Phys. Rev. Lett.}\ }\textbf
  {\bibinfo {volume} {110}},\ \bibinfo {pages} {1--5} (\bibinfo {year}
  {2013})}\BibitemShut {NoStop}%
\bibitem [{\citenamefont {Kerman}(2010)}]{Kerman2010}%
  \BibitemOpen
  \bibfield  {author} {\bibinfo {author} {\bibfnamefont {Andrew~J.}\
  \bibnamefont {Kerman}},\ }\bibfield  {title} {\enquote {\bibinfo {title}
  {Metastable superconducting qubit},}\ }\href {\doibase
  10.1103/PhysRevLett.104.027002} {\bibfield  {journal} {\bibinfo  {journal}
  {Phys. Rev. Lett.}\ }\textbf {\bibinfo {volume} {104}},\ \bibinfo {pages}
  {027002} (\bibinfo {year} {2010})}\BibitemShut {NoStop}%
\bibitem [{\citenamefont {Manucharyan}\ \emph {et~al.}(2009)\citenamefont
  {Manucharyan}, \citenamefont {Koch}, \citenamefont {Glazman},\ and\
  \citenamefont {Devoret}}]{Manucharyan2009}%
  \BibitemOpen
  \bibfield  {author} {\bibinfo {author} {\bibfnamefont {Vladimir~E.}\
  \bibnamefont {Manucharyan}}, \bibinfo {author} {\bibfnamefont {Jens}\
  \bibnamefont {Koch}}, \bibinfo {author} {\bibfnamefont {Leonid~I.}\
  \bibnamefont {Glazman}}, \ and\ \bibinfo {author} {\bibfnamefont {Michel~H.}\
  \bibnamefont {Devoret}},\ }\bibfield  {title} {\enquote {\bibinfo {title}
  {Fluxonium: Single cooper-pair circuit free of charge offsets},}\ }\href
  {\doibase 10.1126/science.1175552} {\bibfield  {journal} {\bibinfo  {journal}
  {Science}\ }\textbf {\bibinfo {volume} {326}},\ \bibinfo {pages} {113--116}
  (\bibinfo {year} {2009})}\BibitemShut {NoStop}%
\bibitem [{\citenamefont {Pop}\ \emph {et~al.}(2014)\citenamefont {Pop},
  \citenamefont {Geerlings}, \citenamefont {Catelani}, \citenamefont
  {Schoelkopf}, \citenamefont {Glazman},\ and\ \citenamefont
  {Devoret}}]{Pop2014}%
  \BibitemOpen
  \bibfield  {author} {\bibinfo {author} {\bibfnamefont {Ioan~M.}\ \bibnamefont
  {Pop}}, \bibinfo {author} {\bibfnamefont {Kurtis}\ \bibnamefont {Geerlings}},
  \bibinfo {author} {\bibfnamefont {Gianluigi}\ \bibnamefont {Catelani}},
  \bibinfo {author} {\bibfnamefont {Robert~J.}\ \bibnamefont {Schoelkopf}},
  \bibinfo {author} {\bibfnamefont {Leonid~I.}\ \bibnamefont {Glazman}}, \ and\
  \bibinfo {author} {\bibfnamefont {Michel~H.}\ \bibnamefont {Devoret}},\
  }\bibfield  {title} {\enquote {\bibinfo {title} {Coherent suppression of
  electromagnetic dissipation due to superconducting quasiparticles},}\ }\href
  {\doibase 10.1038/nature13017} {\bibfield  {journal} {\bibinfo  {journal}
  {Nature}\ }\textbf {\bibinfo {volume} {508}},\ \bibinfo {pages} {1476--4687}
  (\bibinfo {year} {2014})}\BibitemShut {NoStop}%
\bibitem [{\citenamefont {Nguyen}\ \emph {et~al.}(2019)\citenamefont {Nguyen},
  \citenamefont {Lin}, \citenamefont {Somoroff}, \citenamefont {Mencia},
  \citenamefont {Grabon},\ and\ \citenamefont {Manucharyan}}]{Nguyen2019}%
  \BibitemOpen
  \bibfield  {author} {\bibinfo {author} {\bibfnamefont {Long~B.}\ \bibnamefont
  {Nguyen}}, \bibinfo {author} {\bibfnamefont {Yen-Hsiang}\ \bibnamefont
  {Lin}}, \bibinfo {author} {\bibfnamefont {Aaron}\ \bibnamefont {Somoroff}},
  \bibinfo {author} {\bibfnamefont {Raymond}\ \bibnamefont {Mencia}}, \bibinfo
  {author} {\bibfnamefont {Nicholas}\ \bibnamefont {Grabon}}, \ and\ \bibinfo
  {author} {\bibfnamefont {Vladimir~E.}\ \bibnamefont {Manucharyan}},\
  }\bibfield  {title} {\enquote {\bibinfo {title} {High-coherence fluxonium
  qubit},}\ }\href {\doibase 10.1103/PhysRevX.9.041041} {\bibfield  {journal}
  {\bibinfo  {journal} {Phys. Rev. X}\ }\textbf {\bibinfo {volume} {9}},\
  \bibinfo {pages} {041041} (\bibinfo {year} {2019})}\BibitemShut {NoStop}%
\bibitem [{\citenamefont {Gyenis}\ \emph {et~al.}(2019)\citenamefont {Gyenis},
  \citenamefont {Mundada}, \citenamefont {Paolo}, \citenamefont {Hazard},
  \citenamefont {You}, \citenamefont {Schuster}, \citenamefont {Koch},
  \citenamefont {Blais},\ and\ \citenamefont {Houck}}]{Gyenis2019}%
  \BibitemOpen
  \bibfield  {author} {\bibinfo {author} {\bibfnamefont {Andras}\ \bibnamefont
  {Gyenis}}, \bibinfo {author} {\bibfnamefont {Pranav~S.}\ \bibnamefont
  {Mundada}}, \bibinfo {author} {\bibfnamefont {Agustin~Di}\ \bibnamefont
  {Paolo}}, \bibinfo {author} {\bibfnamefont {Thomas~M.}\ \bibnamefont
  {Hazard}}, \bibinfo {author} {\bibfnamefont {Xinyuan}\ \bibnamefont {You}},
  \bibinfo {author} {\bibfnamefont {David~I.}\ \bibnamefont {Schuster}},
  \bibinfo {author} {\bibfnamefont {Jens}\ \bibnamefont {Koch}}, \bibinfo
  {author} {\bibfnamefont {Alexandre}\ \bibnamefont {Blais}}, \ and\ \bibinfo
  {author} {\bibfnamefont {Andrew~A.}\ \bibnamefont {Houck}},\ }\href@noop {}
  {\enquote {\bibinfo {title} {Experimental realization of an intrinsically
  error-protected superconducting qubit},}\ } (\bibinfo {year} {2019}),\
  \Eprint {http://arxiv.org/abs/1910.07542} {arXiv:1910.07542 [quant-ph]}
  \BibitemShut {NoStop}%
\bibitem [{\citenamefont {Barends}\ \emph
  {et~al.}(2014{\natexlab{a}})\citenamefont {Barends}, \citenamefont {Kelly},
  \citenamefont {Megrant}, \citenamefont {Veitia}, \citenamefont {Sank},
  \citenamefont {Jeffrey}, \citenamefont {White}, \citenamefont {Mutus},
  \citenamefont {Fowler}, \citenamefont {Campbell}, \citenamefont {Chen},
  \citenamefont {Chen}, \citenamefont {Chiaro}, \citenamefont {Dunsworth},
  \citenamefont {Neill}, \citenamefont {O’Malley}, \citenamefont {Roushan},
  \citenamefont {Vainsencher}, \citenamefont {Wenner}, \citenamefont
  {Korotkov}, \citenamefont {Cleland},\ and\ \citenamefont
  {Martinis}}]{Barends_Martinis_nature2014}%
  \BibitemOpen
  \bibfield  {author} {\bibinfo {author} {\bibfnamefont {R.}~\bibnamefont
  {Barends}}, \bibinfo {author} {\bibfnamefont {J.}~\bibnamefont {Kelly}},
  \bibinfo {author} {\bibfnamefont {A.}~\bibnamefont {Megrant}}, \bibinfo
  {author} {\bibfnamefont {A.}~\bibnamefont {Veitia}}, \bibinfo {author}
  {\bibfnamefont {D.}~\bibnamefont {Sank}}, \bibinfo {author} {\bibfnamefont
  {E.}~\bibnamefont {Jeffrey}}, \bibinfo {author} {\bibfnamefont {T.~C.}\
  \bibnamefont {White}}, \bibinfo {author} {\bibfnamefont {J.}~\bibnamefont
  {Mutus}}, \bibinfo {author} {\bibfnamefont {A.~G.}\ \bibnamefont {Fowler}},
  \bibinfo {author} {\bibfnamefont {B.}~\bibnamefont {Campbell}}, \bibinfo
  {author} {\bibfnamefont {Y.}~\bibnamefont {Chen}}, \bibinfo {author}
  {\bibfnamefont {Z.}~\bibnamefont {Chen}}, \bibinfo {author} {\bibfnamefont
  {B.}~\bibnamefont {Chiaro}}, \bibinfo {author} {\bibfnamefont
  {A.}~\bibnamefont {Dunsworth}}, \bibinfo {author} {\bibfnamefont
  {C.}~\bibnamefont {Neill}}, \bibinfo {author} {\bibfnamefont
  {P.}~\bibnamefont {O’Malley}}, \bibinfo {author} {\bibfnamefont
  {P.}~\bibnamefont {Roushan}}, \bibinfo {author} {\bibfnamefont
  {A.}~\bibnamefont {Vainsencher}}, \bibinfo {author} {\bibfnamefont
  {J.}~\bibnamefont {Wenner}}, \bibinfo {author} {\bibfnamefont {A.~N.}\
  \bibnamefont {Korotkov}}, \bibinfo {author} {\bibfnamefont {A.~N.}\
  \bibnamefont {Cleland}}, \ and\ \bibinfo {author} {\bibfnamefont {John~M.}\
  \bibnamefont {Martinis}},\ }\bibfield  {title} {\enquote {\bibinfo {title}
  {Superconducting quantum circuits at the surface code threshold for fault
  tolerance},}\ }\href@noop {} {\bibfield  {journal} {\bibinfo  {journal}
  {Nature}\ }\textbf {\bibinfo {volume} {508}},\ \bibinfo {pages} {500--503}
  (\bibinfo {year} {2014}{\natexlab{a}})}\BibitemShut {NoStop}%
\bibitem [{\citenamefont {Gambetta}\ \emph {et~al.}(2006)\citenamefont
  {Gambetta}, \citenamefont {Blais}, \citenamefont {Schuster}, \citenamefont
  {Wallraff}, \citenamefont {Frunzio}, \citenamefont {Majer}, \citenamefont
  {Devoret}, \citenamefont {Girvin},\ and\ \citenamefont
  {Schoelkopf}}]{Gambetta2006}%
  \BibitemOpen
  \bibfield  {author} {\bibinfo {author} {\bibfnamefont {Jay}\ \bibnamefont
  {Gambetta}}, \bibinfo {author} {\bibfnamefont {Alexandre}\ \bibnamefont
  {Blais}}, \bibinfo {author} {\bibfnamefont {D.~I.}\ \bibnamefont {Schuster}},
  \bibinfo {author} {\bibfnamefont {A.}~\bibnamefont {Wallraff}}, \bibinfo
  {author} {\bibfnamefont {L.}~\bibnamefont {Frunzio}}, \bibinfo {author}
  {\bibfnamefont {J.}~\bibnamefont {Majer}}, \bibinfo {author} {\bibfnamefont
  {M.~H.}\ \bibnamefont {Devoret}}, \bibinfo {author} {\bibfnamefont {S.~M.}\
  \bibnamefont {Girvin}}, \ and\ \bibinfo {author} {\bibfnamefont {R.~J.}\
  \bibnamefont {Schoelkopf}},\ }\bibfield  {title} {\enquote {\bibinfo {title}
  {Qubit-photon interactions in a cavity: Measurement-induced dephasing and
  number splitting},}\ }\href {\doibase 10.1103/PhysRevA.74.042318} {\bibfield
  {journal} {\bibinfo  {journal} {Phys. Rev. A}\ }\textbf {\bibinfo {volume}
  {74}},\ \bibinfo {pages} {042318} (\bibinfo {year} {2006})}\BibitemShut
  {NoStop}%
\bibitem [{\citenamefont {Chiorescu}\ \emph {et~al.}(2003)\citenamefont
  {Chiorescu}, \citenamefont {Nakamura}, \citenamefont {Harmans},\ and\
  \citenamefont {Mooij}}]{Chiorescu2003}%
  \BibitemOpen
  \bibfield  {author} {\bibinfo {author} {\bibfnamefont {I.}~\bibnamefont
  {Chiorescu}}, \bibinfo {author} {\bibfnamefont {Y.}~\bibnamefont {Nakamura}},
  \bibinfo {author} {\bibfnamefont {C.~J. P.~M.}\ \bibnamefont {Harmans}}, \
  and\ \bibinfo {author} {\bibfnamefont {J.~E.}\ \bibnamefont {Mooij}},\
  }\bibfield  {title} {\enquote {\bibinfo {title} {Coherent quantum dynamics of
  a superconducting flux qubit},}\ }\href {\doibase 10.1126/science.1081045}
  {\bibfield  {journal} {\bibinfo  {journal} {Science}\ }\textbf {\bibinfo
  {volume} {299}},\ \bibinfo {pages} {1869--1871} (\bibinfo {year}
  {2003})}\BibitemShut {NoStop}%
\bibitem [{\citenamefont {DiCarlo}\ \emph {et~al.}(2009)\citenamefont
  {DiCarlo}, \citenamefont {Chow}, \citenamefont {Gambetta}, \citenamefont
  {Bishop1}, \citenamefont {Johnson}, \citenamefont {Schuster}, \citenamefont
  {Majer3}, \citenamefont {Blais}, \citenamefont {Frunzio}, \citenamefont
  {Girvin},\ and\ \citenamefont {Schoelkopf}}]{DiCarlo_nature2009}%
  \BibitemOpen
  \bibfield  {author} {\bibinfo {author} {\bibfnamefont {L.}~\bibnamefont
  {DiCarlo}}, \bibinfo {author} {\bibfnamefont {J.~M.}\ \bibnamefont {Chow}},
  \bibinfo {author} {\bibfnamefont {J.~M.}\ \bibnamefont {Gambetta}}, \bibinfo
  {author} {\bibfnamefont {Lev~S.}\ \bibnamefont {Bishop1}}, \bibinfo {author}
  {\bibfnamefont {B.~R.}\ \bibnamefont {Johnson}}, \bibinfo {author}
  {\bibfnamefont {D.~I.}\ \bibnamefont {Schuster}}, \bibinfo {author}
  {\bibfnamefont {J.}~\bibnamefont {Majer3}}, \bibinfo {author} {\bibfnamefont
  {A.}~\bibnamefont {Blais}}, \bibinfo {author} {\bibfnamefont
  {L.}~\bibnamefont {Frunzio}}, \bibinfo {author} {\bibfnamefont {S.~M.}\
  \bibnamefont {Girvin}}, \ and\ \bibinfo {author} {\bibfnamefont {R.~J.}\
  \bibnamefont {Schoelkopf}},\ }\bibfield  {title} {\enquote {\bibinfo {title}
  {Demonstration of two-qubit algorithms with a superconducting quantum
  processor},}\ }\href@noop {} {\bibfield  {journal} {\bibinfo  {journal}
  {Nature}\ }\textbf {\bibinfo {volume} {460}},\ \bibinfo {pages} {240--244}
  (\bibinfo {year} {2009})}\BibitemShut {NoStop}%
\bibitem [{\citenamefont {Martinis}\ and\ \citenamefont
  {Geller}(2014)}]{Martinis2014}%
  \BibitemOpen
  \bibfield  {author} {\bibinfo {author} {\bibfnamefont {John~M.}\ \bibnamefont
  {Martinis}}\ and\ \bibinfo {author} {\bibfnamefont {Michael~R.}\ \bibnamefont
  {Geller}},\ }\bibfield  {title} {\enquote {\bibinfo {title} {Fast adiabatic
  qubit gates using only ${\ensuremath{\sigma}}_{z}$ control},}\ }\href
  {\doibase 10.1103/PhysRevA.90.022307} {\bibfield  {journal} {\bibinfo
  {journal} {Phys. Rev. A}\ }\textbf {\bibinfo {volume} {90}},\ \bibinfo
  {pages} {022307} (\bibinfo {year} {2014})}\BibitemShut {NoStop}%
\bibitem [{\citenamefont {Kjaergaard}\ \emph
  {et~al.}(2020{\natexlab{b}})\citenamefont {Kjaergaard}, \citenamefont
  {Schwartz}, \citenamefont {Greene}, \citenamefont {Samach}, \citenamefont
  {Bengtsson}, \citenamefont {O'Keeffe}, \citenamefont {McNally}, \citenamefont
  {Braumüller}, \citenamefont {Kim}, \citenamefont {Krantz}, \citenamefont
  {Marvian}, \citenamefont {Melville}, \citenamefont {Niedzielski},
  \citenamefont {Sung}, \citenamefont {Winik}, \citenamefont {Yoder},
  \citenamefont {Rosenberg}, \citenamefont {Obenland}, \citenamefont {Lloyd},
  \citenamefont {Orlando}, \citenamefont {Marvian}, \citenamefont
  {Gustavsson},\ and\ \citenamefont {Oliver}}]{Kjaergaard_DME2020}%
  \BibitemOpen
  \bibfield  {author} {\bibinfo {author} {\bibfnamefont {M.}~\bibnamefont
  {Kjaergaard}}, \bibinfo {author} {\bibfnamefont {M.E.}\ \bibnamefont
  {Schwartz}}, \bibinfo {author} {\bibfnamefont {A.}~\bibnamefont {Greene}},
  \bibinfo {author} {\bibfnamefont {G.O.}\ \bibnamefont {Samach}}, \bibinfo
  {author} {\bibfnamefont {A.}~\bibnamefont {Bengtsson}}, \bibinfo {author}
  {\bibfnamefont {M.}~\bibnamefont {O'Keeffe}}, \bibinfo {author}
  {\bibfnamefont {C.M.}\ \bibnamefont {McNally}}, \bibinfo {author}
  {\bibfnamefont {J.}~\bibnamefont {Braumüller}}, \bibinfo {author}
  {\bibfnamefont {D.K.}\ \bibnamefont {Kim}}, \bibinfo {author} {\bibfnamefont
  {P.}~\bibnamefont {Krantz}}, \bibinfo {author} {\bibfnamefont
  {M.}~\bibnamefont {Marvian}}, \bibinfo {author} {\bibfnamefont
  {A.}~\bibnamefont {Melville}}, \bibinfo {author} {\bibfnamefont {B.M.}\
  \bibnamefont {Niedzielski}}, \bibinfo {author} {\bibfnamefont
  {Y.}~\bibnamefont {Sung}}, \bibinfo {author} {\bibfnamefont {R.}~\bibnamefont
  {Winik}}, \bibinfo {author} {\bibfnamefont {J.L.}\ \bibnamefont {Yoder}},
  \bibinfo {author} {\bibfnamefont {D.}~\bibnamefont {Rosenberg}}, \bibinfo
  {author} {\bibfnamefont {K.}~\bibnamefont {Obenland}}, \bibinfo {author}
  {\bibfnamefont {S.}~\bibnamefont {Lloyd}}, \bibinfo {author} {\bibfnamefont
  {T.P.}\ \bibnamefont {Orlando}}, \bibinfo {author} {\bibfnamefont
  {I.}~\bibnamefont {Marvian}}, \bibinfo {author} {\bibfnamefont
  {S.}~\bibnamefont {Gustavsson}}, \ and\ \bibinfo {author} {\bibfnamefont
  {W.D.}\ \bibnamefont {Oliver}},\ }\bibfield  {title} {\enquote {\bibinfo
  {title} {A quantum instruction set implemented on a superconducting quantum
  computer},}\ }\href@noop {} {\bibfield  {journal} {\bibinfo  {journal}
  {arXiv:2001.08838}\ } (\bibinfo {year} {2020}{\natexlab{b}})}\BibitemShut
  {NoStop}%
\bibitem [{\citenamefont {Chang}\ \emph {et~al.}(2020)\citenamefont {Chang},
  \citenamefont {Chakram}, \citenamefont {Roy}, \citenamefont {Earnest},
  \citenamefont {Lu}, \citenamefont {Huang}, \citenamefont {Weiss},
  \citenamefont {Koch},\ and\ \citenamefont {Schuster}}]{Zhang2020}%
  \BibitemOpen
  \bibfield  {author} {\bibinfo {author} {\bibfnamefont {Helin}\ \bibnamefont
  {Chang}}, \bibinfo {author} {\bibfnamefont {Srivatsan}\ \bibnamefont
  {Chakram}}, \bibinfo {author} {\bibfnamefont {Tanay}\ \bibnamefont {Roy}},
  \bibinfo {author} {\bibfnamefont {Nathan}\ \bibnamefont {Earnest}}, \bibinfo
  {author} {\bibfnamefont {Yao}\ \bibnamefont {Lu}}, \bibinfo {author}
  {\bibfnamefont {Ziwen}\ \bibnamefont {Huang}}, \bibinfo {author}
  {\bibfnamefont {Daniel}\ \bibnamefont {Weiss}}, \bibinfo {author}
  {\bibfnamefont {Jens}\ \bibnamefont {Koch}}, \ and\ \bibinfo {author}
  {\bibfnamefont {David~I.}\ \bibnamefont {Schuster}},\ }\href@noop {}
  {\enquote {\bibinfo {title} {Universal fast flux control of a coherent,
  low-frequency qubit},}\ } (\bibinfo {year} {2020}),\ \Eprint
  {http://arxiv.org/abs/2002.10653} {arXiv:2002.10653 [quant-ph]} \BibitemShut
  {NoStop}%
\bibitem [{\citenamefont {Brooks}\ \emph {et~al.}(2013)\citenamefont {Brooks},
  \citenamefont {Kitaev},\ and\ \citenamefont {Preskill}}]{Brooks2013}%
  \BibitemOpen
  \bibfield  {author} {\bibinfo {author} {\bibfnamefont {Peter}\ \bibnamefont
  {Brooks}}, \bibinfo {author} {\bibfnamefont {Alexei}\ \bibnamefont {Kitaev}},
  \ and\ \bibinfo {author} {\bibfnamefont {John}\ \bibnamefont {Preskill}},\
  }\bibfield  {title} {\enquote {\bibinfo {title} {Protected gates for
  superconducting qubits},}\ }\href {\doibase 10.1103/PhysRevA.87.052306}
  {\bibfield  {journal} {\bibinfo  {journal} {Phys. Rev. A}\ }\textbf {\bibinfo
  {volume} {87}},\ \bibinfo {pages} {052306} (\bibinfo {year}
  {2013})}\BibitemShut {NoStop}%
\bibitem [{\citenamefont {Petta}\ \emph {et~al.}(2010)\citenamefont {Petta},
  \citenamefont {Lu},\ and\ \citenamefont {Gossard}}]{Petta2010}%
  \BibitemOpen
  \bibfield  {author} {\bibinfo {author} {\bibfnamefont {J.~R.}\ \bibnamefont
  {Petta}}, \bibinfo {author} {\bibfnamefont {H.}~\bibnamefont {Lu}}, \ and\
  \bibinfo {author} {\bibfnamefont {A.~C.}\ \bibnamefont {Gossard}},\
  }\bibfield  {title} {\enquote {\bibinfo {title} {A coherent beamsplitter for
  electronic spins states},}\ }\href@noop {} {\bibfield  {journal} {\bibinfo
  {journal} {Science}\ }\textbf {\bibinfo {volume} {327}},\ \bibinfo {pages}
  {669--672} (\bibinfo {year} {2010})}\BibitemShut {NoStop}%
\bibitem [{\citenamefont {Andrews}\ \emph {et~al.}(2019)\citenamefont
  {Andrews}, \citenamefont {Jones}, \citenamefont {Reed}, \citenamefont
  {Jones}, \citenamefont {Ha}, \citenamefont {Jura}, \citenamefont {Kerckhoff},
  \citenamefont {Levendorf}, \citenamefont {Meenehan}, \citenamefont {Merkel},
  \citenamefont {Smith}, \citenamefont {Sun}, \citenamefont {Weinstein},
  \citenamefont {Rakher}, \citenamefont {Ladd},\ and\ \citenamefont
  {Borselli}}]{Andrews2019}%
  \BibitemOpen
  \bibfield  {author} {\bibinfo {author} {\bibfnamefont {Reed~W.}\ \bibnamefont
  {Andrews}}, \bibinfo {author} {\bibfnamefont {Cody}\ \bibnamefont {Jones}},
  \bibinfo {author} {\bibfnamefont {Matthew~D.}\ \bibnamefont {Reed}}, \bibinfo
  {author} {\bibfnamefont {Aaron~M.}\ \bibnamefont {Jones}}, \bibinfo {author}
  {\bibfnamefont {Sieu~D.}\ \bibnamefont {Ha}}, \bibinfo {author}
  {\bibfnamefont {Michael~P.}\ \bibnamefont {Jura}}, \bibinfo {author}
  {\bibfnamefont {Joseph}\ \bibnamefont {Kerckhoff}}, \bibinfo {author}
  {\bibfnamefont {Mark}\ \bibnamefont {Levendorf}}, \bibinfo {author}
  {\bibfnamefont {Sean}\ \bibnamefont {Meenehan}}, \bibinfo {author}
  {\bibfnamefont {Seth~T.}\ \bibnamefont {Merkel}}, \bibinfo {author}
  {\bibfnamefont {Aaron}\ \bibnamefont {Smith}}, \bibinfo {author}
  {\bibfnamefont {Bo}~\bibnamefont {Sun}}, \bibinfo {author} {\bibfnamefont
  {Aaron~J.}\ \bibnamefont {Weinstein}}, \bibinfo {author} {\bibfnamefont
  {Matthew~T.}\ \bibnamefont {Rakher}}, \bibinfo {author} {\bibfnamefont
  {Thaddeus~D.}\ \bibnamefont {Ladd}}, \ and\ \bibinfo {author} {\bibfnamefont
  {Matthew~G.}\ \bibnamefont {Borselli}},\ }\bibfield  {title} {\enquote
  {\bibinfo {title} {Quantifying error and leakage in an encoded si/sige
  triple-dot qubit},}\ }\href@noop {} {\bibfield  {journal} {\bibinfo
  {journal} {Nature Nanotechnology}\ }\textbf {\bibinfo {volume} {14}},\
  \bibinfo {pages} {747--750} (\bibinfo {year} {2019})}\BibitemShut {NoStop}%
\bibitem [{\citenamefont {Bernien}\ \emph {et~al.}(2017)\citenamefont
  {Bernien}, \citenamefont {Schwartz}, \citenamefont {Keesling}, \citenamefont
  {Levine}, \citenamefont {Omran}, \citenamefont {Pichler}, \citenamefont
  {Choi}, \citenamefont {Zibrov}, \citenamefont {Endres}, \citenamefont
  {Greiner} \emph {et~al.}}]{Bernien2017}%
  \BibitemOpen
  \bibfield  {author} {\bibinfo {author} {\bibfnamefont {Hannes}\ \bibnamefont
  {Bernien}}, \bibinfo {author} {\bibfnamefont {Sylvain}\ \bibnamefont
  {Schwartz}}, \bibinfo {author} {\bibfnamefont {Alexander}\ \bibnamefont
  {Keesling}}, \bibinfo {author} {\bibfnamefont {Harry}\ \bibnamefont
  {Levine}}, \bibinfo {author} {\bibfnamefont {Ahmed}\ \bibnamefont {Omran}},
  \bibinfo {author} {\bibfnamefont {Hannes}\ \bibnamefont {Pichler}}, \bibinfo
  {author} {\bibfnamefont {Soonwon}\ \bibnamefont {Choi}}, \bibinfo {author}
  {\bibfnamefont {Alexander~S}\ \bibnamefont {Zibrov}}, \bibinfo {author}
  {\bibfnamefont {Manuel}\ \bibnamefont {Endres}}, \bibinfo {author}
  {\bibfnamefont {Markus}\ \bibnamefont {Greiner}},  \emph {et~al.},\
  }\bibfield  {title} {\enquote {\bibinfo {title} {Probing many-body dynamics
  on a 51-atom quantum simulator},}\ }\href@noop {} {\bibfield  {journal}
  {\bibinfo  {journal} {Nature}\ }\textbf {\bibinfo {volume} {551}},\ \bibinfo
  {pages} {579--584} (\bibinfo {year} {2017})}\BibitemShut {NoStop}%
\bibitem [{\citenamefont {C\^{o}t\'{e}}\ \emph {et~al.}(2009)\citenamefont
  {C\^{o}t\'{e}}, \citenamefont {Ylein},\ and\ \citenamefont
  {DeMille}}]{DeMille2009}%
  \BibitemOpen
  \bibfield  {author} {\bibinfo {author} {\bibfnamefont {R.}~\bibnamefont
  {C\^{o}t\'{e}}}, \bibinfo {author} {\bibfnamefont {S.}~\bibnamefont {Ylein}},
  \ and\ \bibinfo {author} {\bibfnamefont {D.}~\bibnamefont {DeMille}},\
  }\bibfield  {title} {\enquote {\bibinfo {title} {Quantum information
  processing with ultracold polar molecules},}\ }in\ \href@noop {} {\emph
  {\bibinfo {booktitle} {Cold molecules: theory, experiment, applications}}},\
  \bibinfo {editor} {edited by\ \bibinfo {editor} {\bibfnamefont {W.~D.}\
  \bibnamefont {Stwalley}}, \bibinfo {editor} {\bibfnamefont {R.~V.}\
  \bibnamefont {Krems}}, \ and\ \bibinfo {editor} {\bibfnamefont
  {B.}~\bibnamefont {Friedrich}}}\ (\bibinfo  {publisher} {CRC Press},\
  \bibinfo {year} {2009})\BibitemShut {NoStop}%
\bibitem [{\citenamefont {Yu}\ \emph {et~al.}(2019)\citenamefont {Yu},
  \citenamefont {Cheuk}, \citenamefont {Kozyryev},\ and\ \citenamefont
  {Doyle}}]{Yu2019}%
  \BibitemOpen
  \bibfield  {author} {\bibinfo {author} {\bibfnamefont {Phelan}\ \bibnamefont
  {Yu}}, \bibinfo {author} {\bibfnamefont {Lawrence~W}\ \bibnamefont {Cheuk}},
  \bibinfo {author} {\bibfnamefont {Ivan}\ \bibnamefont {Kozyryev}}, \ and\
  \bibinfo {author} {\bibfnamefont {John~M}\ \bibnamefont {Doyle}},\ }\bibfield
   {title} {\enquote {\bibinfo {title} {A scalable quantum computing platform
  using symmetric-top molecules},}\ }\href {\doibase 10.1088/1367-2630/ab428d}
  {\bibfield  {journal} {\bibinfo  {journal} {New Journal of Physics}\ }\textbf
  {\bibinfo {volume} {21}},\ \bibinfo {pages} {093049} (\bibinfo {year}
  {2019})}\BibitemShut {NoStop}%
\bibitem [{\citenamefont {Avinadav}\ \emph {et~al.}(2014)\citenamefont
  {Avinadav}, \citenamefont {Fischer}, \citenamefont {London},\ and\
  \citenamefont {Gershoni}}]{Avinadav2014}%
  \BibitemOpen
  \bibfield  {author} {\bibinfo {author} {\bibfnamefont {Chen}\ \bibnamefont
  {Avinadav}}, \bibinfo {author} {\bibfnamefont {Ran}\ \bibnamefont {Fischer}},
  \bibinfo {author} {\bibfnamefont {Paz}\ \bibnamefont {London}}, \ and\
  \bibinfo {author} {\bibfnamefont {David}\ \bibnamefont {Gershoni}},\
  }\bibfield  {title} {\enquote {\bibinfo {title} {Time-optimal universal
  control of two-level systems under strong driving},}\ }\href {\doibase
  10.1103/PhysRevB.89.245311} {\bibfield  {journal} {\bibinfo  {journal}
  {Physical Review B}\ }\textbf {\bibinfo {volume} {89}},\ \bibinfo {pages}
  {245311} (\bibinfo {year} {2014})}\BibitemShut {NoStop}%
\bibitem [{\citenamefont {Chow}\ \emph {et~al.}(2011)\citenamefont {Chow},
  \citenamefont {C\'orcoles}, \citenamefont {Gambetta}, \citenamefont
  {Rigetti}, \citenamefont {Johnson}, \citenamefont {Smolin}, \citenamefont
  {Rozen}, \citenamefont {Keefe}, \citenamefont {Rothwell}, \citenamefont
  {Ketchen},\ and\ \citenamefont {Steffen}}]{Cross-resonance2}%
  \BibitemOpen
  \bibfield  {author} {\bibinfo {author} {\bibfnamefont {Jerry~M.}\
  \bibnamefont {Chow}}, \bibinfo {author} {\bibfnamefont {A.~D.}\ \bibnamefont
  {C\'orcoles}}, \bibinfo {author} {\bibfnamefont {Jay~M.}\ \bibnamefont
  {Gambetta}}, \bibinfo {author} {\bibfnamefont {Chad}\ \bibnamefont
  {Rigetti}}, \bibinfo {author} {\bibfnamefont {B.~R.}\ \bibnamefont
  {Johnson}}, \bibinfo {author} {\bibfnamefont {John~A.}\ \bibnamefont
  {Smolin}}, \bibinfo {author} {\bibfnamefont {J.~R.}\ \bibnamefont {Rozen}},
  \bibinfo {author} {\bibfnamefont {George~A.}\ \bibnamefont {Keefe}}, \bibinfo
  {author} {\bibfnamefont {Mary~B.}\ \bibnamefont {Rothwell}}, \bibinfo
  {author} {\bibfnamefont {Mark~B.}\ \bibnamefont {Ketchen}}, \ and\ \bibinfo
  {author} {\bibfnamefont {M.}~\bibnamefont {Steffen}},\ }\bibfield  {title}
  {\enquote {\bibinfo {title} {Simple all-microwave entangling gate for
  fixed-frequency superconducting qubits},}\ }\href {\doibase
  10.1103/PhysRevLett.107.080502} {\bibfield  {journal} {\bibinfo  {journal}
  {Phys. Rev. Lett.}\ }\textbf {\bibinfo {volume} {107}},\ \bibinfo {pages}
  {080502} (\bibinfo {year} {2011})}\BibitemShut {NoStop}%
\bibitem [{\citenamefont {Magesan}\ \emph {et~al.}(2011)\citenamefont
  {Magesan}, \citenamefont {Gambetta},\ and\ \citenamefont
  {Emerson}}]{Magesan2011}%
  \BibitemOpen
  \bibfield  {author} {\bibinfo {author} {\bibfnamefont {Easwar}\ \bibnamefont
  {Magesan}}, \bibinfo {author} {\bibfnamefont {J.~M.}\ \bibnamefont
  {Gambetta}}, \ and\ \bibinfo {author} {\bibfnamefont {Joseph}\ \bibnamefont
  {Emerson}},\ }\bibfield  {title} {\enquote {\bibinfo {title} {Scalable and
  robust randomized benchmarking of quantum processes},}\ }\href {\doibase
  10.1103/PhysRevLett.106.180504} {\bibfield  {journal} {\bibinfo  {journal}
  {Phys. Rev. Lett.}\ }\textbf {\bibinfo {volume} {106}},\ \bibinfo {pages}
  {180504} (\bibinfo {year} {2011})}\BibitemShut {NoStop}%
\bibitem [{\citenamefont {Barends}\ \emph
  {et~al.}(2014{\natexlab{b}})\citenamefont {Barends}, \citenamefont {Kelly},
  \citenamefont {Megrant}, \citenamefont {Veitia}, \citenamefont {Sank},
  \citenamefont {Jeffrey}, \citenamefont {White}, \citenamefont {Mutus},
  \citenamefont {Fowler}, \citenamefont {Campbell}, \citenamefont {Chen},
  \citenamefont {Chen}, \citenamefont {Chiaro}, \citenamefont {Dunsworth},
  \citenamefont {Neill}, \citenamefont {O'Malley}, \citenamefont {Roushan},
  \citenamefont {Vainsencher}, \citenamefont {Wenner}, \citenamefont
  {Korotkov}, \citenamefont {Cleland},\ and\ \citenamefont
  {Martinis}}]{Barends2014a}%
  \BibitemOpen
  \bibfield  {author} {\bibinfo {author} {\bibfnamefont {R.}~\bibnamefont
  {Barends}}, \bibinfo {author} {\bibfnamefont {J.}~\bibnamefont {Kelly}},
  \bibinfo {author} {\bibfnamefont {A.}~\bibnamefont {Megrant}}, \bibinfo
  {author} {\bibfnamefont {A.}~\bibnamefont {Veitia}}, \bibinfo {author}
  {\bibfnamefont {D.}~\bibnamefont {Sank}}, \bibinfo {author} {\bibfnamefont
  {E.}~\bibnamefont {Jeffrey}}, \bibinfo {author} {\bibfnamefont {T.~C.}\
  \bibnamefont {White}}, \bibinfo {author} {\bibfnamefont {J.}~\bibnamefont
  {Mutus}}, \bibinfo {author} {\bibfnamefont {A.~G.}\ \bibnamefont {Fowler}},
  \bibinfo {author} {\bibfnamefont {B.}~\bibnamefont {Campbell}}, \bibinfo
  {author} {\bibfnamefont {Y.}~\bibnamefont {Chen}}, \bibinfo {author}
  {\bibfnamefont {Z.}~\bibnamefont {Chen}}, \bibinfo {author} {\bibfnamefont
  {B.}~\bibnamefont {Chiaro}}, \bibinfo {author} {\bibfnamefont
  {A.}~\bibnamefont {Dunsworth}}, \bibinfo {author} {\bibfnamefont
  {C.}~\bibnamefont {Neill}}, \bibinfo {author} {\bibfnamefont
  {P.}~\bibnamefont {O'Malley}}, \bibinfo {author} {\bibfnamefont
  {P.}~\bibnamefont {Roushan}}, \bibinfo {author} {\bibfnamefont
  {A.}~\bibnamefont {Vainsencher}}, \bibinfo {author} {\bibfnamefont
  {J.}~\bibnamefont {Wenner}}, \bibinfo {author} {\bibfnamefont {A.~N.}\
  \bibnamefont {Korotkov}}, \bibinfo {author} {\bibfnamefont {A.~N.}\
  \bibnamefont {Cleland}}, \ and\ \bibinfo {author} {\bibfnamefont {John~M.}\
  \bibnamefont {Martinis}},\ }\bibfield  {title} {\enquote {\bibinfo {title}
  {{Superconducting quantum circuits at the surface code threshold for fault
  tolerance}},}\ }\href {\doibase 10.1038/nature13171} {\bibfield  {journal}
  {\bibinfo  {journal} {Nature}\ }\textbf {\bibinfo {volume} {508}},\ \bibinfo
  {pages} {500--503} (\bibinfo {year} {2014}{\natexlab{b}})}\BibitemShut
  {NoStop}%
\bibitem [{\citenamefont {Magesan}\ \emph {et~al.}(2012)\citenamefont
  {Magesan}, \citenamefont {Gambetta}, \citenamefont {Johnson}, \citenamefont
  {Ryan}, \citenamefont {Chow}, \citenamefont {Merkel}, \citenamefont
  {da~Silva}, \citenamefont {Keefe}, \citenamefont {Rothwell}, \citenamefont
  {Ohki}, \citenamefont {Ketchen},\ and\ \citenamefont
  {Steffen}}]{Magesan2012}%
  \BibitemOpen
  \bibfield  {author} {\bibinfo {author} {\bibfnamefont {Easwar}\ \bibnamefont
  {Magesan}}, \bibinfo {author} {\bibfnamefont {Jay~M.}\ \bibnamefont
  {Gambetta}}, \bibinfo {author} {\bibfnamefont {B.~R.}\ \bibnamefont
  {Johnson}}, \bibinfo {author} {\bibfnamefont {Colm~A.}\ \bibnamefont {Ryan}},
  \bibinfo {author} {\bibfnamefont {Jerry~M.}\ \bibnamefont {Chow}}, \bibinfo
  {author} {\bibfnamefont {Seth~T.}\ \bibnamefont {Merkel}}, \bibinfo {author}
  {\bibfnamefont {Marcus~P.}\ \bibnamefont {da~Silva}}, \bibinfo {author}
  {\bibfnamefont {George~A.}\ \bibnamefont {Keefe}}, \bibinfo {author}
  {\bibfnamefont {Mary~B.}\ \bibnamefont {Rothwell}}, \bibinfo {author}
  {\bibfnamefont {Thomas~A.}\ \bibnamefont {Ohki}}, \bibinfo {author}
  {\bibfnamefont {Mark~B.}\ \bibnamefont {Ketchen}}, \ and\ \bibinfo {author}
  {\bibfnamefont {M.}~\bibnamefont {Steffen}},\ }\bibfield  {title} {\enquote
  {\bibinfo {title} {Efficient measurement of quantum gate error by interleaved
  randomized benchmarking},}\ }\href {\doibase 10.1103/PhysRevLett.109.080505}
  {\bibfield  {journal} {\bibinfo  {journal} {Phys. Rev. Lett.}\ }\textbf
  {\bibinfo {volume} {109}},\ \bibinfo {pages} {080505} (\bibinfo {year}
  {2012})}\BibitemShut {NoStop}%
\end{thebibliography}%

\clearpage
 
\onecolumngrid
\setcounter{figure}{0}
\setcounter{equation}{0}
\setcounter{table}{0}
\renewcommand\theequation{S\arabic{equation}}
\renewcommand\thefigure{S\arabic{figure}}
\renewcommand\thetable{S\arabic{table}}

\setcounter{section}{0}

\section*{Supplementary Material}
\section{CQB Architecture and Measured Parameters}
\begin{center}
\begin{table}[h!]
\begin{tabular}{l l l || l l}
\multicolumn{3}{c}{Transmon} & \multicolumn{2}{c}{CQB}\\
\hline
quantity & value & dependencies & quantity & value\\
\hline
\hline
                            &               &                       &                           &                   \\
\underline{\textbf{Q1}}     &               &                       & \underline{\textbf{CQB-A}}&                   \\
    $f_{\mathrm{readout}}$  & 7.173 GHz     & @ $f_{\mathrm{max}}$  & $f_{\mathrm{target}}$     & 3.610(1) GHz      \\
    $\kappa/2\pi$           & 0.26 MHz      &                       & $\Delta/2\pi$             & 65.4 MHz          \\
    $\chi/2\pi$             & 0.28(3) MHz   & @ $f_{\mathrm{max}}$  & $T_1$                     & $> 2\text{ ms}$   \\
    $f_{\mathrm{max}}$      & 3.825 GHz     &                       & $T_{\textrm{1,leakage from }\ket{1}}$  & 27(1) $\mu s$  \\
    $f_{\mathrm{min}}$      & 3.540 GHz     &                       & $T_{\textrm{1,leakage from }\ket{0}}$  & 41(1) $\mu s$   \\
    $T_1$                   & 40(10) $\mu s$& @ $f_{\mathrm{max}}$  & $T_{\mathrm{2R}}$         & 7.7(5) $\mu s$    \\
    $T_{\mathrm{2R}}$       & 28.2(9)$\mu s$& @ $f_{\mathrm{max}}$  & $T_{\mathrm{2E}}$         & 23(2) $\mu s$     \\
    $T_{\mathrm{2E}}$       & 60(1) $\mu s$ & @ $f_{\mathrm{max}}$  & $T_{\textrm{2E,leakage}}$       & 31(3) $\mu s$     \\
    $E_C/h$               & 199.5(1) MHz          &                                  & $\mathcal{F}$             & 0.9983(1)         \\
\cline{1-3}
                            &               &                       & $\mathcal{F}_{\textrm{leakage}}$               & 0.9990(2)                   \\
\underline{\textbf{Q2}}     &                           &                       &                &                                         \\
    $f_{\mathrm{readout}}$  & 7.203 GHz     & @ $f_{\mathrm{max}}$  &                           &                   \\
    $\kappa/2\pi$           & 0.29 MHz      &                       &                           &                   \\
    $\chi/2\pi$             & 0.31(2) MHz   & @ $f_{\mathrm{max}}$  &                           &                   \\
    $f_{\mathrm{max}}$      & 3.822 GHz     &                       &                           &                   \\
    $f_{\mathrm{min}}$      & 3.536 GHz     &                       &                           &                   \\
    $T_1$                   & 50(13)$\mu s$ & @ $f_{\mathrm{max}}$  &                           &                   \\
    $T_{\mathrm{2R}}$       & 24(1) $\mu s$ & @ $f_{\mathrm{min}}$  &                           &                   \\
    $T_{\mathrm{2E}}$       & 17(1) $\mu s$ & @ $f_{\mathrm{min}}$  &                           &                   \\
    $T_{\mathrm{2E}}$       & 3.2(3)$\mu s$ & @ $f = 3.68$ GHz      &                           &                   \\
    $E_C/h$               & 199.5(1) MHz          &                        &                                    &                     \\
    \hline
    \hline
                            &               &                       &                           &                   \\
\underline{\textbf{Q3}}     &               &                       & \underline{\textbf{CQB-B}}&                   \\
    $f_{\mathrm{readout}}$  & 7.231 GHz     & @ $f_{\mathrm{max}}$  & $f_{\mathrm{target}}$     & 4.120(1) GHz      \\
    $\kappa/2\pi$           & 0.32 MHz      &                       & $\Delta/2\pi$             & 70.2 MHz          \\
    $f_{\mathrm{max}}$      & 4.381 GHz     &                       & $\mathcal{F}$             & 0.9973(1)         \\
    $f_{\mathrm{min}}$      & 4.0427 GHz    &                       & $\mathcal{F}_{\textrm{leakage}}$   & 0.9974(2)\\
    $T_1$                   & 18(4) $\mu s$ & @ $f_{\mathrm{max}}$  &                           &                   \\
    $T_1$                   & 44(7) $\mu s$ & @ $f_{\mathrm{min}}$  &                           &                   \\
    $T_{\mathrm{2R}}$       & 11(2) $\mu s$ & @ $f_{\mathrm{max}}$  &                           &                   \\
    $E_C/h$               & 199.5(1) MHz          &                        &                                    &                     \\
    \cline{1-3}
                            &               &                       &                           &                   \\
\underline{\textbf{Q4}}     &               &                       &                           &                   \\
    $f_{\mathrm{readout}}$  & 7.262 GHz     & @ $f_{\mathrm{max}}$  &                           &                   \\
    $\kappa/2\pi$           & 0.40 MHz      &                       &                           &                   \\
    $f_{\mathrm{max}}$      & 4.338 GHz     &                       &                           &                   \\
    $f_{\mathrm{min}}$      & 4.010 GHz     &                       &                           &                   \\
    $T_1$                   & 1.68(5) $\mu s$& @ $f_{\mathrm{max}}$ &                           &                   \\
    $T_1$                   & 7.0(6) $\mu s$& @ $f_{\mathrm{min}}$  &                           &                   \\
    $E_C/h$               & 199.5(1) MHz          &                        &                                    &                     \\
\end{tabular}
\label{table:S1}
\caption{Device parameter summary.
    $T_{\mathrm{2R}}$ indicates the inhomogeneous $T_2$ measured using a Ramsey experiment.
    $T_{\mathrm{2E}}$ indicates the $T_2$ measured using a Hahn echo experiment.
    $T_{\textrm{1,leakage from }\ket{1}}$ measures the decay time for leakage from the CQB $\ket{1}$ state to states outside the CQB subspace.
    $T_{\textrm{1,leakage from }\ket{0}}$ measures the decay time for leakage from the CQB $\ket{0}$ state to states outside the CQB subspace.
    $T_{\textrm{2E,leakage}}$ measures the $T_{\mathrm{2E}}$ accounting only for relaxation events due to leakage.
    $\mathcal{F}_{\textrm{leakage}}$ measures the simultaneous randomized benchmarking fidelity accounting for decay due to leakage outside the CQB subspace.
    }
\end{table}
\end{center}

The test architecture consists of four frequency-tunable transmon superconducting qubits with an asymmetric flux-tunable SQUID~\cite{Hutchings2017} that interact via nearest-neighbor-fixed-capacitive coupling as shown in Fig.~1\textbf{a}.
The transmon layout is of the ``Xmon style'', with a fixed, always-on capacitive coupling between transmons~\cite{Barends_Martinis_nature2014}.
We set pairs of nearest-neighbor qubits (transmons 1 and 2, and also 3 and 4) to be energy-degenerate.
CQB-A comprises transmons 1 and 2, and CQB-B comprises transmons 3 and 4.

The fixed capacitive coupling between degenerate qubits determines the frequency splitting of CQB-A and CQB-B's hybridized states, labelled $\ket{0}$ and $\ket{1}$ for CQB-A as shown in Fig.~1\textbf{b}.
These hybridized states constitute the CQB subspace, the computational states of our composite qubits.
Hanging resonators allow frequency multiplexed local readout of each physical transmon.
Our system has no local microwave drive lines, which represents an overall decrease in the infrastructure needed to control two CQBs relative to two transmons.
A microwave tone is only needed for initialization and for readout (described in detail below).
All transmon, CQB, and resonator parameters are provided in Table~S1.

\section{Reduced Microwave Resources}

\begin{figure}[b!]
\includegraphics[width=5in]{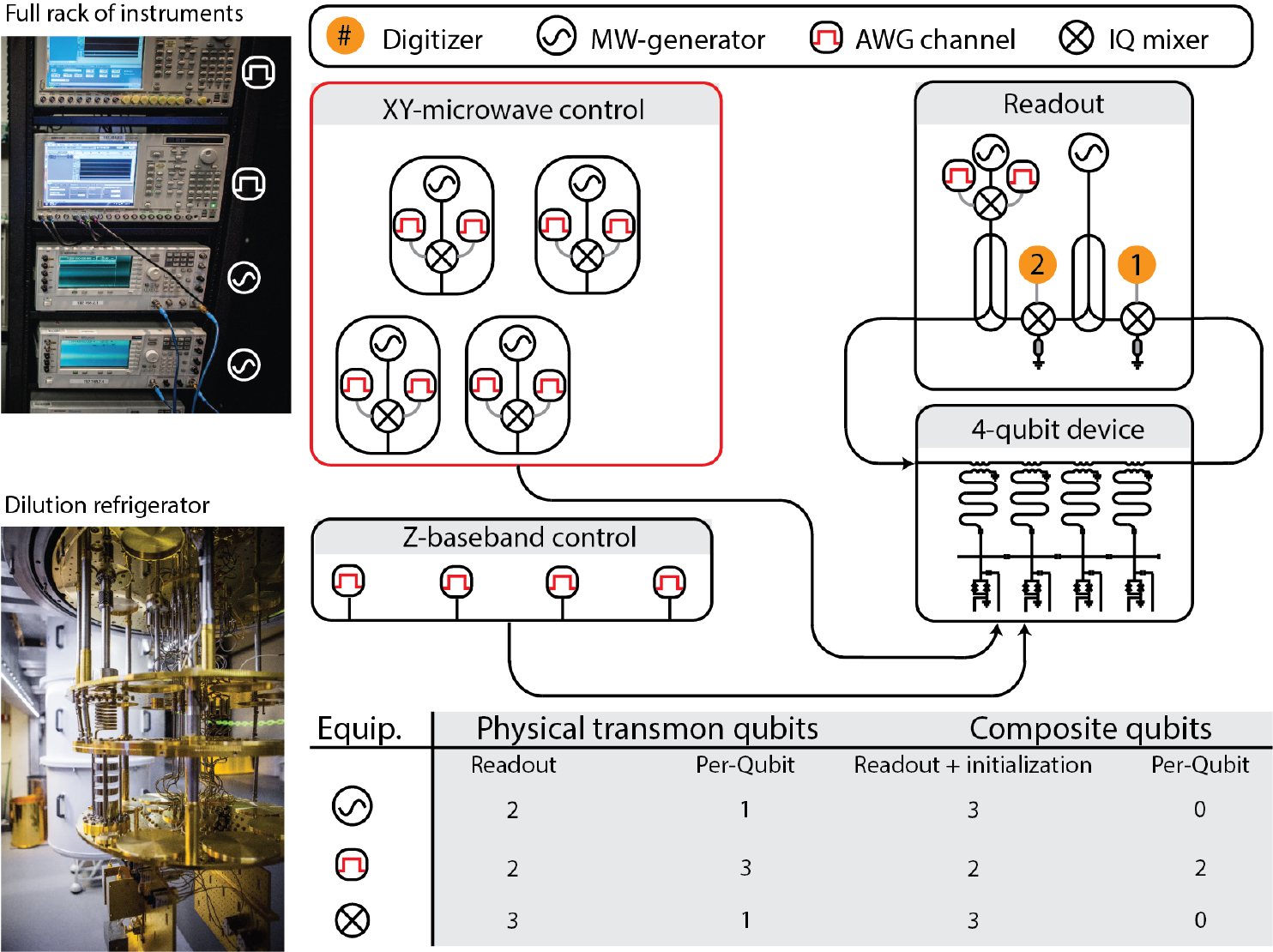}
\caption{\label{fig:resources} A side-by-side comparison of transmon verses CQB qubit architectures shows reduced per-qubit cost for CQB operation.}
\end{figure}

The CQB architecture doubles the number of transmons in a computational unit but does not necessitate a corresponding doubling of the number of phase-stable microwave generators, arbitrary waveform generators (AWGs), and calibration routines.
In fact, it required \textit{less} equipment to perform the experiments in this work.

Fig.~\ref{fig:resources} provides a side-by-side comparison of the per-qubit cost to operate transmons vs CQBs.
For comparison, we consider transmons that require both local $xy$-control for single-qubit operations and flux-based $z$-control for two-qubit gates (IBM's cross resonant gate is a notable exception to this rule~\cite{Cross-resonance2}).
By contrast, CQBs require only baseband $z$-control for all gates.
The caveat to this claim is that initialization of the CQB must be performed using a global microwave drive, which is discussed further in the next section.

Eliminating microwave control of traditional local $xy$-gates 
dramatically reduces the microwave generator cost per-qubit.
In the above transmon case, two AWG channels are mixed with a microwave generator to synthesize a pulse envelop for $xy$-control.
It is necessary to apply precision calibration techniques to eliminate non-linearities introduced by the mixer.

In our experiments, control is implemented by a single AWG channel per qubit.
The calibration procedure is described in detail below.
%

\section{Initialization and Readout}
\label{sec:initreadout}
\begin{figure}
\includegraphics[width=7in]{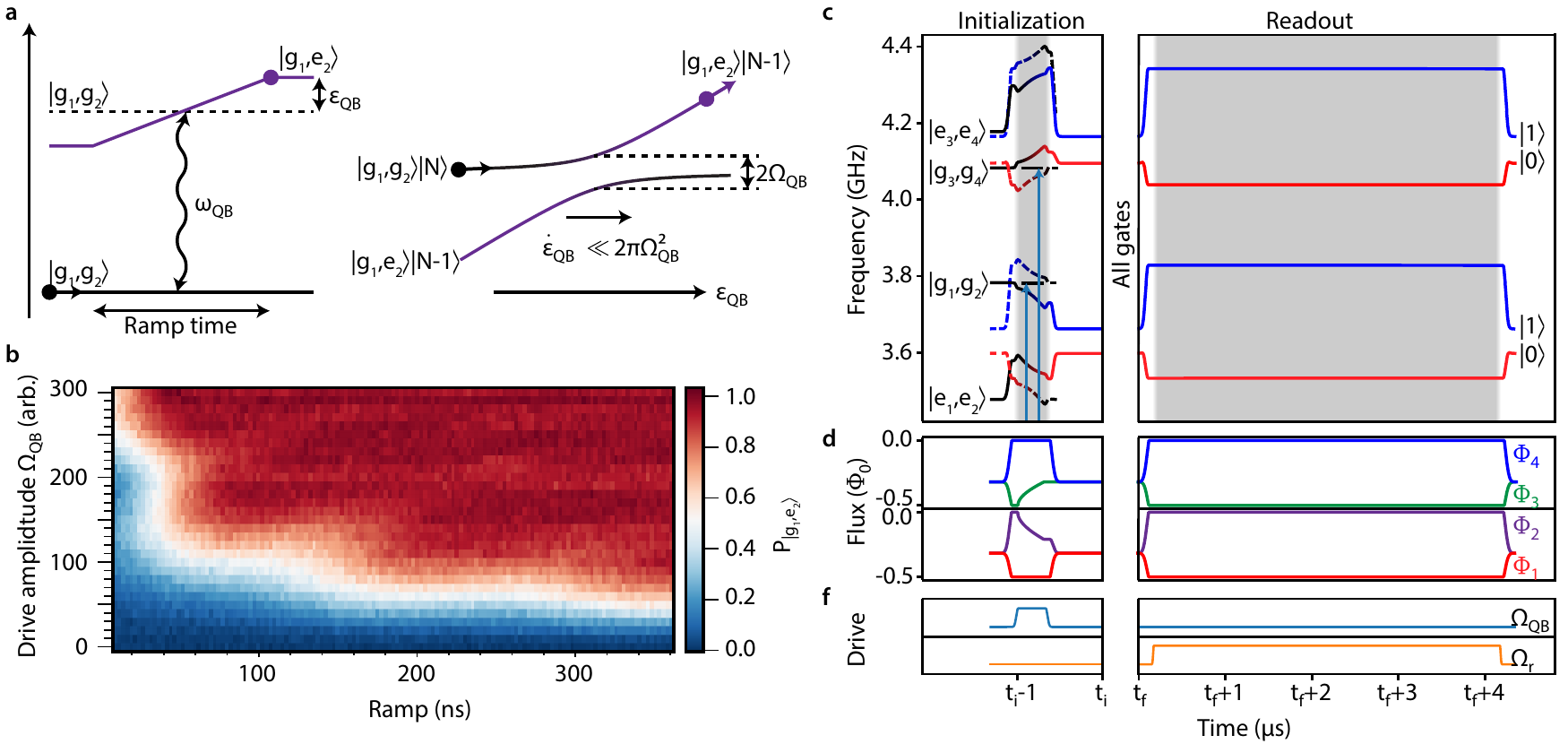}
\caption{\label{fig:LZ_schematic} \textbf{a)} Adiabatic ramps through a driving field can mediate state-changing operations.  \textbf{b)} This approach can robustly prepare a qubit in its excited state over a wide selection of ramp rates and drive amplitudes. \textbf{c)} We show the simulated state structure during our initialization and readout ramp sequences alongside \textbf{d)} the flux and \textbf{e)} microwave control sequences.}
\end{figure}

High fidelity state preparation is a key requirement for quantum computation.
Transmon qubits naturally relax into their ground state, which can be used for passive initialization provided $k_B T \ll \hbar\omega_{1}$.

Unlike most qubit modalities, the computational subspace of a CQB does not include the absolute ground state of the system.
State initialization could be accomplished using microwave pulses tailored to each qubit.
However, this approach would require single-qubit microwave control, which is resource and calibration intensive for an operation that need only be performed once per computation.
Instead, we take an approach described in Figs.~\ref{fig:LZ_schematic}\textbf{a} and \ref{fig:LZ_schematic}\textbf{b}, where a common microwave drive is applied to all the physical qubits through the readout resonators.
This technique does not require local microwave control for each CQB, including IQ mixers and microwave pulse tune-up.
Instead, the frequencies of the physical qubits are swept through a static driving field.
Additionally, this technique has the benefit of being insensitive to small variations in fabrication and the effective driving strength seen by each physical qubit and, therefore, requires minimal tuneup to implement.
The increased time requirement for adiabaticity can be mitigated by using a more intense driving field (a ``fast-adiabatic'' approach).

For example, we consider the initialization of CQB-A, with both transmons 1 and 2 in their ground states, $\ket{g_1,g_2}$.
A microwave drive field is turned on, and the transmon-2 frequency is swept through it, where the qubit-drive detuning is $\epsilon_{QB} = \omega_{2} - \omega_{QB}$.
The states $\ket{g_1,g_2}$ and $\ket{g_1,e_2}$ hybridize via the coherent driving field while sweeping through the resulting Autler-Townes splitting of magnitude $2\hbar\Omega_{QB}$.
The probability of exciting transmon-2 by linearly ramping $\epsilon_{QB}$ is unity if the ramp is adiabatic, and is generally given by the Landau-Zener (LZ) formula
\begin{equation}
P_{e_1g_2} = 1-e^{-2\pi\Omega_{QB}^2 / \dot{\epsilon}_{QB}},
\end{equation}
where $\dot{\epsilon}_{QB}$ is the sweep rate.
In the limit $2\pi\Omega_{QB}^2\gg \dot{\epsilon}_{QB}$, transmon-2 is brought from its ground state $\ket{g_1,g_2}$ to its excited state $\ket{g_1,e_2}$ with high fidelity.

We then adiabatically tune the transmons in flux to the CQB operating point, so that the eigenstate of CQB-A adiabatically evolves 
from $\ket{g_1,e_2}$ to $(1/\sqrt{2})(\ket{g_1,e_2}+\ket{e_1,g_2}) \equiv \ket{1}$. Likewise, the eigenstate of CQB-B evolves from $\ket{e_3,g_4}$ to $(1/\sqrt{2})(\ket{g_3,e_4}-\ket{e_3,g_4}) \equiv \ket{0}$. 
Although the speed limit for this operation ($\sim 2\pi/\Delta \approx 15\text{ ns}$) can be formally realized using a Slepian ramp~\cite{Martinis2014}, or Landau-Zener interferometry of the type described here, 
for simplicity we elect to use here a Gaussian ramp with a $50$~ns $1/e^2$~time constant.
In total, our state preparation takes $\sim 250$~ns.

For readout, we performed a $50$~ns time constant Gaussian ramp, adiabatically bringing transmon-2 to its frequency maximum and transmon-1 to its frequency minimum, as shown in Fig.~\ref{fig:LZ_schematic}c-e.
This operation adiabatically evolves the upper hybridized state $\ket{1}$ onto $\ket{g_1 ,e_2}$ and the lower hybridized state $\ket{0}$ onto  $\ket{e_1, g_2}$.
The transmons rest for $20\text{ ns}$ before a coherent drive is applied to each readout resonator for $4\text{ }\mu\text{s}$ to readout the transmons.

Since the basis states of the CQB are entangled at $\varepsilon = 0$, the information is joint between the two physical qubits.
The readout is designed such that $1/\Delta$ is much smaller than the time it takes to extract information and readout the qubits via the resonator.
As a result, our readout process is sensitive to the energy eigenstate basis, but not to the individual transmon diabatic states.
This is useful for measuring the presence of leakage, while leaving the CQB quantum information essentially unaffected.
However, adiabatically ramping the individual transmons from the CQB degeneracy points to flux bias points where the system eigenstates (solid lines in Figure 1b, 1c) and the constituent diabatic states (dashed lines in Figure 1b, 1c) are essentially indistinguishable enables state-readout.
As mentioned in the main text, this is similar to the SQUID-based measurement of persistent-current flux qubits, where a relatively slow SQUID magnetometer cannot distinguish the circulating currents at the flux qubit degeneracy point due to the relatively large $\Delta$ at gigahertz frequencies~\cite{Orlando1999}.
There, too, the flux qubit was ramped away from degeneracy to allow for qubit readout~\cite{Chiorescu2003}

\section{Flux Control of the CQB}

Gate operations on a CQB are performed by tuning the frequencies of its transmons via baseband flux control.
We fabricated transmons with highly asymmetric tunable Josephson junctions.
In this regime of high asymmetry, the frequencies of the two transmons are given by~\cite{Gramajo2019}
\begin{align}
    \omega_1(\varphi_1) &= \delta\omega_1\cos(2\pi\varphi_1) + \bar{\omega}_1 ~, \\
    \omega_2(\varphi_1) &= \delta\omega_2\cos(2\pi\varphi_2) + \bar{\omega}_2 ~,
\end{align}
where $\delta\omega_i = \left(\omega_i^{(max)}-\omega_i^{(min)}\right)/2$ and $\bar{\omega}_i = \left(\omega_i^{(max)}+\omega_i^{(min)}\right)/2$.
Taking CQB A as an example, the two transmons comprising a CQB are nominally identically designed such that (to a good approximation)  $\delta\omega_1=\delta\omega_2 \equiv \delta\omega_{\textrm{A}}$ and $\bar{\omega}_1=\bar{\omega}_2 \equiv \bar{\omega}_{\textrm{A}}$.
The transmons are energy degenerate when biased at a $\varphi_1=-\varphi_{\textrm{A}}^*$ and $\varphi_2=\varphi_{\textrm{A}}^*$ (see Fig.~\ref{fig:fig1} in the main text). This occurs at transmon frequencies:
\begin{align}
    \omega_{\textrm{A}}^* \equiv \omega_1(-\varphi_{\textrm{A}}^*) &= \omega_2(\varphi_{\textrm{A}}^*) = \delta \omega \cos(2\pi\varphi_{\textrm{A}}^*) + \bar{\omega}_{\textrm{A}}.
\end{align}
The frequency $\omega_{\textrm{A}}^*$ is the frequency at the CQB ``degeneracy point'' that would occur if the transmons were uncoupled, in other words, in the diabatic energy basis. This frequency may be selected by the bias $\varphi_{\textrm{A}}^*$, and we note that it is not unique, but may in principle take any value in the range of accessible transmon frequencies. In our experiment, we chose a value approximately halfway between the maximum and minimum transmon frequencies (see Fig.~\ref{fig:fig1}). Once the specific value for $\varphi_{\textrm{A}}^*$, and thereby $\omega_{\textrm{A}}^*$, is chosen, the coupling of the transmons opens an avoided crossing with strength $\Delta$, determined by the strength of the capacitive coupling between transmons 1 and 2. 

We detune the effective flux $\delta f_{\textrm{A}}$ in this two-level system by biasing the individual transmon $\varphi_i$ away from the degeneracy point $\varphi_{\textrm{A}}^*$ such that $\varphi_1 = -\varphi_{\textrm{A}}^* + \delta f_{\textrm{A}}$ and $\varphi_2 = \varphi_{\textrm{A}}^* + \delta f_{\textrm{A}}$.
Then, the detuning $\delta f_{\textrm{A}}$ is used to set the diabatic energy separation $\varepsilon$:
\begin{align}
    \varepsilon \left(\delta f_{\textrm{A}} \right) &\equiv \omega_1 - \omega_2 \nonumber\\
    &= \delta\omega \left( \cos\left(2\pi\varphi_1\right) - \cos\left(2\pi\varphi_2\right) \right) \nonumber \\
    &= 2 \delta\omega_{\textrm{A}} \sin\left(2\pi\varphi_{\textrm{A}}^*\right) \sin\left(2\pi \delta f_{\textrm{A}}\right) ~.
\end{align}
For small $\delta f_{\textrm{A}} (\ll 1)$, the energy separation $\varepsilon$ is linear with the flux $\delta f_{\textrm{A}}$,
\begin{equation}
    \varepsilon\left(\delta f_{\textrm{A}}\right) \approx 4\pi\delta\omega_{\textrm{A}} \sin\left(2\pi\varphi_{\textrm{A}}^*\right)\delta f_{\textrm{A}}~.
\end{equation}

Although the frequencies $\omega_{\textrm{A}}*$ and $\omega_{\textrm{B}}*$ are generally fixed for single qubit operation, they are tuned to implement the two-qubit CZ gate, as described in the main text. By tuning the value of $\varphi_{\textrm{A}}*$ and $\varphi_{\textrm{B}}*$, the frequencies $\omega_{\textrm{A}}*$ and $\omega_{\textrm{B}}*$ can be tuned relative to one another while maintaining each individual CQB at its noise-insensitive point (at the avoided crossing). This may be contrasted with conventional transmon implementations of a CZ gate, for which the transmons must leave the noise-insensitive operating point to realize the two-qubit gate.    


\section{Single-CQB Gates}
\label{sec:scqb}
\begin{figure}
\includegraphics[width=3.3in]{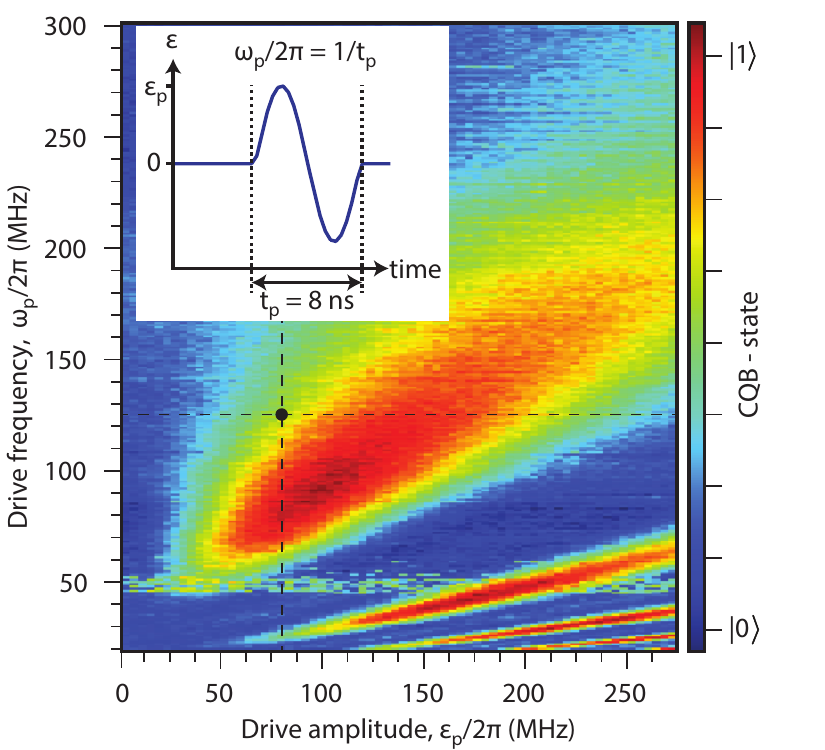}
\caption{\label{fig:S4} The color contour plot shows the excitation probability as a function of driving amplitude $\varepsilon_p$ and frequency $\omega_p/2\pi = 1/t_p$, which are defined again in the inset. The dotted lines indicate the operating point for $X(\pi/2)$-gates on CQB-A. In principle, the gate can be parameterized anywhere along the color contour. While it is possible to obtain $X(\pi)$-gates from such a 2D scan, it is not straightforward to then infer a robust parametrization of $X(\pi/2)$, requiring a separate scan to complete a universal gate set. The features at $50$~MHz are imaging artifacts.}
\end{figure}

The class of waveforms (pulses) that produce high fidelity $xy$ gates can be inferred from a close look at the three key properties of small gapped systems and Landau-Zener interference.
First, idling the CQB at its avoided level crossing $\varepsilon = 0$ has many advantages (discussed further in the noise immunity section of this supplement).
Therefore, each pulse in a CQB's set of gates starts and ends at $\varepsilon = 0$.
Second, the waveform must induce a degree of non-adiabaticity when sweeping away, through, or towards the avoided crossing to induce Landau-Zener state transitions and quantum interference.
In general, a pulse needs a fast changing (short timescale) component near $\varepsilon = 0$ to mediate transitions, and a phase evolution component when $\varepsilon \ne 0$ that mediates constructing and destructive interference. 
The avoided crossing acts as a beamsplitter for the qubit state, and non-adiabiatically leaving or traversing the avoided crossing mixes the diabatic states.
And, as in an interferometer~\cite{Oliver_Science2005}, the phase evolution is responsible for the resulting state 
when the pulse returns to $\varepsilon = 0$.
Noise on this phase evolution brings us to our third property of small gapped systems: ambient noise in the environment and the control typically take the form of $1/f$ noise which couple primarily to the (often slowly changing) phase evolution component of a pulse.
Using zero-average symmetric flux pulses is a technique that samples the noise twice, with $\varepsilon >0$ and again with $\varepsilon < 0$, such that extra phase evolution on one excursion due to noise is canceled by reduced phase evolution on the opposite excursion away from $\varepsilon =0$.
This approach also mitigates memory effects in the flux bias from eddy currents by making the time averaged flux a constant.

In our setup, the baseband flux control has a sharp dropoff at $400\text{ MHz}$.
As a consequence, pulse waveforms with high frequency components, such as square pulses, are undesirable.
In contrast, the higher-frequency components of a single-period sinusoidal pulse fall off relatively quickly, and the pulse is still relatively fast changing near $\varepsilon = 0$.
Fig.~\ref{fig:S4} plots the CQB state as a function of frequency and amplitude for such a sinusoid.
The sinusoid used in this work has an $8$~ns period (125~MHz) and an amplitude $\varepsilon_p \sim 80$~MHz.

The deep red regions of Fig.~\ref{fig:S4} correspond to the implementation of a $X(\pi)$ gate, which rotates the qubit state by $\pi$ around $x$-axis.
However, a $\pi$ rotation would require additional gates to form a universal gate set, e.g., to form Hadamard gate.
%
For this reason, and for the practicality of calibration, we elect to form $X(\pi/2)$ and $Y(\pi/2)$ gates.

\subsection{Motivating the need for extra idling time during a gate}

In this subsection, we mathematically describe, as a unitary gate $\mathcal{G}$, the operation of an arbitrary set of excursions from $\varepsilon = 0$ (the form of these excursions is a sinusoid in our work) that transforms a CQB from the poles of the Bloch sphere onto the equator.
\begin{equation}
\mathcal{G}\left(
\begin{array}{c}
1 \\
0
\end{array}
\right)
= \frac{1}{\sqrt{2}}\left(
\begin{array}{c}
1 \\
e^{i\phi}
\end{array}
\right),
\quad
\mathcal{G}\left(
\begin{array}{c}
0 \\
1
\end{array}
\right)
= \frac{e^{i\phi'}}{\sqrt{2}}\left(
\begin{array}{c}
1 \\
-e^{i\phi}
\end{array}
\right)
\end{equation}
can be written as
\begin{equation}
\mathcal{G}\left(\phi,\phi'\right) = \frac{1}{\sqrt{2}}
\left(
\begin{array}{cc}
1 & e^{i\phi'} \\
e^{i\phi} & e^{i\left(\phi+\phi'+\pi\right)}
\end{array}
\right)~.
\end{equation}
In general, $\mathcal{G}\left(\phi,\phi'\right)$ will not be an $X(\pi/2)$-gate, and applying $\mathcal{G}$ a multiple of four times will not return the system to its original state, where an $X(\pi/2)$-gate is a fixed angle version of the more general $X(\theta)$-gate, which rotates a qubit about the $x$-axis of its Bloch sphere by $\theta$
\begin{align}
X(\theta) \equiv e^{-i\frac{\sigma_x}{2} \theta}.
\end{align}
The mathematical properties of an $X(\pi/2)$ gate may be recovered by adding corrective $Z(\alpha)$-gates before and after $\mathcal{G}\left(\phi,\phi'\right)$.
Defined by analogy with $X(\theta)$-gates, $Z(\alpha)$ gates perform a free rotation about the $z$-axis by angle $\alpha$.

To perform this correction we solve for the following $Z$-gate angles $\alpha$ and $\beta$,
\begin{equation}
Z(\alpha) \mathcal{G}\left(\phi,\phi'\right) Z(\beta)
=\frac{1}{\sqrt{2}}e^{-i\frac{\alpha+\beta}{2}}
\left(
\begin{array}{cc}
1 & e^{i\left(\phi'+\beta\right)} \\
e^{i\left(\phi+\alpha\right)} & e^{i\left(\phi+\phi'+\alpha+\beta+\pi\right)}
\end{array}
\right)
=\frac{1}{\sqrt{2}}
\left(
\begin{array}{cc}
1 & -i \\
-i & 1
\end{array}
\right)
= X(\pi/2) ~,
\end{equation}
with
\begin{equation}
\alpha = -\phi - \frac{\pi}{2} + 2m\pi , \quad\quad
\beta = -\phi' - \frac{\pi}{2} + 2n\pi
\end{equation}
for any integers $m$ and $n$.
Similarly, we can realize $Y(\pi/2)$ gate by choosing
\begin{equation}
\alpha = -\phi + 2m\pi , \quad\quad
\beta = -\phi' + \left(2n+1\right)\pi ~,
\end{equation}
which has the sinusoidal pulse shifted by $\pi/2$ $z$-rotation from the $X(\pi/2)$ gate.
In practice, we don't need to scan $\alpha$ and $\beta$ independently to optimize $X(\pi/2)$ gate.
We only need to scan the total idle $z$-rotation, $\alpha+\beta$, and set $\alpha$ (or $\beta$) to define the $x$-axis.
In this work, we set $\alpha$ for $X(\pi/2)$ such that the sinusoid is time-shifted from the center of the total duration ($t_d/2$) by $t_{xy}/2=t_{\Delta}/8$.
Then the $Y(\pi/2)$ can be achieved by advancing the sinusoid by $t_{xy}=t_{\Delta}/4$.

Once we have tuned up a $X(\pi/2)$ gate, we can implement any arbitrary single qubit gate using $X(\pi/2)$ and $Z(\alpha)$.
Any single qubit unitary gate can be written as
\begin{equation}
\mathcal{U}\left(\theta,\phi,\psi\right) = \left(
\begin{array}{cc}
\cos\frac{\theta}{2} & -\sin\frac{\theta}{2}e^{-i\phi}e^{i\psi} \\
\sin\frac{\theta}{2}e^{i\phi} & \cos\frac{\theta}{2}e^{i\psi}
\end{array}
\right)~,
\end{equation}
which can be realized as
\begin{equation}
\mathcal{U}\left(\theta,\phi,\psi\right) = Z(\alpha) X(\pi/2) Z(\beta) X(\pi/2) Z(\gamma)
\end{equation}
with
\begin{eqnarray}
\alpha &=& \phi + 2m\pi ~,\\
\beta &=& \pi - \theta ~,\\
\gamma &=& -\phi + \psi + \left(2n+1\right)\pi ~,
\end{eqnarray}
for any integers $m$ and $n$.

\subsection{Tune-up procedure}

The discrete $X(\pi/2)$ and $Y(\pi/2)$ gates and a continuously parameterizable $Z(\alpha)$ can be used to form a complete universal single-CQB gate set.
Strictly speaking, one of the $X(\pi/2)$ and $Y(\pi/2)$ is not necessary, but we include it for convenience.
The $X(\pi/2)$ gate consists two parts: a simple sinusoidal excursion in $\epsilon(\delta f)$ (we choose to fix the sinusoid frequency $\omega_p/2\pi$); and the duration of the gate $t_d$, which ensures that consecutive identical gates constructively interfere.
The second gate $Z(\alpha)$ arises from the relative time evolution of the CQB's eigenenergies, with a period $t_{\Delta} = 2\pi/\Delta$ such that $\alpha = 2\pi t_d/t_{\Delta}$ at $\varepsilon = 0$.
We outline the tune-up procedure for these gates in what follows.

\begin{figure}
\includegraphics[width=5.1in]{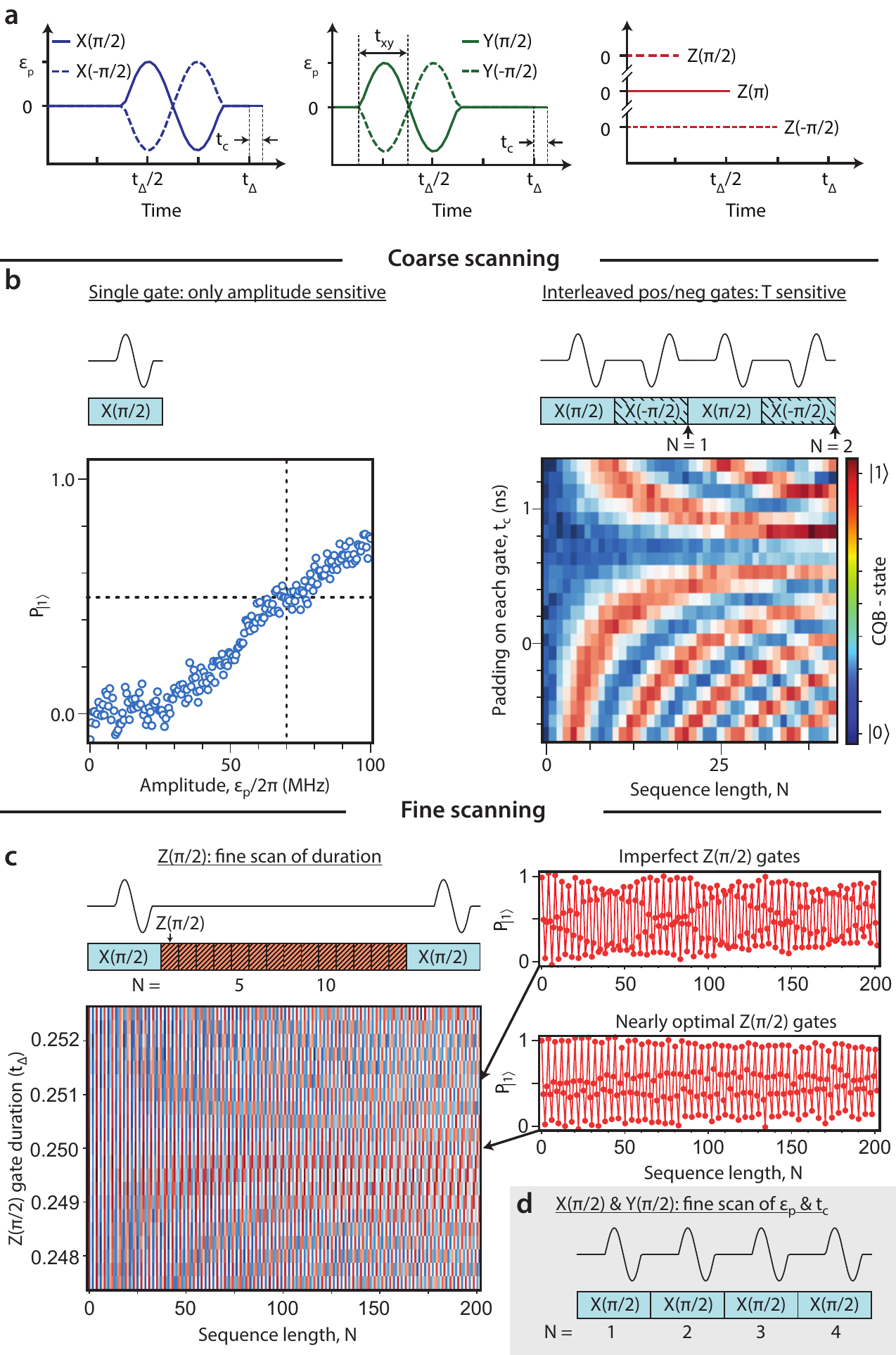}
    \caption{\label{fig:S3} Tune up procedure.
    \textbf{a)} $X(\pi/2)$ and $Y(\pi/2)$ gates are produced using a single period sinusoidal drive while the $Z(\pi/2)$ gate is simply time evolution for $t_d = t_{\Delta}/4 = 2\pi/4\Delta$.
    \textbf{b)} With fixed driving period, our tuneup procedure first identifies the amplitude at half excitation and the gate duration $t_d$ that reveals the expected interference pattern (see text).
    \textbf{c)} Fine scanning involves stepping gates in quarter rotations about the Bloch sphere (or an odd multiple of quarter rotations) while scanning a parameter. This takes the form of consecutive $X(\pi/2)$ or $Z(\pi/2)$ gates. These traces can be robustly Fourier transformed to identify the ideal operating parameter.
    \textbf{d)} Similarly, concatenating many $X(\pi/2)$ gates allows fine scanning of $\varepsilon_p$ and $t_c$.}
\end{figure}

We first set up transmon readout and perform spectroscopy while sweeping $\delta f$ to obtain an estimate for $\Delta$.
We then pick a sinusoid frequency $\omega_p/2\pi$ and scan for a driving amplitude $\varepsilon_p$ that produces a half excitation in the CQB, as shown in Fig~\ref{fig:S3}b.
Next, we want consecutive operations of $X(\pi/2)$ gates to rotate about the same axis, such that they constructively interfere.
To obtain this, we apply $X(\pi/2)$ and $X(-\pi/2)$ pulses, one after the other, such that the sequence is the identity when consecutive pulses rotate about the same axis.
We then vary a correction time $t_c$, which is padded to the end of gate as shown in Fig.~\ref{fig:S3}a until this condition is satisfied, as demonstrated in Fig.~\ref{fig:S3}b.
We then chain many $X(\pi/2)$ gates to increase the measurement sensitivity to small errors and perform fine tuning in $\varepsilon_p$ and $t_d$.

$Z(\alpha)$ is tuned up by performing a Ramsey measurement. Figure~\ref{fig:S3} shows the Ramsey measurement with four samples per period $2\pi/\Delta$, speeding up the measurement while avoiding imaging artifacts.
This measurement obtains a precise value for $\Delta$.

Finally, the distinction between $X(\pi/2)$ and $Y(\pi/2)$ is given by the time shift $t_{xy} = 1/4 \times 2\pi/\Delta = t_{\Delta}/4$.
This moves the $X(\pi/2)$ ($Y(\pi/2)$) gate sinusoid to the right (left) in its window by an eighth of a period.
The total shift between the two sinusoids is exactly a quarter period.

\section{Measuring CQB coherence and energy relaxation}

\begin{figure}
\includegraphics[width=6.3in]{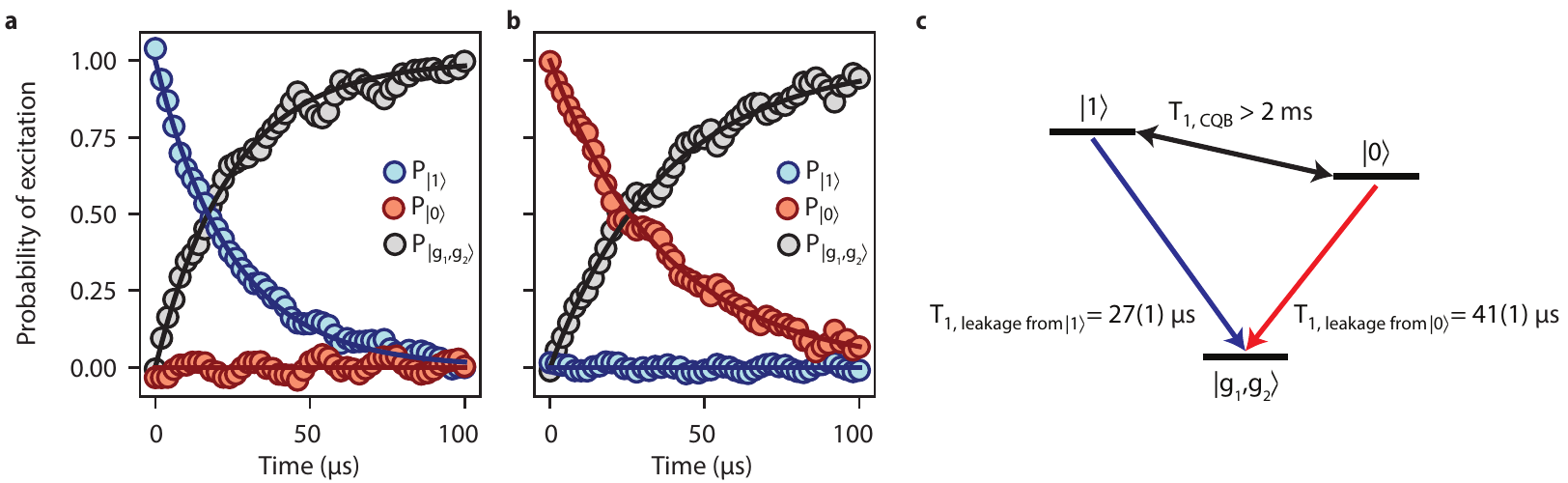}
\caption{\label{fig:leakage} We measure the energy relaxation within the CQB-A subspace and leakage from $\ket{0}$ and $\ket{1}$ to the overall ground state $\ket{g_1g_2}$. \textbf{a)} CQB-A is initially prepared in $\ket{0}$ and monitored as a function of time. \textbf{b)} CQB-A is instead prepared in $\ket{1}$. \textbf{c)} Population switching from the initially populated CQB-A state to the unpopulated one is evidence for T1 relaxation within the CQB subspace. Statistically, this process is likely $T1 > 2$~ms. Leakage lifetimes from $\ket{1}$ and $\ket{0}$ to the ground state are given in the figure.}
\end{figure}

Energy relaxation measurements are performed by preparing a CQB eigenstate, either $\ket{0}$ or $\ket{1}$, and monitoring the average state populations as a function of time to obtain the data shown in Fig.~\ref{fig:leakage}.
At each time step, we are able measure the population of the system eigenstates $\ket{0}$, $\ket{1}$, and $\ket{|g_1,g_2}$ using the readout techniques outlined in the section ``Initialiation and Readout''.
To separate out the relaxation within the CQB subspace from the leakage out of the CQB subspace we simultaneously fit the functions:
\begin{align}
\label{eq:relaxation}
P_{\ket{1}}(t) &= [0.5 + 0.5\exp{(-\Gamma_{CQB}t)}]\exp{(-\Gamma_{\textrm{leakage}}t)}P_{\ket{1}}(0)\nonumber\\
P_{\ket{0}}(t) &= [0.5 - 0.5\exp{(-\Gamma_{CQB}t)}]\exp{(-\Gamma_{\textrm{leakage}}t)}P_{\ket{0}}(0)\nonumber\\
P_{\ket{g_1,g_2}}(t) &= [1-\exp{(-\Gamma_{\textrm{leakage}}t)}]P_{\ket{g_1,g_2}}(0).
\end{align}
For example, in Fig.~\ref{fig:leakage}\textbf{b}, the CQB is prepared in $P_{\ket{1}}(0)$, which could relax into $P_{\ket{0}}(0)$ or leak into $P_{\ket{g_1,g_2}}(t) = 1 - P_{\ket{1}}(t) - P_{\ket{0}}(t)$.
Eq.~\ref{eq:relaxation} implicitly presumes an equal up and down rate to the relaxation in the CQB subspace, essentially asserting that for gaps $\omega/2\pi \approx 70$ MHz, the Boltzmann factor $\exp[-\hbar \omega / k_{\mathrm{B}}T] \approx \exp[-70/800] =0.92$ is approximately 1, where we take $T \approx 40$ mK (equivalent to about 800 MHz) as the qubit temperature.
However, as shown in Fig.~\ref{fig:leakage}, we were not able to detect \textit{any} relaxation within the CQB subspace, and statistically bound the relaxation times $T_{1, CQB} = 1/\Gamma_{CQB}$ above $2$~ms.
As explained in the main text, this is due in part to (1) the small gap, and (2) the need for a correlated two-photon interaction with the environment to cause such transitions (up or down).

Ramsey and Echo decoherence measurements are obtained by preparing the CQB in $\ket{0}$ and then performing the appropriate sequence of single-CQB gates, which are defined in the section~``Sinlge-CQB Gates''.
For Ramsey, we apply $X(\pi/2)$, implement increasing numbers $N$ of $Z(\pi/2)$ idling gates in odd-numbered increments $m$ (e.g., increment $m=1$ would be $1,2,3 \ldots$, or $m=3$ would be $3,6,9, \ldots$, etc.) -- where increasing the number of idling gates is equivalent to scanning the free-evolution time $t_N = N m t_{\Delta}/4$ -- followed by a final $X(\pi/2)$ gate.
The use of an odd-increment $m$ generates an oscillating pattern similar to that seen with a detuned Ramsey experiment (see Fig.~\ref{fig:S3}\textbf{c} for an example), with a detuning ``frequency'' of $1/m t_{\Delta}$.
Similarly, a Hahn echo sequence implements $X(\pi/2)$ followed by $N/2$ $Z(2\pi)$ gates, $Y(\pi/2)Y(\pi/2)$, another $N/2$ $Z(2\pi)$ gates, and is completed by a final $X(\pi/2)$ gate.
In this case, because we are applying finite-time identity gates, any choice of increment $m$ leads ideally to a monotonic decay envelope.
The total duration of the Hahn echo sequence is then $N t_{\Delta}$.
We fit these sequences to Eq.~\ref{eq:relaxation}, with an additional sinusoidal term in the first two lines for Ramsey, and compute the in-CQB $T_{2R}$ and $T_{2E}$ as well as the sequences' leakage timeconstants $T_{2R\textrm{, leakage}}$ and $T_{2E\textrm{, leakage}}$.

\section{Noise immunity of the composite qubit}

\subsection{Noise sensitivity of the transmon and the composite qubit}
\subsubsection{Flux Noise}
Flux noise is one of the main sources of decoherence for transmons with tunable Josephson junctions.
For a transmon with frequency
\begin{equation}
    E(\varphi) = \delta\omega\cos(2\pi\varphi) + \bar{\omega} ~,
\end{equation}
the sensitivity of the qubit frequency with respect to some fluctuating noisy flux $\varphi=\varphi_0 + \delta\varphi$ is
\begin{eqnarray}
    \delta E &\equiv& E(\varphi) - E(\varphi_0) \nonumber\\
    &\approx& -2\pi \delta\omega \sin\left(2\pi\varphi_0\right)\delta\varphi - 2\pi^2\delta\omega\cos\left(2\pi\varphi_0\right) \delta\varphi^2 ~.
\end{eqnarray}
At the flux sweet spot $\varphi_0 = 0, 1/2, 1, \ldots$,
\begin{equation}
    \left| \delta E \right| = 2\pi^2\delta\omega \delta\varphi^2 ~.
\end{equation}

The frequency of the CQB is $f_{CQB}=\sqrt{\Delta^2 + \varepsilon^2}$,
and its sensitivity at the CQB sweet spot $\varepsilon=0$ with respect to the fluctuating flux $\varphi_1 = -\varphi^* + \delta\varphi_1$ and $\varphi_2 = \varphi^* + \delta\varphi_2$ is
\begin{equation}\label{eq:flux_noise}
    \delta f_{CQB} \approx \frac{2\pi^2\delta\omega^2\sin^2\left(2\pi\varphi^*\right)}{\Delta} \left(\delta\varphi_1+\delta\varphi_2\right)^2 ~.
\end{equation}
The CQB is insensitive to fluctuations in $\varphi_1$ and $\varphi_2$ to first order, even though its constituent transmons are not biased at their conventional flux sweet spots.
This is the usual protection afforded by an avoided crossing.
For the CQB-A studied in this work ($\delta\omega$ = 143 MHz, $\Delta$ = 65 MHz, $\varphi^*$ = 0.28),
\begin{equation}
    \left| \frac{\delta f_{CQB}}{\delta E} \right| \approx \sin^2\left(2\pi\varphi^*\right) \frac{\delta\omega}{\Delta} = 2.108 ~.
\end{equation}

\subsubsection{Photon Shot Noise}
Another major source of decoherence is photon shot noise - photon number fluctuations in the readout resonators that lead to the fluctuation of the transmon qubit frequencies.
Individual transmon qubits are not protected from this type of decoherence at any frequency, whereas the CQB is first-order insensitive to this type of noise.
For fluctuating transmon frequencies, $E_i = \bar{E}_i + \delta E$, the sensitivity of the CQB frequency is
\begin{eqnarray}
    \delta f_{CQB} &=&  f_{CQB} - \bar{f}_{CQB} \nonumber \\
    &\approx& \varepsilon \left(\Delta^2+\varepsilon^2\right)^{-1/2} \left(\delta E_1 - \delta E_2 \right) \nonumber\\
    && + \frac{1}{2} \left(  \left(\Delta^2+\varepsilon^2\right)^{-1/2} - \varepsilon^2 \left(\Delta^2+\varepsilon^2\right)^{-3/2} \right) \left(\delta E_1 - \delta E_2 \right) ^2 ~.
\end{eqnarray}
At the CQB sweet spot $\varepsilon=0$, the first order term vanishes and
\begin{equation}\label{eq:photon_shot_noise}
    \delta f_{CQB} \approx \frac{1}{2\Delta} \left(\delta E_1 - \delta E_2 \right) ^2 ~.
\end{equation}
Therefore CQB is first-order insensitive to photon shot noise, unlike the individual transmons, as demonstrated in the main text.
This again arises due to the avoided crossing, but it is manifest in the CQB, because $\varepsilon$ is related to any fluctuation of the individual transmon frequency, and not just those due to flux noise.

The CQB's immunity to \textit{any} frequency fluctuations on the constituent transmons can be further mitigated by increasing the capacitive coupling $\Delta$ between the transmons.
This effectively broadens the curvatures of the avoided crossing region and extends the region of first-order insensitivity.
This is mathematically captured for the examples of flux noise and photon shot noise explored here from Eqs. (\ref{eq:flux_noise}) and (\ref{eq:photon_shot_noise}).

\subsection{Photon shot noise experimental procedure}

The scattering response of photons driven through a readout resonator gives information about the state of a transmon coupled to that resonator: this physics underlies dispersive readout of superconducting circuits.
During readout, those same photons map noise onto the transmon, such that transmon coherence is lost.
As a consequence, transmon coherence is sensitive to \textit{unwanted} thermal or coherent photon fluctuations in the readout resonator.

Given the importance of transmon-resonator interactions for the purpose of readout, the spectral profile and amplitude of photon shot noise due to a coherently driven resonator and its influence on a transmon is well understood~\cite{Gambetta2006}.
We make use of this relationship to make a well controlled study of CQB-A's sensitivity to noise relative to its constituent transmons.

We used Ramsey measurements to obtain the coherence times of transmon-1, transmon-2, and CQB-A, while a variable amplitude drive was applied to the readout resonator.
For the transmons, the probability of measuring the qubit in the excited state after driving for time $\tau$ has in general the complicated functional form given by Ref.~\cite{Gambetta2006}
\begin{align}
\label{eq:gam}
    P_e(\tau) = \exp{[-(1/T_{2Ramsey}+\Gamma)t-i(\omega_a+B)t]} \exp{[A(1-\exp{[-(\kappa/2+i\chi+i\Delta_r)t]})]},
\end{align}
where we attempted to zero the detuning from the bare resonator frequency $\Delta_r = 0$.
The native decoherence rate $1/T_{\textrm{2R}}$ is enhanced by the Lorentzian $\Gamma$ term, which is linearly proportional to photon number in the resonator.
The parameter $\omega_a$ corresponds to the average frequency of the qubit relative to the Ramsey ``clock'', and $B$ is a photon-number-dependent frequency shift (AC Stark shift).
Finally, the term with $A$ in the argument gives rise to a non-exponential decay profile and reflects the relaxation of the resonator to a steady state driving field.
See Ref.~\cite{Gambetta2006} for further details about Eq.~\ref{eq:gam}.
In our case, the temporal profile of the CQB Ramsey coherence fits reasonably well to a simple exponential decay.
Due to the differing functional forms of the Ramsey coherence decay profile between bare transmons and the CQB, we explicitly define the decoherence rate as the $1/e$ time of the measurement.

Because the resonator tuneup varies slightly with qubit frequency, we first prepare transmon-1 at a flux insensitive point, with transmon-2 detuned.
We then finely scan for transmon-1's resonator, to maximize dephasing, and measure the qubit dephasing as a function of the drive power applied to its resonator. We repeat this procedure for transmon-2.

Finally, we prepare CQB-A at its operating point and find the resonator frequency for which the readout histogram separation between the CQBs $\ket{0}$ and $\ket{1}$ states is largest.
The fits to the decay profiles of the constituent transmons are used to infer the number of photons in the resonator and the value of $\chi$.

\section{Randomized Benchmarking}

Randomized benchmarking (RB) allows characterization of a gate's fidelity averaged over many initial conditions and reduces the contribution of state-preparation and measurement (SPAM) errors in the computed gate fidelity relative to gate set tomography and process tomography~\cite{Magesan2011,Barends2014a}.
Clifford-based RB uses a random sequence of transformations that are intended to evenly sample the Bloch sphere, namely by using gates from the Clifford group.
The Clifford group is the set of transformations closed under any combination of Pauli operators.
Before running the sequence, we classically compute the result of that sequence and concatenate a ``recovery gate'' such that the CQB returns to the initial state if no errors occur.
We first compute and measure the recovery probability for a given number of Cliffords by averaging over randomly sequences of Clifford gates.
We then vary the length of these sequences and fit the resulting trace to a discrete exponential decay function to obtain the average Clifford fidelity.
Interleaved RB interleaves a gate of interest with random Clifford gates.
The decay profile of an interleaved gate set is compared with that of a reference gate set to compute the fidelity of the interleaved gate.

The recovery probability $p_0$ of the reference follows a simple exponential decay with number of Cliffords $m$ assuming gate independent and time-independent errors
\begin{align}
    p_0 &= 0.5 + 0.5\lambda_{1}^m\\
    \mathcal{F}_{1} &= \frac{1}{d}[(d-1)\lambda_{1}+1]\text{, }d=2\nonumber\\
\end{align}
where the fidelity is rescaled by the dimensionality of the Hilbert space $d$, to give the standard gate fidelity $\mathcal{F}_{CQB}$.

The CQB also experiences leakage to the overall ground state $\ket{g_1,g_2}$, which can modify the extracted gate fidelities. %
Sophisticated techniques have been developed to characterize leakage in transmon systems, however they appear ill-suited for describing the relatively simple leakage in a CQB architecture.
The CQB's leakage is one-way, independent of $\mathcal{F}_{CQB}$, and independent of CQB-state.
Therefore, we use a multiplicative model in the recovery probability, the validity of which we verify with a Monte Carlo simulation. %
Because we readout both transmons, we can compute the population of the three states of interest ($\ket{0}, \ket{1}, \ket{g_1,g_2}$)
\begin{align}
    p_0 &= (0.5 + 0.5\lambda_{1-CQB}^m)\lambda_{1-leakage}^m\\
    p_1 &= (0.5 - 0.5\lambda_{1-CQB}^m)\lambda_{1-leakage}^m\nonumber\\
    p_0+p_1 &= \lambda_{1-leakage}^m\nonumber\\
    \mathcal{F}_{1-CQB} &= \frac{1}{d}[(d-1)\lambda_{1-CQB}+1]\text{, }d=2\nonumber\\
    \mathcal{F}_{1-leakage} &= \lambda_{1-leakage}.\nonumber
\end{align}
The fidelities $\mathcal{F}_{1-CQB}$ and $\mathcal{F}_{1-leakage}$, extracted from the reference RB, give the average Clifford fidelities: we did not further break down the single-CQB gates by performing interleaved RB.

For 2-CQB gate fidelity, we use interleaved RB~\cite{Magesan2012}.
We compute the 2-CQB RB recovery probabilities for the reference dataset in the following way
\begin{align}
    p_0p_0 &= (0.25 + 0.75\lambda_{2-CQB}^m)\lambda_{2-leakage}^m\\
    p_0p_0+p_1p_0+p_0p_1+p_1p_1 &= \lambda_{2-leakage}^m.\nonumber\\
\end{align}
Likewise, we compute the recovery probabilities for the dataset that interleaves CZ gates
\begin{align}
    p_0p_0 &= (0.25 + 0.75\rho_{2-CQB}^m)\rho_{2-leakage}^m\\
    p_0p_0+p_1p_0+p_0p_1+p_1p_1 &= \rho_{2-leakage}^m.\nonumber\\
\end{align}
And we may compute the resulting CZ-specific fidelity using the following formula
\begin{align}
    \mathcal{F}_{CZ} &= \frac{1}{d}[(d-1)\left(\frac{\rho_{2-CQB}}{\lambda_{2-CQB}}\right)+1]\text{, } d=4\nonumber\\
    \mathcal{F}_{CZ-leakage} &= \frac{\rho_{2-leakage}}{\lambda_{2-leakage}}.\nonumber
\end{align}

\subsection{Single-CQB Clifford primitives}

CQBs experience an always-on $Z$-gate when they idle, which is a high-fidelity operation. We minimize the average number of primitives per Clifford, and the fidelity of each Clifford, by increasing the use of $Z$-gates relative to other works~\cite{Barends_Martinis_nature2014}
\begin{center}
\begin{tabular}{l l | l l | l l}
Clifford & Gate primitives & Clifford & Gate primitives & Clifford & Gate primitives\\
\hline
\hline
1 & $I$ & 9 & $Y(\pi/2)Z(\pi/2)$ & 17 & $Z(\pi/2)$\\
2 & $X(\pi/2)X(\pi/2)$ & 10 & $Y(\pi/2) Z(3\pi/2)$ & 18 & $Z(3\pi/2)$\\
3 & $Y(\pi/2)Y(\pi/2)$ & 11 & $Y(\pi/2)Z(3\pi/2)$ & 19 & $Z(\pi)Y(\pi/2)$\\
4 & $Z(\pi)$ & 12 & $Y(\pi/2)Z(\pi/2)$ & 20 & $Z(\pi)Y(-\pi/2)$\\
5 & $X(\pi/2)Z(3\pi/2)$ & 13 & $X(\pi/2)$ & 21 & $X(-\pi/2)Z(\pi)$\\
6 & $X(\pi/2)Z(\pi/2)$ & 14 & $X(-\pi/2)$ & 22 & $X(\pi/2)Z(\pi)$\\
7 & $X(-\pi/2)Z(\pi/2)$ & 15 & $Y(\pi/2)$ & 23 & $X(\pi/2)X(\pi/2)Z(3\pi/2)$\\
8 & $X(-\pi/2)Z(3\pi/2)$ & 16 & $Y(-\pi/2)$ & 24 & $Z(3\pi/2)X(-\pi/2)X(-\pi/2)$\\
\end{tabular}\\
Table S2: Clifford gate primitives
\end{center}

\subsection{Example randomized benchmarking sequences}
\begin{figure}
\includegraphics[width=7 in]{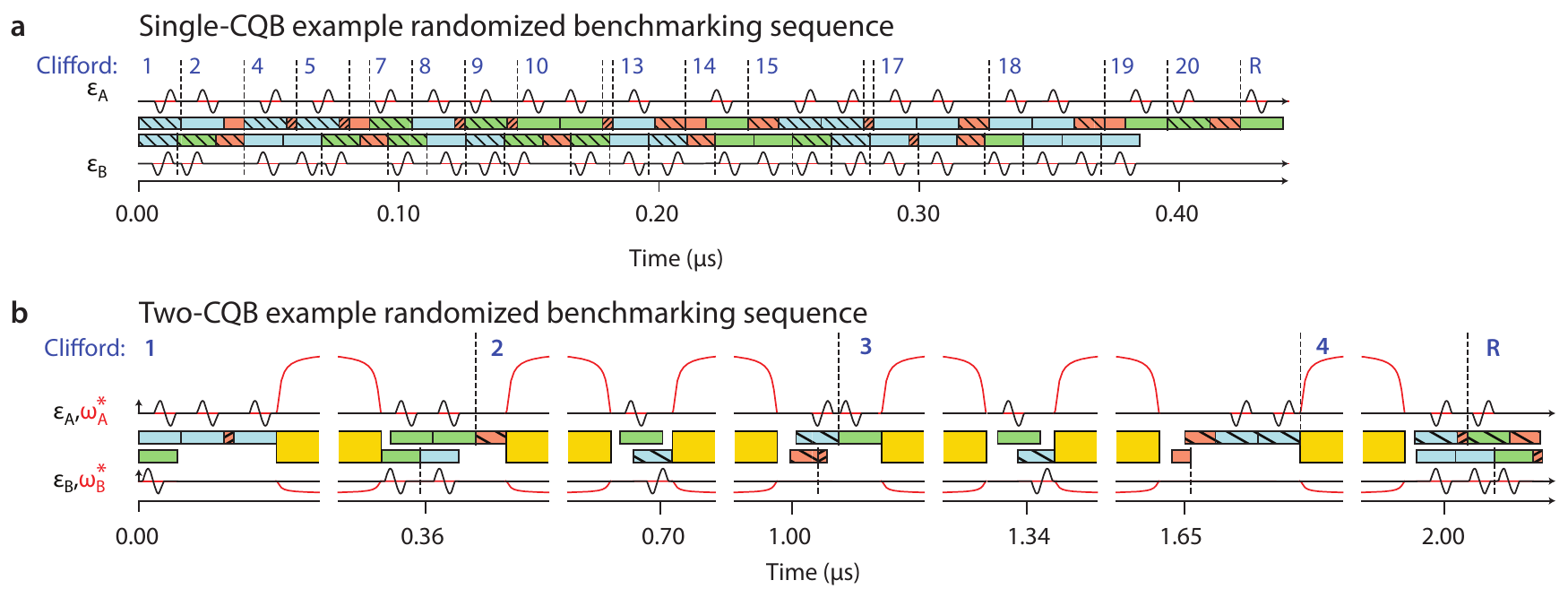}
\label{fig:RBexample}
\caption{A sample randomized benchmarking
    \textbf{a)} 20~Clifford sequence followed by a recovery Clifford is given for a single-CQB architecture and a
    \textbf{b)} four Clifford sequence followed by a recovery gate is given for a two-CQB architecture. Gate primitives are encoded as colored and patterned rectangles (definitions are provided by Figs.~2 and 4 in the main text) lain back-to-back. Above and below these are traces of the baseband flux control that were applied to CQB-A and CQB-B to produce these sequences of Cliffords. The boundaries between Cliffords are given by vertical dotted lines. \label{fig:RBexample}}
\end{figure}
The peculiarities of implementing gates on a CQB architecture become apparent when performing randomized benchmarking.
Using the mapping for Clifford gate primitives in Table~S2, random Cliffords applied sequentially to both CQBs do not synchronize in time, as shown in Fig.~S6\textbf{a}.
For traditional microwave gates, the phase of the pulse's carrier defines the axis of the rotation and is adjusted dynamically to change axes.
Here, the axis of an $XY$ pulse is defined by the relative timing between the single-period sinusoidal drives in CQB pulses.
While the relative timing has periodicity with $t_{\Delta}$, there is some unavoidable granularity in the timing of gates. %
Granularity in the timing of pulses is solved by compensatory $Z$ pulses and enables CZ gates on two CQBs as discussed in the main text and in Fig.~4\textbf{b}.
An example of a sequence of two CQB Cliffords is given in Fig.~S6\textbf{b}.

\end{document}